\documentclass[12pt,preprint]{aastex}

\newcommand{\bvec}[1]{\textbf{\textit{#1}}}

\shorttitle{A Study of MS 1054-0321}
\shortauthors{Jee et al.}

\begin{document}

\title{HST/ACS WEAK-LENSING AND CHANDRA X-RAY STUDIES OF THE HIGH-REDSHIFT CLUSTER MS 1054-0321} 
\author{M.J. JEE\altaffilmark{1}, R.L. WHITE\altaffilmark{2},
        H.C. FORD\altaffilmark{1},
        J.P. BLAKESLEE\altaffilmark{1}, 
        G.D. ILLINGWORTH\altaffilmark{3}, D.A. COE\altaffilmark{1}, AND
	K.-V.H. TRAN\altaffilmark{4}	}

\altaffiltext{1}{Department of Physics and Astronomy, Johns Hopkins
University, 3400 North Charles Street, Baltimore, MD 21218; mkjee@pha.jhu.edu}
\altaffiltext{2}{Space Telescope Science Institute, 3700 San Martin Drive, Baltimore, MD 21218.}
\altaffiltext{3}{University of California Observatories/Lick Observatory, University of California, Santa
Cruz, CA 95064.}
\altaffiltext{4}{Institute for Astronomy, ETH Z\"{u}rich, CH-8093 Z\"{u}rich, Switzerland}

\begin{abstract}
We present $Hubble$ $Space$ $Telescope$/Advanced
Camera for Surveys (ACS) weak-lensing and $Chandra$ X-ray analyses of MS 1054-0321 at z=0.83, 
the most distant and X-ray 
luminous cluster in the $Einstein$ Extended Medium-Sensitivity Survey (EMSS).
The high-resolution mass reconstruction through ACS weak-lensing
reveals the complicated dark matter substructure in unprecedented detail, characterized
by the three dominant mass clumps with the four or more minor satellite groups within
the current ACS field.
The direct comparison of the mass map with 
the $Chandra$ X-ray image shows that the eastern weak-lensing substructure 
is not present in the X-ray image and, more interestingly,
the two X-ray peaks are displaced away from the hypothesized merging direction
with respect to the corresponding central and western mass clumps,
possibly because of ram pressure.
In addition, as observed in our previous weak-lensing study of another
high-redshift cluster CL 0152-1357 at $z=0.84$,
the two dark matter clumps of MS 1054-0321 seem to be offset from the galaxy counterparts.
We examine the significance of these offsets and discuss a possible
scenario, wherein the dark matter clumps
might be moving ahead of the cluster galaxies. 
The non-parametric weak-lensing mass modeling gives a projected mass of
$M(r<1~\mbox{Mpc})=(1.02\pm0.15)\times10^{15} M_{\sun}$, where the uncertainty reflects both
the statistical error and the cosmic shear effects.
Our temperature measurement of $T=8.9_{-0.8}^{+1.0}$ keV utilizing the
newest available low-energy quantum efficiency degradation prescription for the
$Chandra$ instrument, together with the isothermal beta description of the cluster
($r_c=16\arcsec\pm15\arcsec$ and $\beta=0.78\pm0.08$),
yields a projected mass of $M(r<1~\mbox{Mpc})=(1.2\pm0.2)\times10^{15} M_{\sun}$, consistent with the weak-lensing result.
\end{abstract}


\section{INTRODUCTION}
Galaxy clusters, the largest gravitationally bound systems in the universe, are believed to be the youngest
physical structures, only recently dissociated from the Hubble expansion. 
They serve as powerful probes of cosmology and large scale structure formation
because they retain a ``memory'' of their assembly history. 
Especially, many known high-redshift
clusters at $z\sim1$ show snapshots of their early stage of formation when the Universe was at approximately
half its present age. Detailed studies of these distant,
still-forming clusters can considerably enhance our understanding of the dynamical evolution of the 
cluster substructure.

Clusters of galaxies are composed of three major components: visible galaxies, a hot intracluster medium (ICM), and dark matter. 
The advent of high-resolution X-ray observatories have enabled remarkable
progress in tracing the complicated thermodynamical structure of the ICM of a galaxy cluster, whereas weak-lensing
particularly empowered by the $Hubble$ $Space$ $Telescope$ (HST) high-resolution observations is starting to yield 
a detailed distribution of dark matter.
And extensive spectroscopic surveys of cluster fields by ground-based facilities with large aperture telescopes provide
invaluable information on the distribution of the member galaxies as well as the
kinematics of the cluster.

Weak-lensing analysis can provide the most direct measurement of the cluster mass distribution
without any assumption about the dynamical phase of the cluster. However, 
under the simplified hypothesis that the cluster is in quasi-equilibrium, 
both the kinematics and the X-ray analyses can also independently 
be used to derive the mass properties of the target cluster.
 
A joint analysis of weak-lensing and X-ray measurements is a unique opportunity
to investigate the interplay between the ICM and dark matter within the cluster.
Because of their different natures, it is expected that these two components
have somewhat dissimilar distributions within the cluster. The ICM, due to
its collisional property, is subject to ram pressure, and 
shocks can be detected between merging sub-clusters where the transverse velocities
exceed the sound speed. In addition, the frequent
collisions between gas particles should evolve the system
into a high degree of virialization on a relatively short time scale.
On the other hand, cold dark matter, due to its hypothesized collisionless
nature, is often believed to be traced by the cluster galaxies whose
approximate dynamical behaviors are also collisionless. 

However, it is very difficult to probe and test these theoretical speculations in detail
for a distant cluster whose angular extent is only a few arcminutes. The
substantial decrease in gravitational distortion of source galaxy images, because of the large
distance between the observer and the lens, also requires a precise
measurement of a galaxy shape with a careful removal of the instrumental artifacts.

Weak-lensing studies of high-redshift clusters benefit remarkably from the recent installation of 
the Advanced Camera for Surveys (ACS) on $HST$, which can resolve faint, more highly 
distorted distant galaxies (Jee et al. 2005, hereafter J05; Lombardi et al. 2005).
Because of a high number density of background galaxies 
in the cluster field, the resulting mass reconstruction is capable of tracing 
the dark matter distribution in great detail.

In the current paper, we present the first weak-lensing study of MS 1054-0321 using the 
ACS observations 
and a reanalysis of the archival $Chandra$ X-ray data. MS 1054-0321, the most distant and X-ray 
luminous cluster in the $Einstein$ Extended Medium-Sensitivity Survey 
(Gioia et al. 1990; Henry et al. 1992; Gioia \& Luppino 1994), has 
received profuse attention during the past decade because of its unusual richness at such a high-redshift
(e.g., Gioia \& Luppino 1994; Luppino \& Kaiser 1997; Donahue et al. 1998; Tran et al. 1999; Clowe et al. 2000; van Dokkum 2000; 
Neumann \& Arnaud 2000; Hoekstra et al. 2000; 
Jeltema et al. 2001; Joy et al. 2001; Vikhlinin et al. 2002; Gioia et al. 2004). 
The mere existence of such a massive cluster at $z\sim0.83$ can impose strong constraints on the cosmological parameters, favoring a low-density
Universe (e.g., Bahcall \& Fan 1998; Donahue et al. 1998; Jeltema et al. 2001) or non-Gaussianity of the primordial density
fluctuations (Willick 2000).
The first $HST$ weak-lensing study of the cluster was
performed by Hoekstra, Franx, \& Kuijken (2000, hereafter HFK00) using WFPC2 observations. 
Their mass reconstruction revealed a complicated
mass distribution characterized by three dominant clumps, in good spatial agreement with the cluster galaxy 
concentrations. 
However, the X-ray studies (e.g., Neumann \& Arnaud 2000; Clowe et al. 2000; Jeltema et al. 2001) have demonstrated there are only 
two distinct X-ray peaks corresponding to the central and western clumps in the HFK00 mass distribution. The question whether
the absence of the third peak is due to the low X-ray emission of the eastern clump or to an artifact in the weak-lensing
mass reconstruction was raised (Jeltema et al. 2001). Because the Wide Field Channel (WFC) of ACS has twice the sampling resolution
of the Wide Field (WF) chips of WFPC2 and a factor of 5 improvement in sensitivity, the use of ACS observations 
in the weak-lensing analysis can substantially improve the signal-to-noise ratio of the resulting mass reconstruction, therefore enabling
us to test the significance of the cluster substructure presented by HFK00. 

The X-ray study of MS1054 has an interesting history. Donahue et al. (1998) determined the temperature of the cluster to be
very high ($T=12.3^{+3.1}_{-2.2}$keV) analyzing both $ASCA$ and $ROSAT$ data. Using $Chandra$ observations, Jeltema et al. (2001)
obtained a slightly lower temperature $T=10.4^{+1.7}_{-1.5}$keV for the entire cluster. 
Joy et al. (2001) estimated the temperature of the cluster to be $T=10.4^{+5}_{-2}$keV from the Sunyaev-Zeldovich measurements.
However, Vikhlinin et al. (2002) and Tozzi et al. (2003), using the same $Chandra$ data, obtained a somewhat lower temperature 
of $T=7.8\pm0.6$keV 
and $T=8.0\pm0.5$keV, respectively. A even lower temperature of $T=7.2^{+0.7}_{-0.6}$ keV is reported by Gioia et al. (2004) from
the XMM-Newton observations.
With the possibility of time variability excluded, these different temperature estimates simply demonstrate that
the X-ray temperature measurement is sensitive to the analysis procedure as well as the maturity of the calibration of 
the instrument.
For example, Gioia et al.(2004) were able to reproduce all the results of Jeltema et al. (2001) by
duplicating the previous analysis procedures of the $Chandra$ data. However, when they altered the flare interval removal 
(resulting in shorter exposure time) 
and corrected for the quantum efficiency degradation, their results became consistent with the XMM-Newton measurements.

Despite these already copious X-ray studies in the literature, we present our reanalysis of the $Chandra$ archival data 
of MS 1054-0321 for the following two reasons. First, our understanding of the $Chandra$ instrument is still evolving
and we find that the application of the new calibration data affects the derived physical quantities of the cluster.
Second, we attempt to make a direct comparison between the results of the weak-lensing and the X-ray studies in terms of
the total gravitating mass as well as the cluster substructure.
Our previous ACS weak-lensing study (J05) of another high-redshift cluster CL 0152-1357 at a very similar redshift ($z\sim0.84$)
revealed a significant substructure interestingly
offset from the X-ray and cluster galaxies, suggestive of an on-going merger of the two main clumps. Because MS 1054-0321
also possesses distinct substructure well traced by $Chandra$ and weak-lensing analyses, the current
study is a good test to see whether we can detect similar offsets between the dark matter, ICM, and cluster galaxies.
 
Throughout the paper we assume a $\Lambda$CDM cosmology where $\Omega_M=0.27, 
\Omega_{\Lambda}=0.73,$ and $H_0 = 71$km $\mbox{s}^{-1}$. All the quoted uncertainties are at the 68\% level.  

\section{OBSERVATION}
\subsection{ACS Data Reduction}
\begin{figure}
\plotone{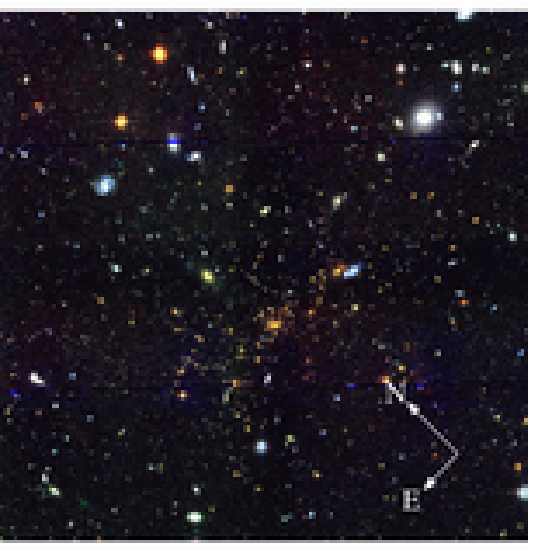}
\caption{Color composite image of MS 1054-0321. The $v_{606}$, $i_{775}$, and $z_{850}$ intensities are
represented by blue, green, and red colors, respectively.
The red early-type 
galaxies of the cluster appear to form a filamentary structure, mainly stretched east-west.
\label{fig_color_composite}}
\end{figure}

The ACS/WFC images of MS 1054-0321 (Goto et al. 2005; Postman et al. 2005) consist of $2\times2$ pointings observed in F606W, F775W, and F850LP 
(hereafter $v_{606}$, $i_{775}$, and $z_{850}$, respectively). The total exposure time per pointing
is 2025 s for $v_{606}$, and 4440 s for $i_{775}$ and $z_{850}$. There is a $\sim1\arcmin$
overlap between pointings. After the raw images were processed with the standard STScI
CALACS pipeline, we used the ``APSIS'' ACS Guaranteed Time Observation (GTO) pipeline (Blakeslee et al. 2003) to perform
cosmic ray rejection, geometric distortion correction, and image registration.
The APSIS pipeline calculates image offsets by matching astronomical objects between pointings after
applying a geometric distortion correction (Meurer et al. 2003). The Lanczos3 interpolation kernel, which
creates less noise correlation and a sharper point-spread-function (PSF), is adopted. We present the 
color composite image of the cluster in Figure~\ref{fig_color_composite}, which shows the $v_{606}$, $i_{775}$, and $z_{850}$ intensities
with blue, green, and red colors, respectively.
The main structure is notably delineated by the swath of the cluster red sequence. This orientation
of the image indicated by the compass is maintained for all the subsequent figures
throughout the paper.

\begin{figure}
\plotone{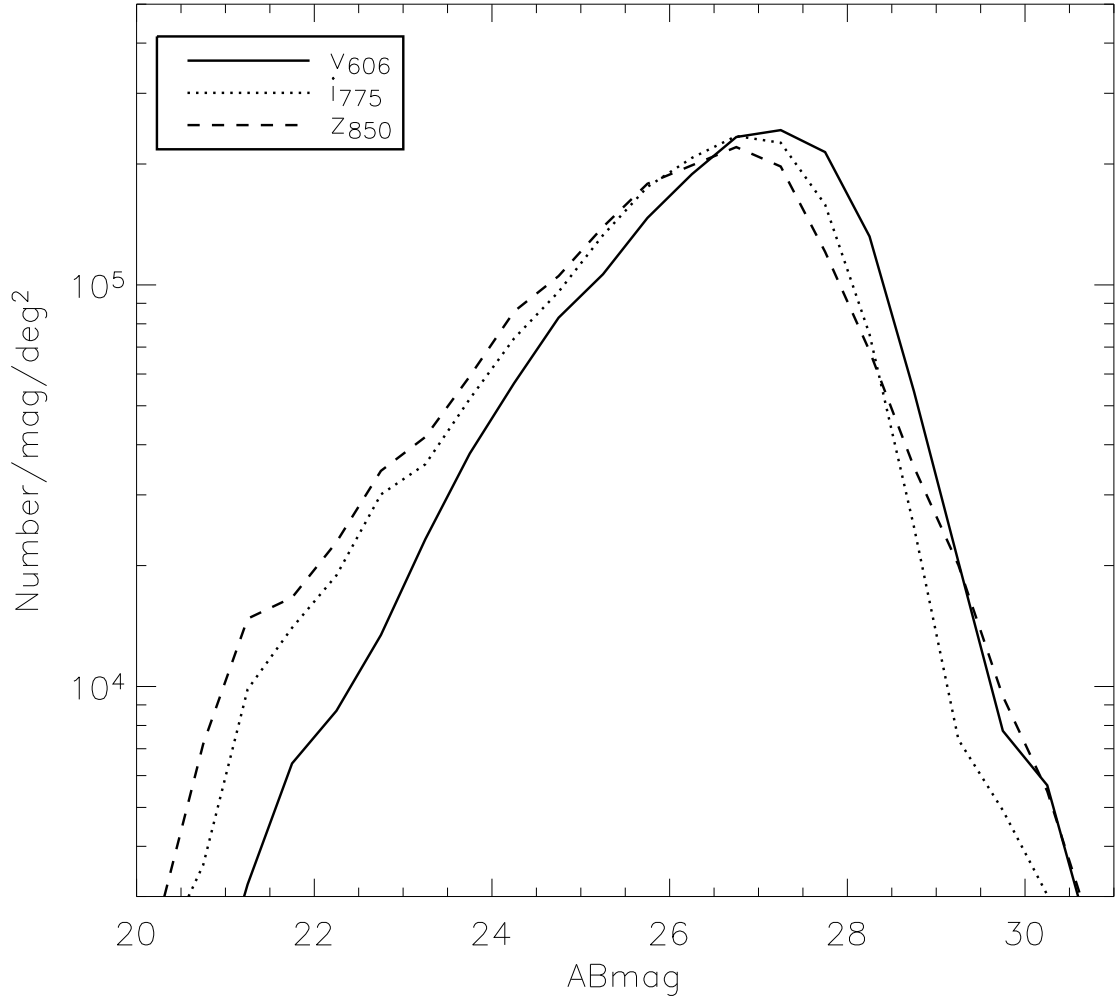}
\caption{Magnitude distribution of detected objects in MS 1054-0321 field. 
Galaxies were detected using
the SExtractor program by searching for
at least five connected pixels above 1.5 times the sky rms. 
The detection looks complete down to $\sim27$th mag in every filter. 
 \label{fig_mag_distribution}}
\end{figure}

Objects were detected using the SExtractor algorithm (Bertin \& Arnouts 1996) by searching for at least five connected pixels above
1.5 times the local sky rms. SExtractor was executed on the detection image created by inverse variance weighting
of all the three passband images. The final catalog is compiled after visually identifying and removing
false objects such as saturated stars and CCD bleeding, diffraction spikes, clipped objects, HII regions in nearby
galaxies, etc. Throughout the paper, we use SExtractor's MAG\_AUTO
and MAG\_ISO magnitudes for object's magnitudes and colors, respectively.
We determined the Galactic extinction (at $\alpha=10^h56^m53^s.59$, $\delta=-03\degr37\arcmin45\arcsec.5$)
to be E(B-V)=0.035 from the dust maps of Schlegel, Finkbeiner, \& Davis (1998) and adjusted the
$v_{606}$, $i_{775}$, and $z_{850}$ magnitudes by 0.103, 0.071, and 0.052, respectively.
Figure~\ref{fig_mag_distribution} shows the magnitude distribution of all the objects in this final catalog.

\subsection{Chandra Data Reduction}
The X-ray observations of MS 1054-0321 were retrieved from the $Chandra$ X-ray Center (at http://cxc.harvard.edu/cda/).
The cluster was observed on 2000 April 21-22 for $\sim90,981$s with the back illuminated S3 chip of the
Advanced CCD Imaging Spectrometer (ACIS-S3) at the focal plane temperature of -120$\degr$C.
We processed the raw X-ray events with the $Chandra$ Interactive Analysis
of Observations (CIAO) software version 3.2 and the Calibration Database version 3.0 (CALDB), discarding 
bad events and $ASCA$ grades 1, 5, and 7. A new level-2 event file was created after
the application of the charge transfer inefficiency (CTI) and the time-dependent gain corrections.

The background light curve was extracted from a circular annulus 
($r=123\arcsec-151\arcsec$) centered at the assumed cluster center 
($\alpha=10^h56^m58^s.6$ and $\delta=-03\degr37\arcmin36\arcsec .7$) after masking out all the point sources
detected by the CIAO wavelet detection program ``wavdetect''. 
We used the $lc\_clean$ script (Markevitch 2003) to identify and remove the flare events, and the net
exposure decreased to a total of $\sim74,456$s. The same annulus was also taken for the background spectrum construction
in spectral analyses.

\section{WEAK-LENSING ANALYSIS}

\subsection{Luminosity of MS 1054-0321 \label{section_luminosity}}

\begin{figure}
\plotone{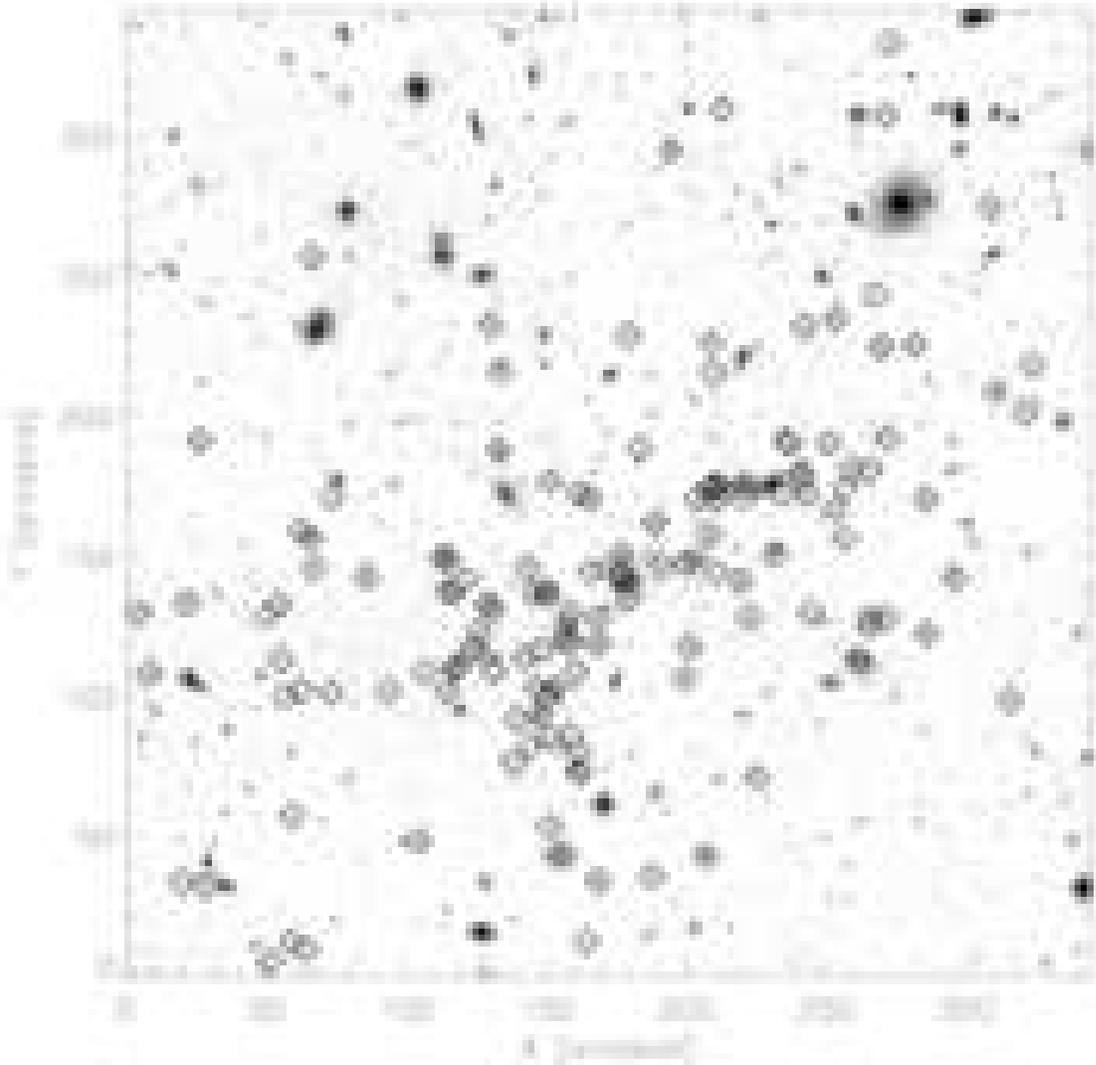}
\caption{Spectroscopically confirmed cluster members. The galaxies whose spectroscopic redshifts are between $0.81 < z < 0.85$ are
marked with squares.
 \label{fig_spec_member}}
\end{figure}

\begin{figure}
\plotone{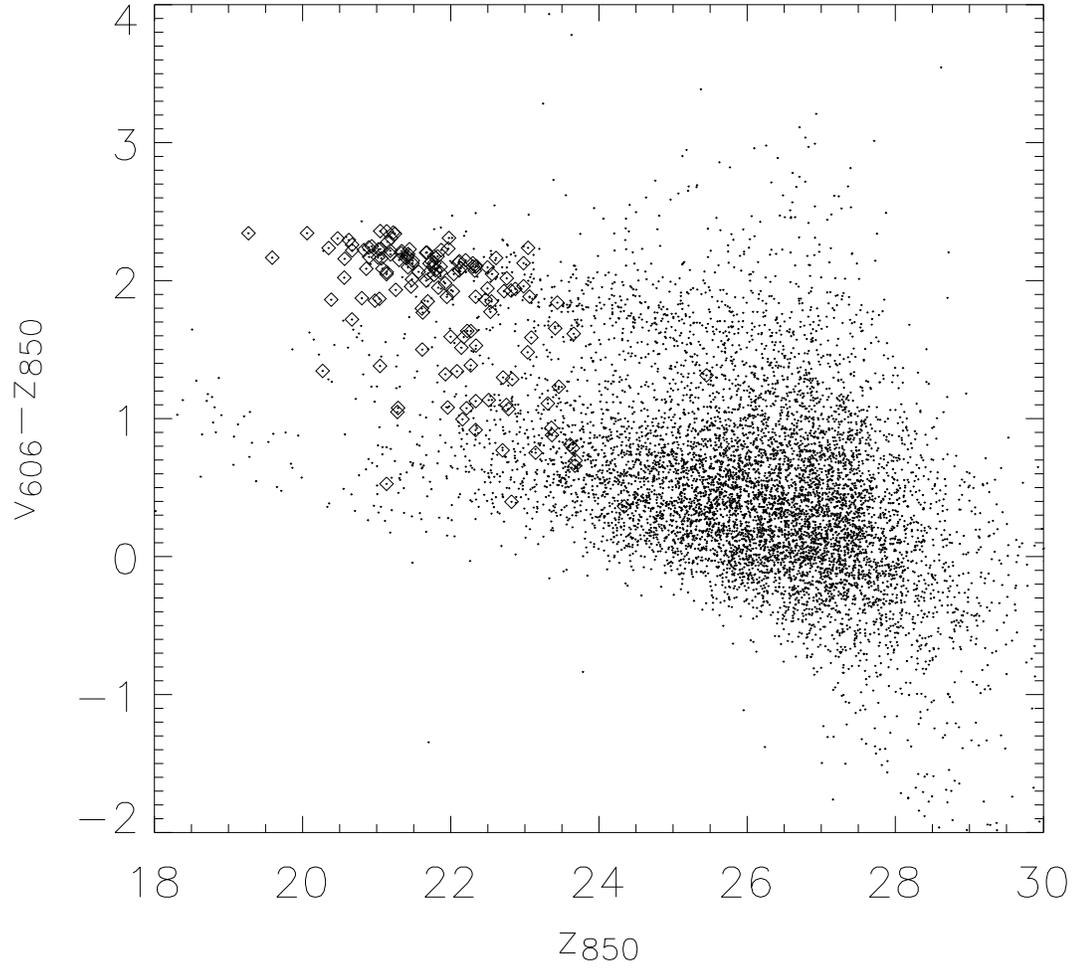}
\caption{Color magnitude diagram. 
Because of the location of the redshifted 4000\AA ~break, the early-type galaxies
are well separated by their $v_{606}-z_{850}$ colors. Square symbols represent the spectroscopically confirmed
cluster members. 
 \label{fig_CM_spec}}
\end{figure}

We combine the extensive spectroscopic catalog of the MS 1054-0321 field (van Dokkum et al. 2000; Tran et al. in preparation) 
with the color-magnitude relation based on the ACS photometry to select the cluster members and 
estimate the rest frame B band luminosity of the cluster. 
We define a spectroscopically confirmed member as a galaxy whose redshift lies between $0.81 < z < 0.85$, which results
in a total of 143 objects (Figure~\ref{fig_spec_member}). We assume that the sample is almost 
complete for galaxies brighter than $z_{850}=22$ 
regardless of galaxy colors. For the intermediate magnitude range ($22<z_{850}<25$), we used the color-magnitude relation of the early type 
galaxies (Figure~\ref{fig_CM_spec}) while rejecting spectroscopically inconsistent members.
Using the spectroscopic catalog of the field,
we estimated that the blue cluster galaxies not included in
this selection comprise nearly 50\% of the cluster luminosity in this magnitude range. 
The contribution from the faint galaxies ($z_{850}>25$) was determined using the best-fit
Schecter luminosity function. Adopting $M^{\star}=-21.47\pm0.29$ and $\alpha=-0.87$ (Goto et al. 2004), we found that
the light from this faint population accounts for $\sim4$\% of the final total luminosity.

The conversion of $i_{775}$ to the rest frame B is derived by performing a synthetic photometry using the Spectral Energy
Distribution (SED) templates of Kinney et al. (1996). J05 established a similar transformation for
Cl 0152-1357 at $z\simeq0.837$, and here we extend the work of J05 by introducing the quadratic color term.

\begin{eqnarray}
B_{rest} &=& i_{775} - (0.159\pm0.163)(i_{775}-z_{850})^2 - (0.224\pm0.177)(i_{775}-z_{850}) 
\nonumber\\
&\:& +(0.728\pm0.039) - DM,
\end{eqnarray}
\noindent
where DM is the distance modulus of 43.605 at $z=0.83$ and the uncertainties are computed by assuming
2\% photometric errors. Then the rest-frame B magnitude is converted to the luminosity by
\begin{equation}
\frac{L_B}{L_{B\sun}} = 10^{0.4(M_{B\sun} - B_{rest})},
\end{equation}
\noindent
where $M_{B\sun}=5.48$ is the absolute B magnitude of the Sun.

\begin{figure}
\plotone{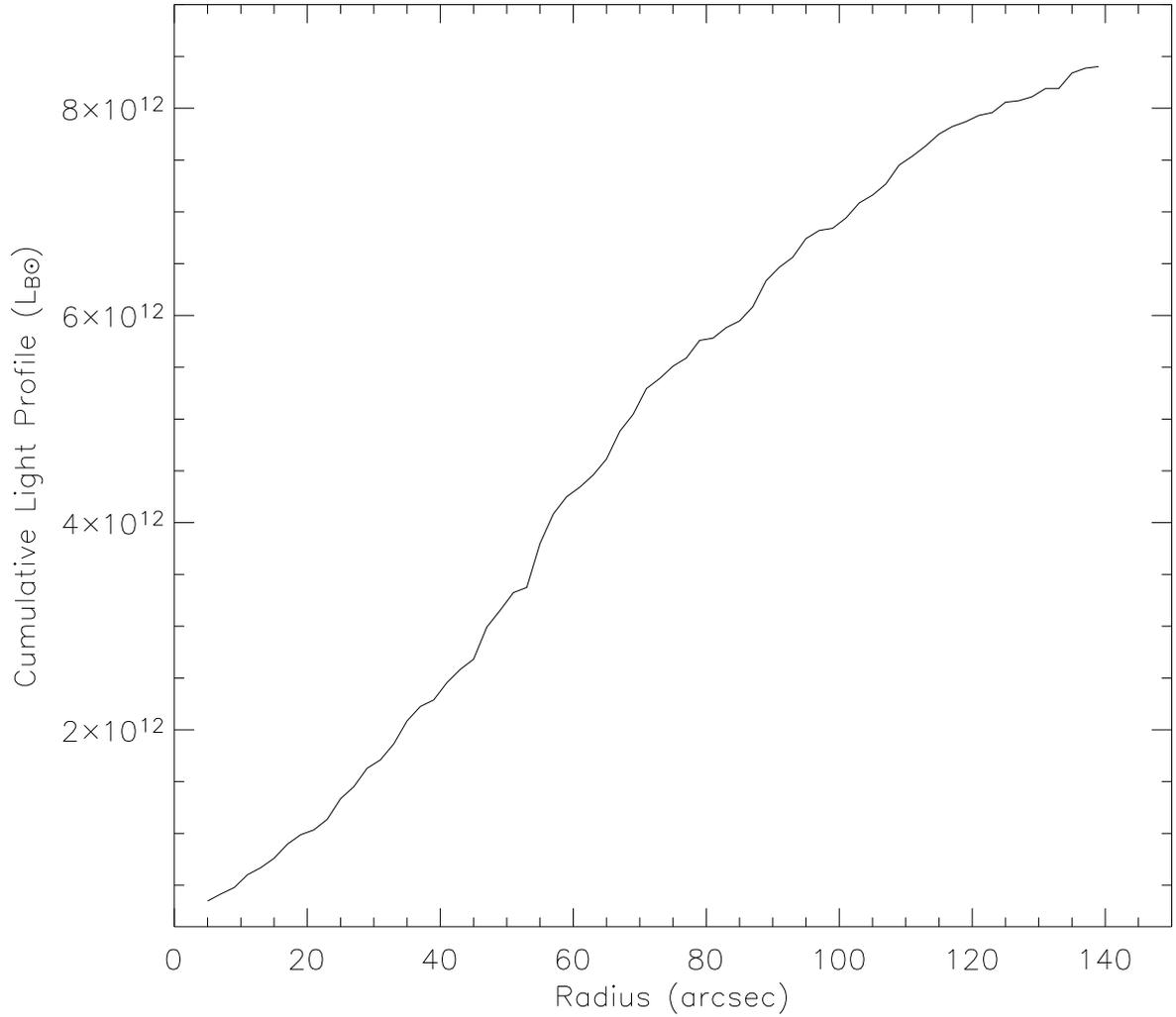}
\caption{Cumulative B band luminosity of MS 1054-0321. We used the spectroscopic catalog in conjunction with
the tight CM relation of the early-type galaxies for member selection (see text for details). Within 1 Mpc ($\sim132\arcsec$)
the cumulative B band luminosity is estimated to be $\sim8.2\times10^{13} L_{B\sun}$.
 \label{fig_lp}}
\end{figure}

Figure~\ref{fig_lp} shows the cumulative light profile of MS 1054-0321. Within 1 Mpc ($\sim132\arcsec$) radius
the total luminosity of the cluster is $8.2\times10^{12} L_{B\sun}$. HFK00 estimated
$1.0\times10^{13} L_{B\sun}$ for $\Omega_M=0.3, \Omega_{\Lambda}=0,$ and $H_0 = 50 \mbox{km~s}^{-1}$ cosmology.
Iterating the above procedure with their cosmological parameters ($DM=44.09$) for comparison yields a total luminosity of $1.1\times10^{13} L_{B\sun}$
within 1 Mpc ($\sim105\arcsec$), which is slightly higher than the HFK00 estimation by $\sim$10\%.

\subsection{Correction for Instrumental Effects}
Because weak-lensing is based on a statistical analysis of a population of very weakly distorted 
galaxy images, any instrumental artifact, usually of a comparable or even higher order of
magnitude than the signal strength one is looking for, must be subtracted from first-hand measurements of object shapes.
The geometric distortion of ACS, primarily due to the location of the instrument far from the telescope axis, is
more significant than any of the previous $HST$ instruments. The uncorrected WFC image possesses an elongation
of $\sim8$\% along the diagonal, thus causing the sky-projected pixel to appear rhombus-shaped. APSIS corrects
the geometric distortion of WFC images using the latest detector distortion model (Meurer et al. 2003; Anderson 2002), which
is derived from the observations of the core of 47 Tucanae. A fourth-order polynomial is used to characterize
the distortion to an accuracy much better than 0.2 pixels over the entire field of view (Pavlovsky et al. 2004).
Precise image alignment is also a principal requirement in a weak-lensing analysis because a slight
registration error can imitate gravitational shears, inducing a coherent elongation of object shapes. APSIS calculates
the image offsets by matching astronomical objects between different exposures. The typical shift uncertainty of
$\sim0.015$ pixels satisfies our analysis requirement. 

The point-spread-function (PSF) of the WFC varies across the field even in the rectified image,
and both the magnitude and orientation of the ellipticity change with the focus offset of the instrument (Krist 2003).
In the previous work (J05), we demonstrated that the PSF model constructed from the 47 Tucanae observations can excellently describe
the PSF pattern of the cluster field when a slight adjustment of the ellipticity is made. 
The PSF of ACS was decomposed via $shapelets$ (Bernstein \& Jarvis 2002, hereafter BJ02; Refregier 2003) and the
spatial variation was modeled by polynomial interpolation of the shapelet coefficients (see J05 for details).

Initial selection of stars in the MS 1054-0321 field was made using the half-light radius versus magnitude plot. After
discarding defective stars by visual inspection, the list contained 94 good stars. We matched these stars in the cluster field
to our model PSF derived from the 47 Tucanae field. The overall agreement is satisfactory, but there exist
tiny, but systematic residuals, which are suspected to arise from the focus breathing of the instrument.
We were able to further reduce these systematics by introducing the following ellipticity fine-tuning:

\begin{equation}
b_{pq}^{\prime} = \mbox{\bf{S}}_{\delta \eta} b_{pq},
\end{equation}
\noindent
where $b_{pq}(b_{pq}^{\prime})$ is the shapelet coefficient of the ACS PSF before (after) the 
adjustment and $\mbox{\bf{S}}_{\delta\eta}$ is the shear operator that
modifies the ellipticity by $\delta \eta$ to improve the agreement.
The evaluation of matrix elements of the shear operator 
$\mbox{\bf{S}}$ can be found in BJ02.
The fine-tuning parameter $\delta\eta$ is held fixed within a field, but allowed to vary between exposures.

\begin{figure}
\plottwo{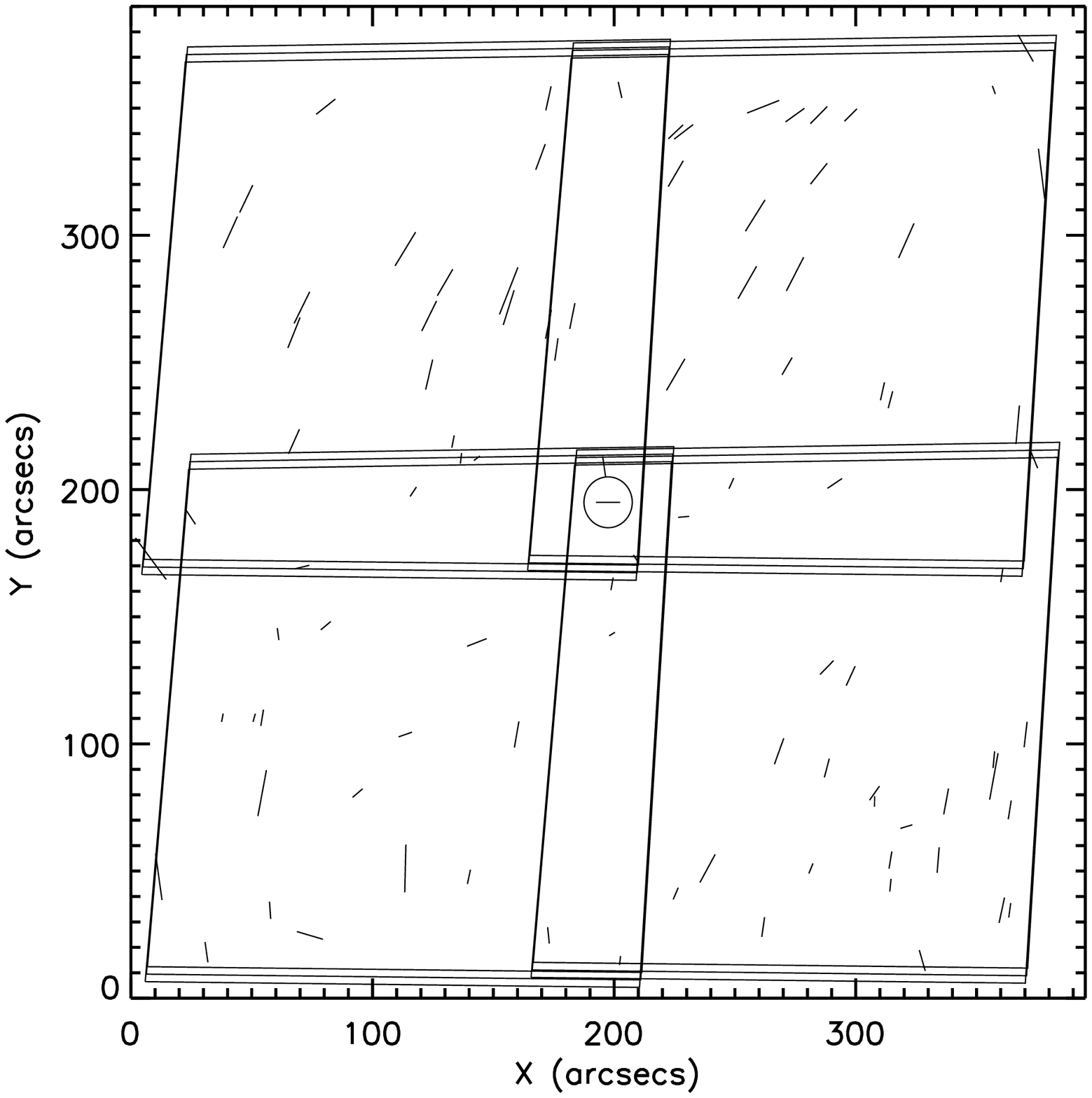}{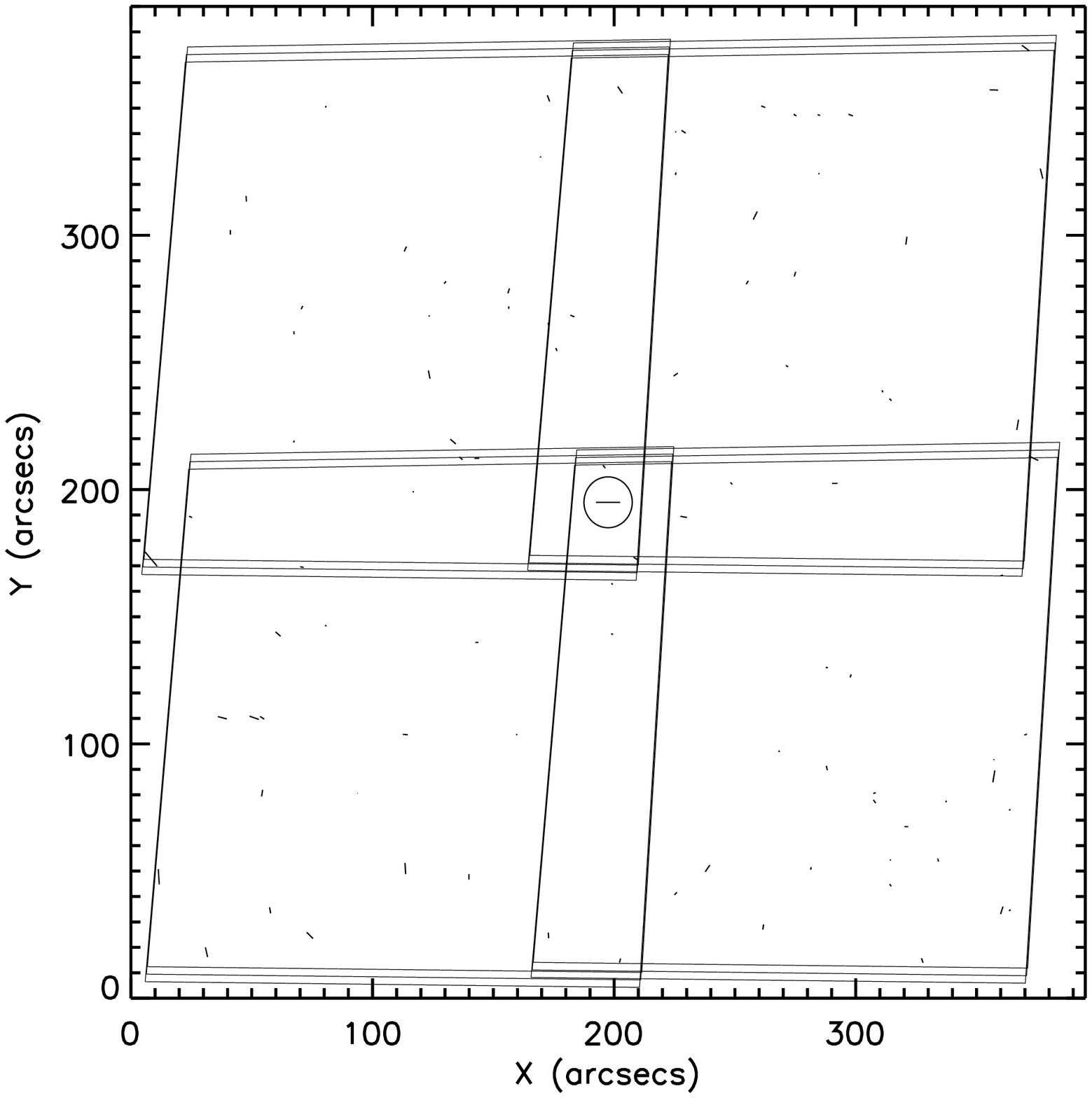}
\caption{PSF anisotropy correction for $v_{606}$ image. The length and orientation of each stick represent
the magnitude and direction of the ellipticity, respectively. The stick in the centers of a) and b)
shows an ellipticity of $\delta=0.1$.
The 2$\times$2 parallelograms illustrate the four pointings of WFC. (a) Stars are elongated
coherently with a typical ellipticity of $\sim$10\%. (b) Rounding kernel reduces the PSF anisotropy remarkably 
($\left<\delta^2\right>^{1/2}\simeq0.02$), 
indicating that our model PSF is an excellent description of the real PSF. 
\label{fig_starfield}}
\end{figure}

\begin{figure}
\plotone{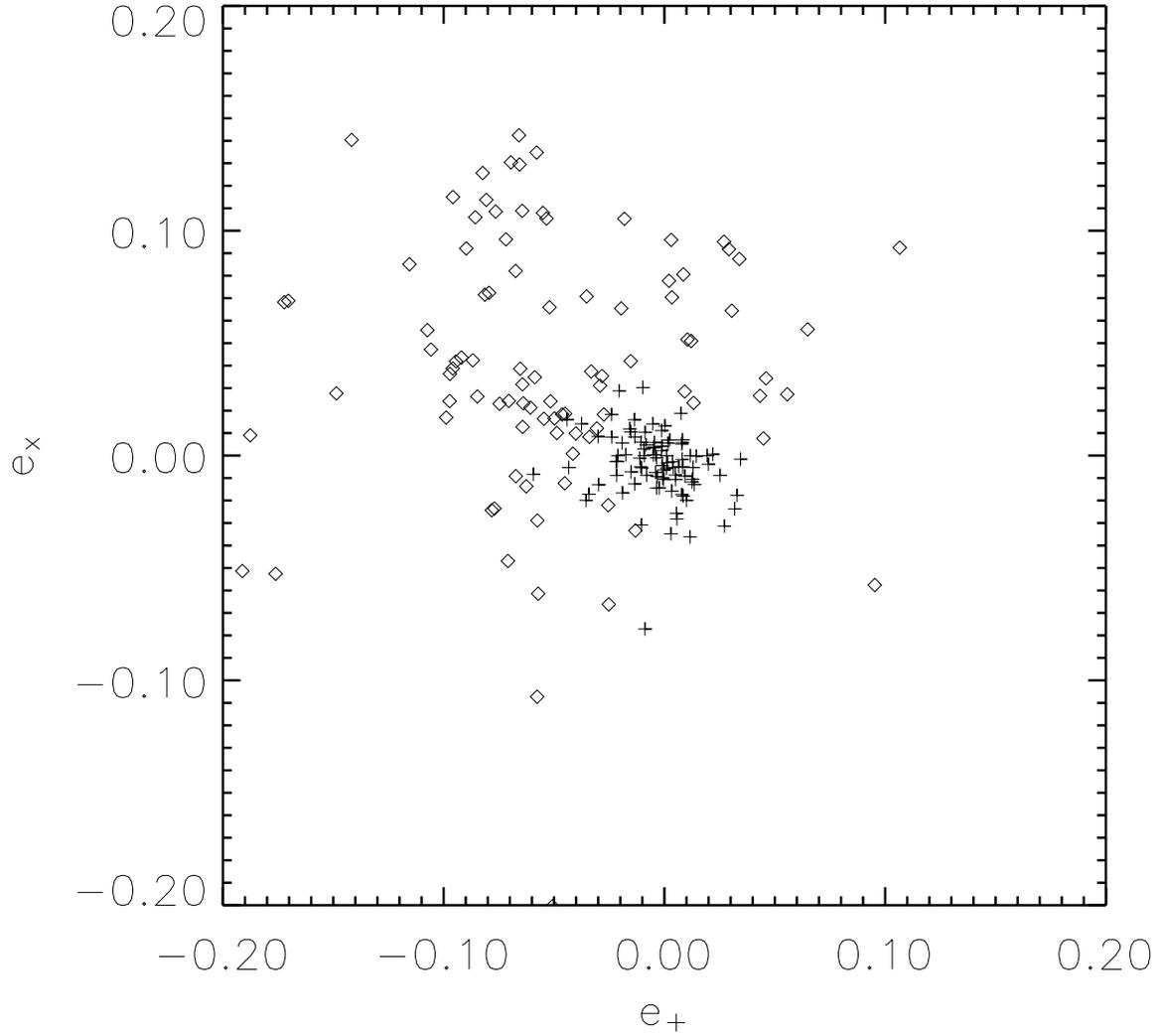}
\caption{PSF anisotropy removal in the $v_{606}$ image. Diamonds show
the initial ellipticities of stars, and plus symbols, the
corrected ellipticities after rounding kernel convolution.
\label{fig_star_anisotropy}}
\end{figure}

Figure~\ref{fig_starfield} and~\ref{fig_star_anisotropy} illustrate the PSF pattern observed in the $v_{606}$ mosaic image of the MS 1054-0321 field
before and after the correction is applied. The correction
is made by constructing a rounding kernel (Fischer \& Tyson 1997; Kaiser 2000; BJ02)
and convolving the stars with this kernel. However, as discussed in J05
(see also BJ02 and Seljak 2004), the 
rounding kernel method has several drawbacks compared to the momentum-based deconvolution. Therefore, we use this
rounding kernel only for the verification of the PSF matching, and the actual correction is
made through the deconvolution in shapelet space.
 
\subsection{Shape Measurement}

Galaxy shapes are decomposed via the following polar shapelets (see Refregier 2003 for Cartesian coordinates):
\begin{equation}
I(r,\theta) = \sum_{p,q \ge 0} b_{pq} \Psi_{pq}^\sigma (r,\theta)
\end{equation}
                                                                      
\begin{equation}
\Psi_{pq}^\sigma (r,\theta) = \frac {(-1)^q }{ \sqrt { \pi} \sigma^2} \sqrt { \frac{q!}{p!}}
   \left ( \frac{r}{\sigma} \right ) ^{(p-q)} e ^{i (p-q) \theta} e^{-r^2 / 2 \sigma ^2} L_q ^{(p-q)}
     ( \frac{r^2}{\sigma^2}) \phantom{xxxxxx} (p \geq q)
\end{equation}

\begin{equation}
\Psi_{qp}^\sigma = \bar{\Psi}_{qp}^\sigma,
\end{equation}
\noindent
where $L_q^{m} (x)$ are the Laguerre polynomials.

Because our model PSF is also compactly described by these shapelets, we are able to
perform the deconvolution through a simple matrix algebra:
                                                                       
\begin{equation}
\bvec{b}^i = \mbox{\bf{P}}^{-1} ~ \bvec{b}^o,
\end{equation}
\noindent
where $\bvec{b}^i$, $\bvec{b}^o$, and $\mbox{\bf{P}}$ are
the deconvolved image, the observed image, and the convolution matrix.

Then, we transform the PSF-corrected image $\bf{\mbox{b}}^i$ by applying
translation, dilation, and shear operators:
\begin{equation}
\bvec{b} ^{i \prime} = ( \mbox{\bf{S}} _\eta \mbox{\bf{D}} _\mu \mbox{\bf{T}}  _z ) \cdot  \bvec{b}^i, \label{eta}
\end{equation}
\noindent
where the $\mbox{\bf{S}}$, $\mbox{\bf{D}}$, and $\mbox{\bf{T}}$ are the shear, dilation, and translation
operators defined in BJ02.

By requiring the above transformation to satisfy the condition $b^{i\prime}_{10}=b^{i\prime}_{11}=b^{i\prime}_{20}=0$, we
can determine the ellipticity $\eta$ when both centroid and significance are optimized (BJ02).                                                   

\subsection{Source Galaxy Selection}

We select our source galaxies whose $v_{606}-z_{850}$ colors are bluer than $-0.123 z_{850} + 4.54$. 
In general, one desires to include as faint galaxies
as possible because fainter galaxies are subject to greater distortion due to their higher cosmological distance
from the lens. However, they also have larger uncertainties in their shape measurement arising from
their poorer photon statistics in addition to greater correction factors for the removal of PSF effect.
In order to establish a cutoff magnitude in our source galaxy catalog, we examined the variation
of the lensing signal for three different magnitude samples: $bright$ ($24<z_{850}<26$), $faint$ ($24<z_{850}<28$), and 
$faintest$ ($24<z_{850}<30$). One useful way to compare the strength of the overall
lensing signal is to inspect the tangential shear defined as:
\begin{equation}
 \gamma_T  = -  \gamma_1 \cos 2\phi - \gamma_2 \sin 2\phi \label{tan_shear},
\end{equation}
\noindent
where $\phi$ is the position angle of the object with respect to the cluster center.
 
\begin{figure}
\plotone{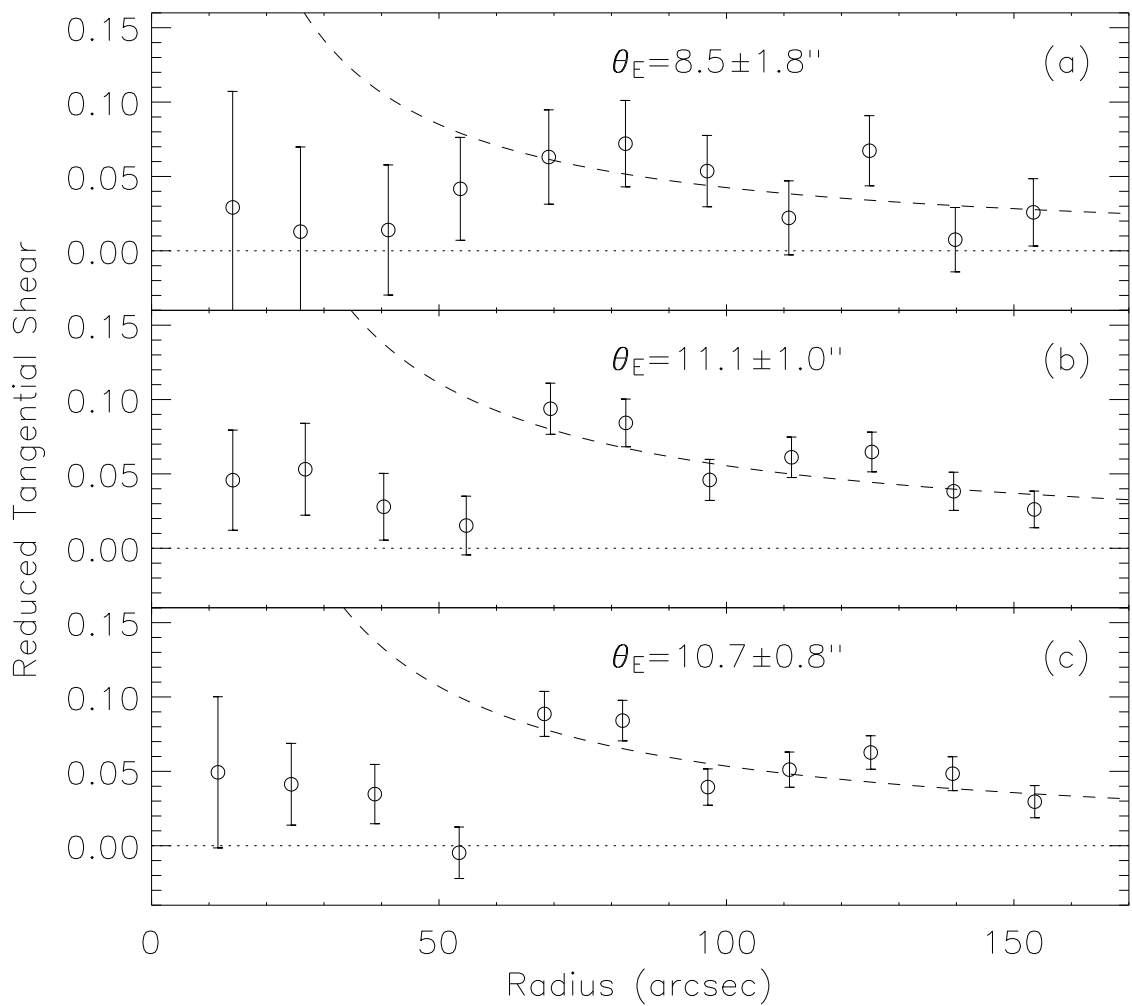}
\caption{Tangential shear for different magnitude samples. Tangential shears are measured for $bright$ (a; $24<z_{850}<26$), $faint$
(b; $24<z_{850}<28$), and $faintest$ (c; $24<z_{850}<30$) source samples. It is found that we can include faint galaxies down
to $z_{850}\sim28$ without diluting the signal.
\label{fig_tan_shear_mag}}
\end{figure}

The amplitude of the tangential shear and the Einstein radius determined from the Singular Isothermal Sphere (SIS) fitting
to each sample suggest
that we can include background galaxies down to $z_{850}=28$ without diluting the signal (Figure~\ref{fig_tan_shear_mag}). 
We excluded the shears $\gamma_T$ at $r<70\arcsec$ in the SIS fitting because we suspect that the measurement
in this region is severely affected by the cluster substructure.

\begin{figure}
\plotone{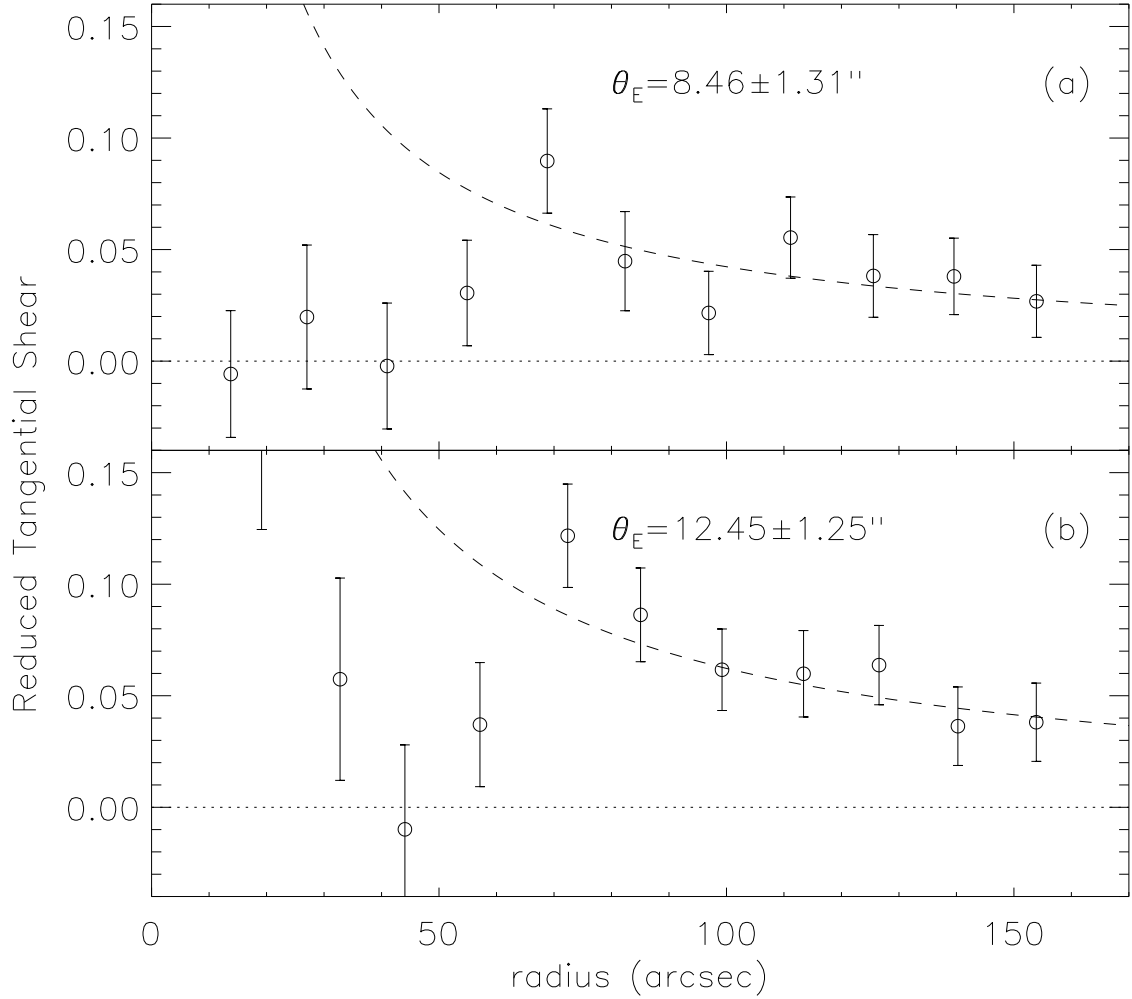}
\caption{Tangential shear for different color samples. The $faint$ sample in Figure~\ref{fig_tan_shear_mag} are further divided
into $faint$ $red$ (top panel) and $faint$ $blue$ (bottom panel) subsamples. There exists a clear dependence of the
signal strength on background galaxy colors.
\label{fig_tan_shear_color}}
\end{figure}

We investigated whether there also exists a strong color dependence in the signal amplitude. The galaxies in the $faint$ sample
are divided into $blue$ $faint$ and $red$ $faint$ galaxies so that the sub-samples contain a roughly equal number of galaxies.
As presented in Figure~\ref{fig_tan_shear_color}, we find the signal from $blue$ $galaxies$ are notably stronger, yielding
an Einstein radius of $\theta_E = 12.45\pm1.25\arcsec$. HFK00 also found that their bluer galaxies create
a larger amplitude in their WFPC2 weak-lensing analysis of MS 1054-0321. HFK00 suggested that their $bluer$ galaxies
have a greater mean redshift. However, it is interesting to
note that in the ACS weak-lensing of CL 0152-1357 
we did not find such a shear excess in the $blue$ sample (J05). Considering the similar depth in both observations,
it appears that the difference originates from the large scale structure of the field. In the following discussion
we regard the $faint$ sample ($24<z_{850}<28$) (i.e., including both red and blue sources) 
as our best source catalog and the results hereafter are based
on this catalog. The catalog contains a total of 5847 galaxies ($\sim154~\mbox{arcmin}^{-2}$).

\subsection{Redshift Distribution of Source Galaxies \label{redshift}}
The redshift distribution of source galaxies must be derived with
extreme care because the critical density $\Sigma_c$ (mass unit)
is a steep function of the source redshift for a high-redshift cluster. 
Consequently, it has been often
considered a major source of uncertainty for weak-lensing mass estimation (e.g., Luppino \& Kaiser 1997). However, the recent availability of a reliable
photometric redshift catalog down to the faint limit of the source sample has remarkably 
stabilized the weak-lensing mass determination up to the cosmic variance.

In the current work, we utilize the photometric redshift
catalog of the Ultra Deep Field (UDF), the deepest available patch of sky to date observed by both ACS and 
the Near Infrared Camera and Multi-Object Spectrometer (NICMOS).
The detailed description of the generation of the photometric redshift catalog will be addressed in a forthcoming
paper (D. Coe et al. in preparation); we give a brief description below.

We performed the PSF matching for different instruments and passbands using the stars present in the UDF field. The
object catalog obtained from the ACS image is complemented with the objects only visible in the NICMOS images. For
the galaxy SEDs, we selected the new templates of the Bayesian Photometric Redshift package (BPZ) (Benitez 2004),
which are composed of the modified E, Sbc, Scd, and Im templates of Coleman, Wu, \& Weedman (1980), two SB2 and SB3 
starburst templates of Kinney et al. (1996), and one synthetic template from Bruzual \& Charlot (2003). The last synthetic
SED was added to account for very young, but faint starburst galaxies, for which a good observed template is not available.

In order to minimize the discrepancy caused by the field variation of the mean redshift of source galaxies, we selected the
UDF galaxies by applying the same criteria as in the cluster field and
determined the following $\beta$ for each magnitude bin:
\begin{equation}
\beta = \left < \mbox{max} ( 0, \frac{D_{ls}} {D_s}) \right > \label{eqn_beta},
\end{equation}
\noindent
where $D_s$ and $D_{ls}$ are the angular diameter distance from the observer to the source
and from the lens to the source, respectively.

Because our cluster observation is shallower than the
UDF, care must be taken not to ignore the incompleteness at faint limits, as well
as the contamination by blue cluster galaxies. Noise was added to
the UDF ACS images to mimic the S/N of the cluster observation, and the ``mock''
catalogs created from this degraded UDF ACS images 
were used to unbias the
estimation of $\left < \beta \right>$ for $z_{850}>26$.
For $z_{850}<26$ galaxies, we used the ACS photometry catalog from the Great Observatories Origins Deep
Survey (GOODS; Giavalisco et al. 2004) as a control field
to estimate the contamination from the blue cluster galaxies.

The final determination of $\left < \beta \right > =0.290$
corresponds to a single source plane at $\left<z\right> \simeq1.325$.
We also found that this single source plane approximation would cause
an overestimation of the reduced shear by $(1 + 0.52 \kappa)$ when 
the first order correction derived by Seitz \& Schneider (1997) and HFK00 was employed.
This correction becomes
important at the dense region of the cluster where the assumption $\kappa<<1$ breaks down.

\subsection{Weak-lensing Mass Estimation \label{mass_estimate}}
First, we consider parameterized models for a description of the mass profile of MS 1054-0321. 
In order to minimize the effect of the cluster substructure,
we use the tangential shears measured from $70\arcsec$ to $140\arcsec$ radii centered at
the location of the BCG.

The SIS fit gives an Einstein radius of $\theta_E = 11\arcsec.1\pm1.\arcsec0$ ($\sim84$ kpc),
yielding an aperture mass of $(9.6\pm0.8) \times 10^{14}M_{\sun}$ within 1 Mpc radius. Under the same SIS assumption, 
this mass is translated to a velocity dispersion of $1150_{-51}^{+49}$ km$s^{-1}$.
A similar
Einstein radius of $11\arcsec.1\pm2\arcsec.4$ 
is obtained when the Singular Isothermal Ellipsoid (SIE) model is fit instead.  
In addition, the position angle of $4\pm11\degr$ with
respect to the x-axis of the image and the axis ratio of $b/a=0.51\pm0.08$ are consistent with
the elongation of the cluster substructure. Fitting an NFW profile to
the tangential shears taken from
$70\arcsec$ to $150\arcsec$ radii yields 
a concentration parameter of $c=3.2\pm2.2$ and a core radius of $r_s = 63\pm28 \arcsec$.
Because these two parameters are not independent (i.e., they can be traded off with each other
and the $\chi^2$ contours are highly elongated in the diagonal direction), a
wide range of the parameter values can describe the observed shear profile without significantly
altering the quality of the fit. Marginalizing with c=3.2, we obtain a projected mass
of $M(r<1 \mbox{Mpc})=(8.9\pm0.4) \times 10^{14} M_{\sun}$.

For parameter-free estimation of the cluster mass profile, we use a conventional aperture mass
densitometry as well as a rescaled ($\kappa \rightarrow \lambda \kappa + (1-\lambda)$)
 mass reconstruction (J05). We constrain
a mean surface mass density of an annulus from $130\arcsec$ to $145\arcsec$ to be $\kappa=0.041+0.005$
utilizing the best-fit SIS parameters. Within 1 Mpc aperture radius, we obtain $(1.02\pm0.04)\times10^{15}M_{\sun}$ 
and $(1.01\pm0.04)\times 10^{15}M_{\sun}$ from aperture mass densitometry and mass reconstruction, respectively. 

\begin{figure}
\plotone{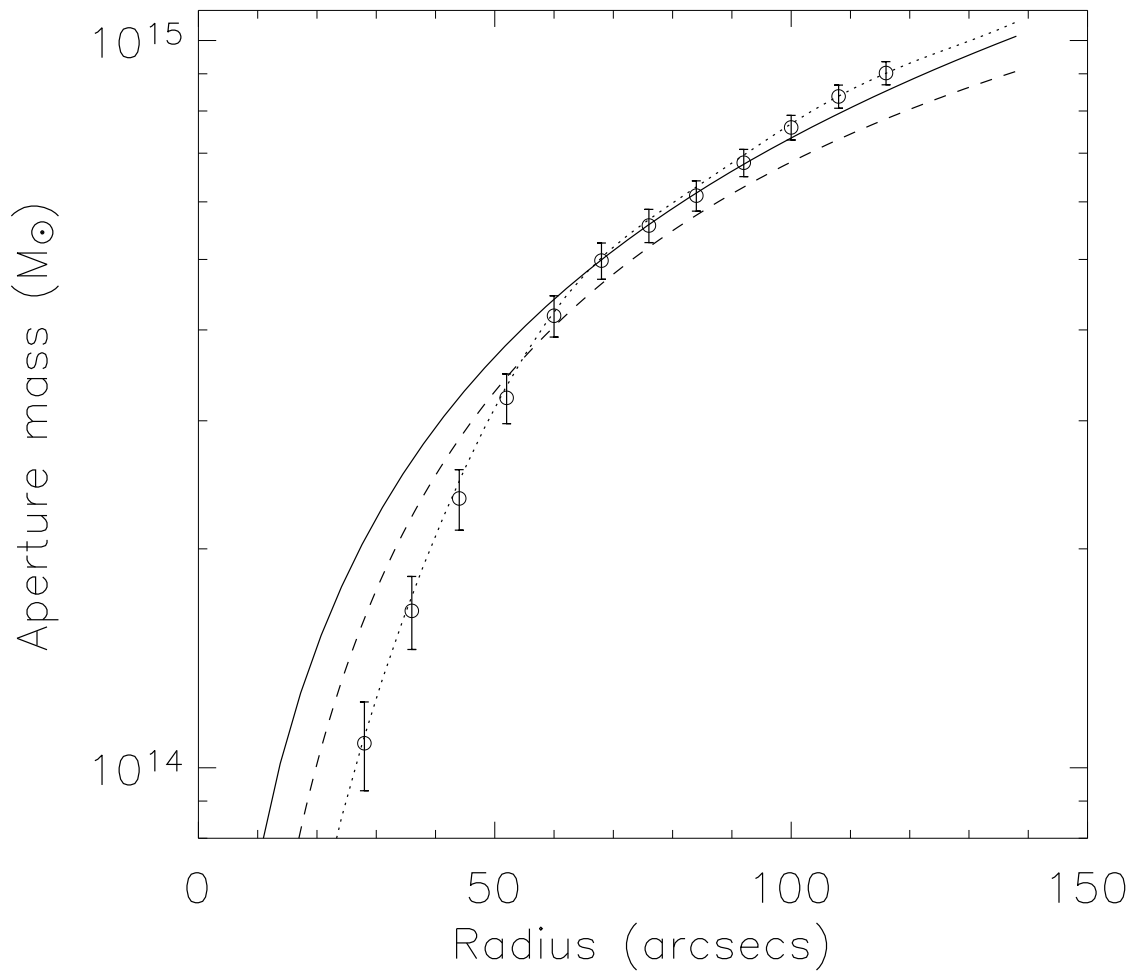}
\caption{Weak-lensing cluster mass estimation. Aperture mass densitometry (open circle with error bars) yields a total aperture mass 
of $(1.02\pm0.04)\times10^{15} M_{\sun}$
within 1 Mpc ($\sim132\arcsec$), which is in good agreement with the measurement $(1.01\pm0.04)\times10^{15} M_{\sun}$
from the rescaled mass map (dotted).
At large radii ($r > 60\arcsec$), these cluster mass profiles are similar to the SIS estimation (solid) nearly out to the field boundary.
The result from the NFW fitting (dashed) predicts a somewhat lower mass ($\sim12$\% at 1 Mpc). 
\label{fig_mass_profile}}
\end{figure}

We compare these various mass profiles in Figure~\ref{fig_mass_profile}. Note that
the mass profile derived from the mass reconstruction agrees nicely with that from aperture mass
densitometry. This remarkable agreement was also observed in the weak-lensing analysis
of CL 0152-1357 in J05, and the use of the rescaled mass map ($\kappa \rightarrow \lambda \kappa + (1-\lambda)$ )
is now encouraged in the mass estimation of the substructure. The best-fit SIS profile describes
the radial distribution of the cluster mass similar to the results from the parameter-free mass estimation
whereas the NFW profile yields somewhat lower mass at large radii ($\sim12$\% at 1 Mpc). 

HFK00 quoted an aperture ($r=1$ Mpc) mass of $(1.07\pm0.12) \times 10^{15} M_{\sun}$ from the measurement
of $\zeta(r)$ statistics in the $(\Omega_M,\Omega_{\Lambda},h)=(0.3,0.7,0.5)$
universe. We find that their result is in good agreement with our  
aperture mass estimate of $(9.91\pm0.42)\times 10^{14} M_{\sun}$ (when our result is reproduced under their cosmological parameters).

A caution must be made regarding the interpretation of the uncertainties quoted above. Although our weak-lensing
analysis benefits considerably from the unprecedentedly high number of background galaxies, the quoted
uncertainties here only reflects the statistical errors. In \textsection~\ref{section_cosmic_shear} we
discuss the importance of the foreground/background cosmic shear effects and estimate how much they
can limit the accuracy of our cluster mass determination.

\subsection{Mass-to-light Ratio Profile}
\begin{figure}
\plotone{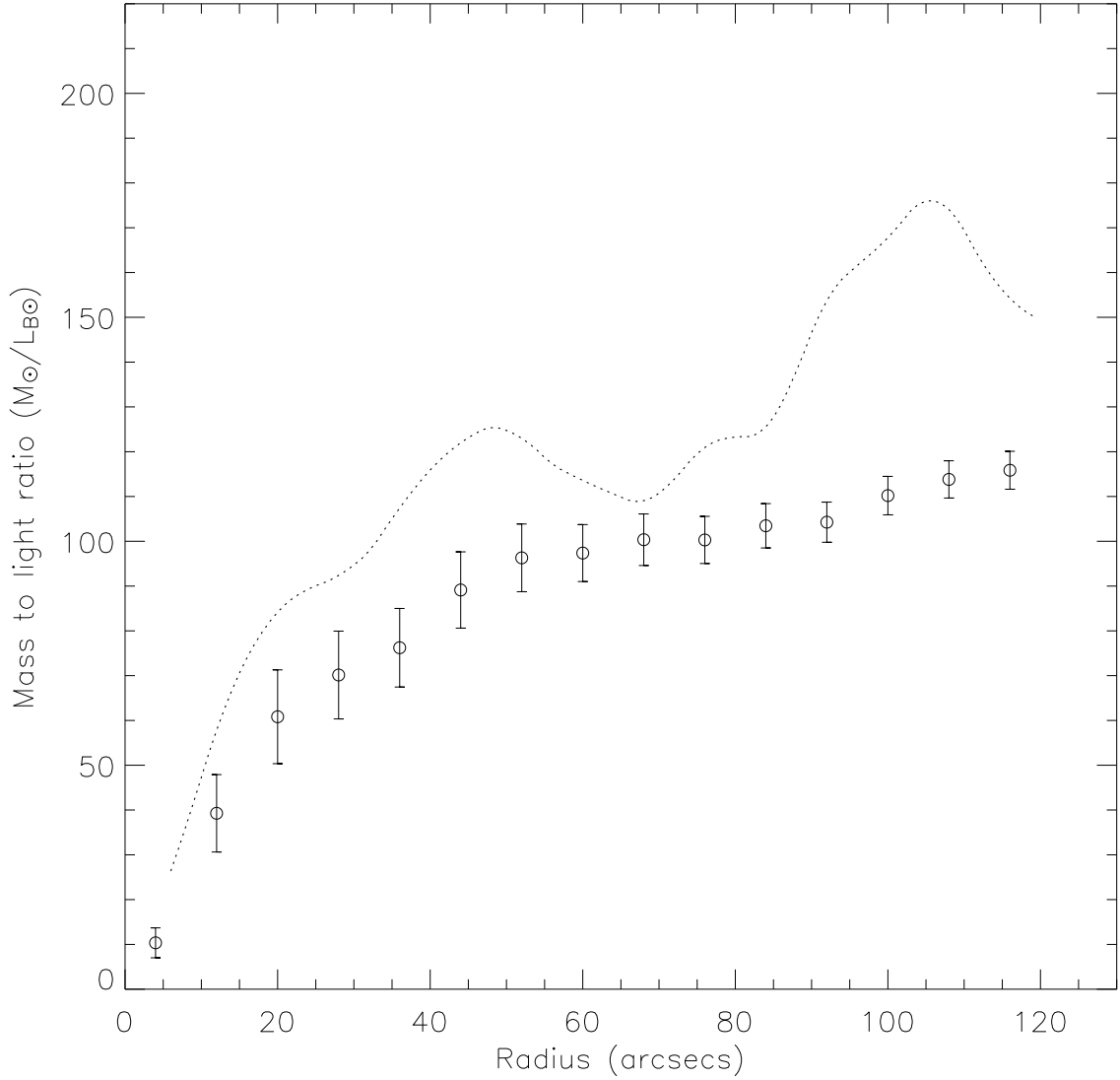}
\caption{Mass-to-light ratio profile of MS 1054. The cumulative $M/L_B$ ratio (circle) continuously increases out to the field limit.
The differential $M/L_B$ ($\delta M(r) / \delta L_B(r)$) plot is computed after smoothing the luminosity profile (Figure~\ref{fig_lp}).
\label{fig_mass2light}}
\end{figure}

We can combine the light and the mass profile to examine the radial behavior of the cluster mass-to-light
ratio (Figure~\ref{fig_mass2light}). The cumulative $M/L_B$ (open circle with error bar) increases
rapidly at small radii ($r<50\arcsec$) and continues to rise rather slowly reaching $124\pm5M_{\sun}/L_{\sun}$
(the uncertainty only reflects the statistical error in mass).
at $r=1$Mpc. This pattern is quite different from that for CL 0152-1357, for which the cumulative
$M/L_B$ arrives at its maximum at $r\sim35\arcsec$ and decreases rather monotonically afterwards.
The $M/L_B$ profile of CL 0152-1357 is very similar to that of an ensemble cluster compiled 
from the extensive kinematic studies of a large cluster sample (Carlberg et al. 1997; Katgert et al 2004).
The outwardly increasing $M/L_B$ profile of MS 1054-0321 is also seen in HFK00 paper.
The smoothed differential $M/L_B$ profile (i.e., $\delta M(r) / \delta L(r)$) shows the radial variation
more markedly and the two $M/L_B$ ``bumps'' at $r\simeq50\arcsec$ and $110\arcsec$ can be associated with the
mass concentrations seen in the weak-lensing mass reconstruction. 

Considering the luminosity evolution $\mbox{ln} (M/L_B) \propto (-1.06\pm0.09)z$ (van Dokkum \& Stanford 2003; Holden et al. 2005),
we can predict the $M/L_B$ value of the cluster at z=0. If the B-band luminosity decreases by $\sim41$\% from z=0.83 to
z=0 (van Dokkum \& Stanford 2003), the corresponding $M/L_B$ of MS 1054-0321 becomes $\sim303 M_{\sun}/L_{\sun}$ ($r<1$Mpc) in the present Universe.
HFK00 obtained a slightly larger $M/L_B$ value of $320\pm35~h_{71} M_{\sun}/L_{B\sun}$, but the result is still consistent with our
$M/L_B$ estimation. They demonstrate that the average $M/L_r$ value  of the 14 near-by clusters originally presented by Carlberg, Yee, \& Ellingson (1997)
becomes similar to the $M/L_B$ of MS 1054-0321 if $B-r\sim1.1$ is taken into account.

\subsection{Substructure \label{substructure}}
\begin{figure}
\plotone{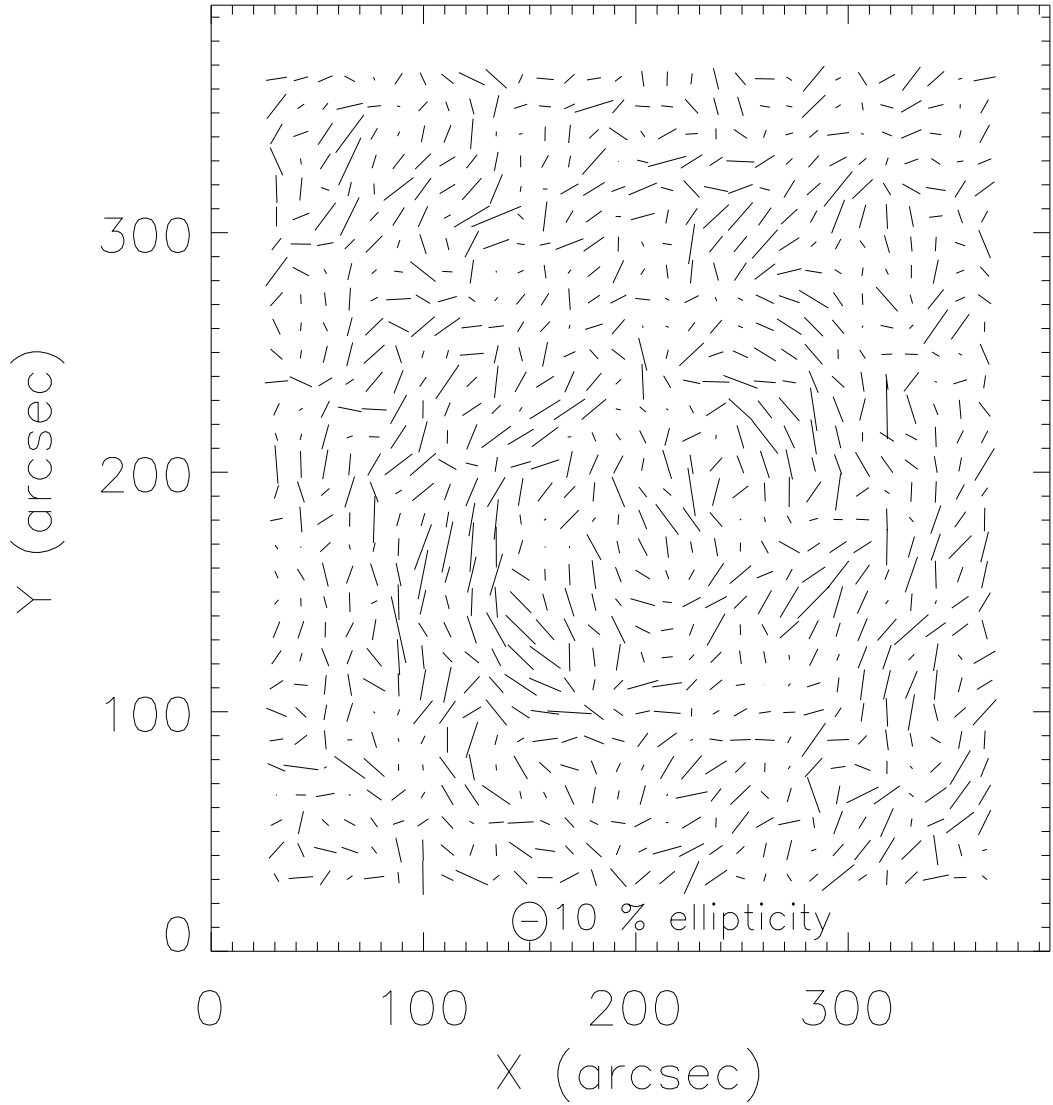}
\caption{Ellipticity distribution of source galaxies. Each stick represents the weighted mean ellipticity of the background galaxies.
Note the tangential alignments of ``whiskers" around the cluster main body. 
\label{fig_whisker}}
\end{figure}

The whisker plot (Figure~\ref{fig_whisker}) demonstrates the weighted mean ellipticity distribution across the
cluster field. The length of each whisker is proportional to the magnitude of the mean ellipticity of the source
galaxies in the region and
the position angle is expected to be aligned with the mean local shear. 
The presence of the lensing signal is obvious from the systematic patterns around the 
cluster galaxies.

\begin{figure}
\plotone{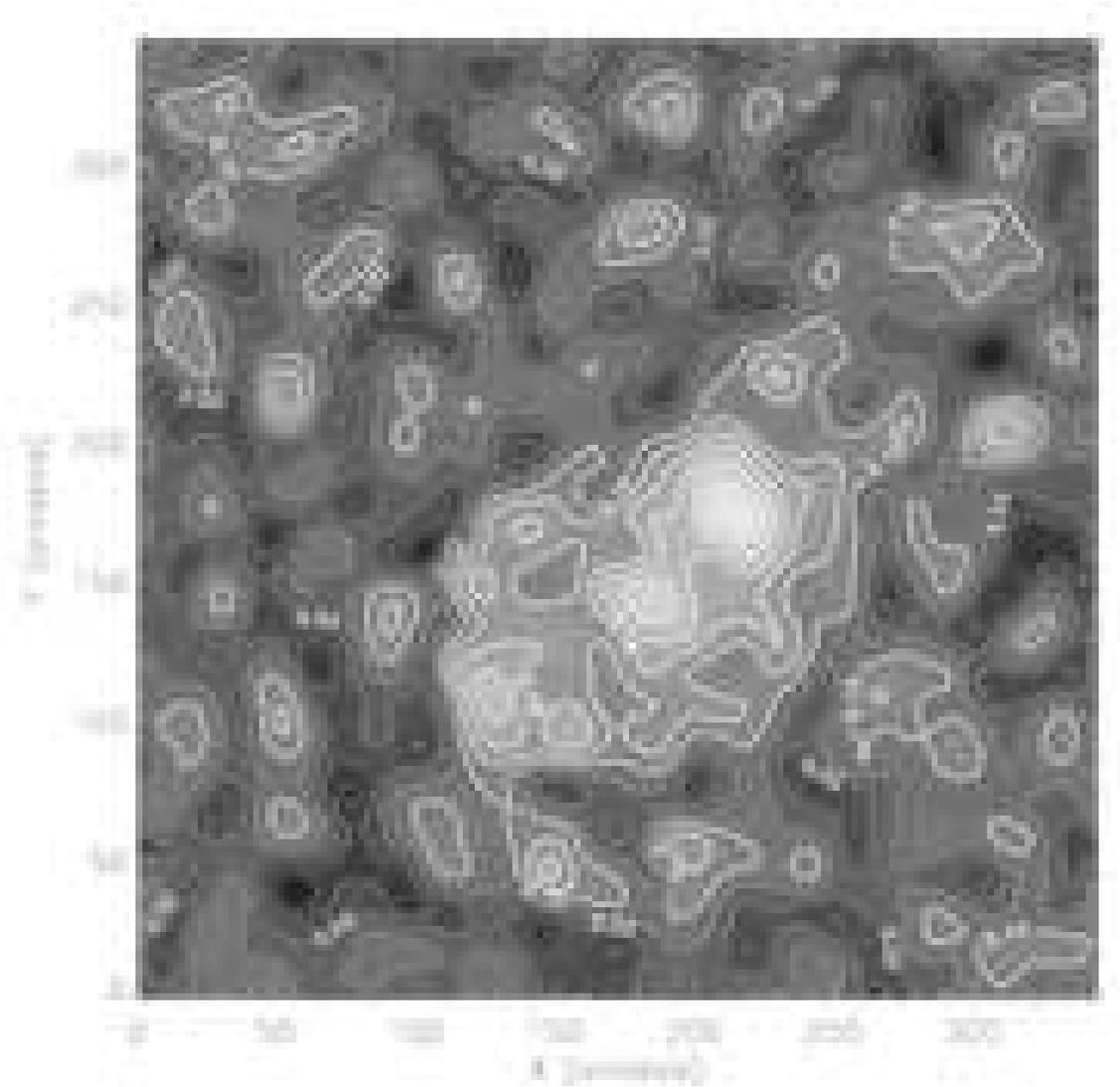}
\caption{Rescaled mass reconstruction. Sheet-mass degeneracies are lifted using the SIS fit result.
The mass map has been smoothed with a Gaussian kernel (FWHM $\sim15\arcsec$). Because we do not vary the kernel size adaptively, the significance
of the substructure is relatively low outside the main cluster body. 
\label{fig_mass_contour}}
\end{figure}

\begin{figure}
\plotone{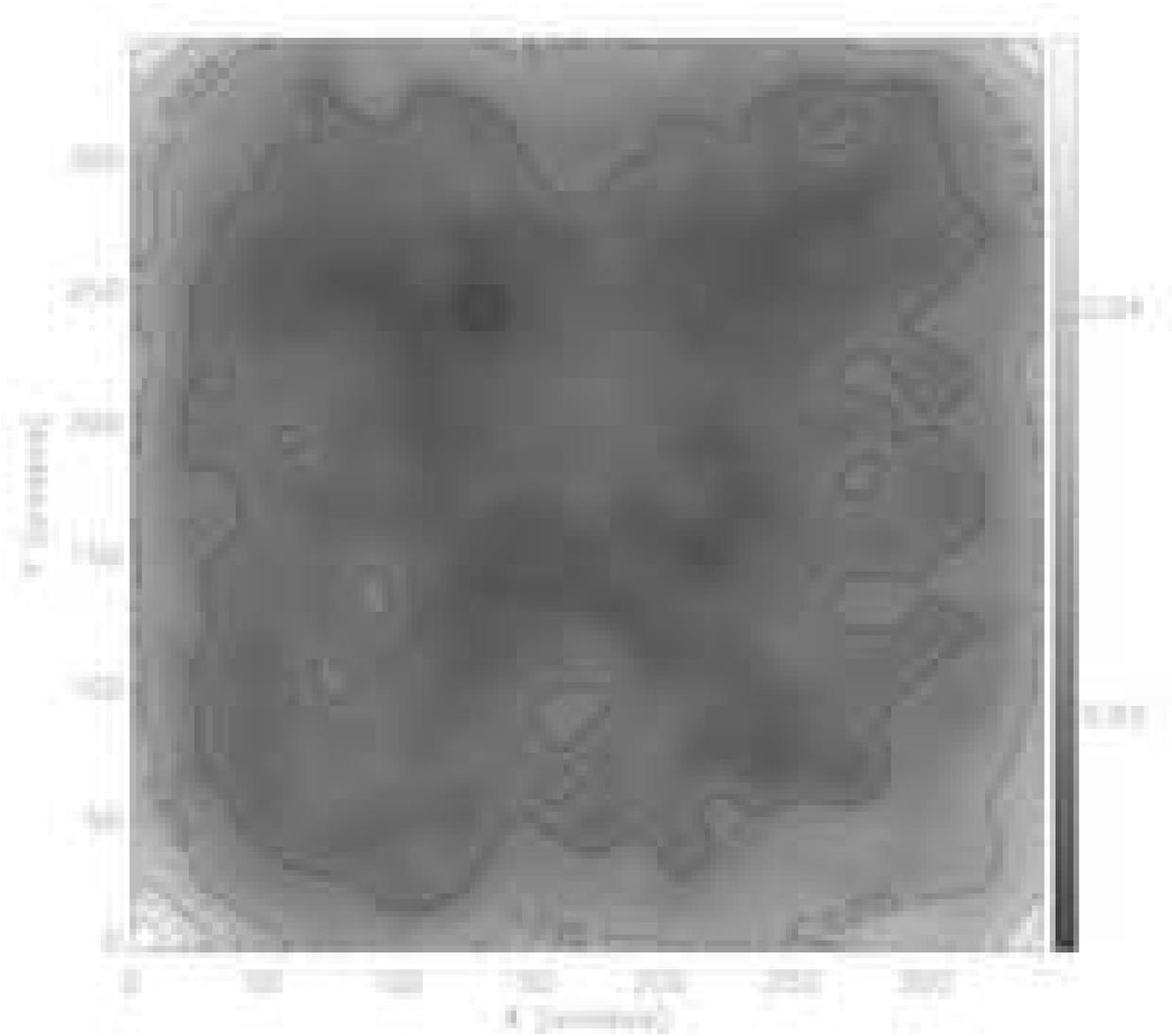}
\caption{RMS map of mass reconstruction presented in Figure~\ref{fig_mass_contour}. The RMS is computed from 5000 bootstrap runs of mass reconstruction.
Note that the RMS increases where the source galaxies are sparse, especially at the four corners.
\label{fig_err_map}}
\end{figure}

\begin{figure}
\plotone{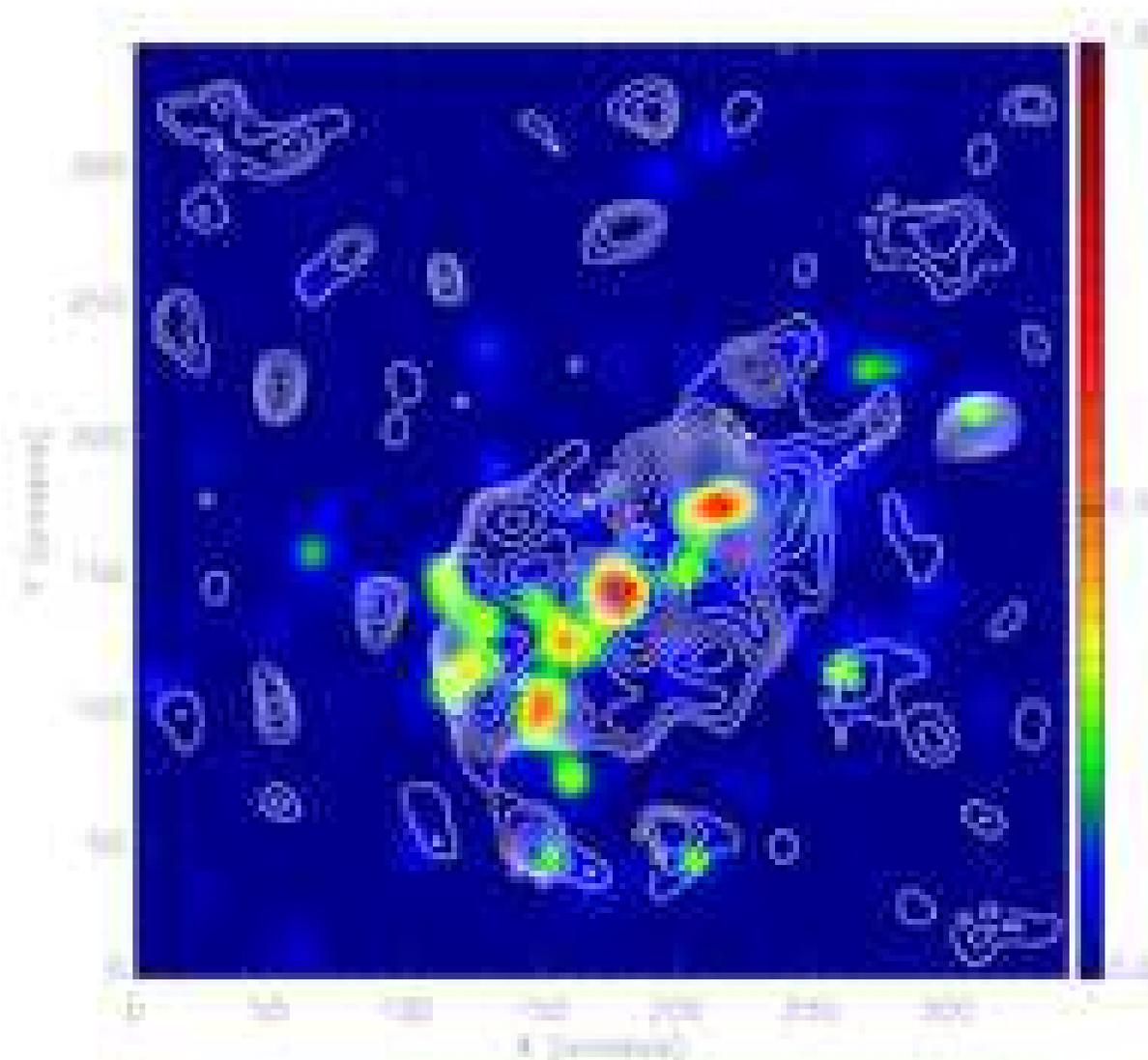}
\caption{Mass contours overlaid on light distribution. The background is color-coded with the rest-frame B band luminosity of the spectroscopically
confirmed cluster galaxies. We only display mass contours whose significance is
approximately above 3 $\sigma$ ($\kappa > 0.1$)
\label{fig_massoverlum}}
\end{figure}

We present the weak-lensing mass reconstruction of MS 1054-0321 and 
its rms in Figure~\ref{fig_mass_contour} and \ref{fig_err_map}, respectively. 
The mass map is generated
via the nonlinear reconstruction (Bartelmann 1995) with a significant modification to
properly account for the rescaling ($\kappa \rightarrow \lambda \kappa + (1-\lambda)$) using
the SIS fit result (J05). We smoothed the result with a $\mbox{FWHM}=10\arcsec$ Gaussian kernel
The rms mass map is
produced from 5000 mass reconstruction maps via Bootstrap resampling. 
The main body of MS 1054-0321 characterized by the three dominant
mass peaks is manifest in the mass reconstruction. These three mass peaks were
first identified by HFK00 in their weak-lensing analysis using WFPC2 observations 
(see \textsection\ref{section_hfk00_compare} for comparison). 
A similar substructure can be obtained when the light distribution of the cluster galaxies is smoothed
with a FWHM$\simeq14\arcsec$ Gaussian kernel (we used a FWHM$\sim10\arcsec$ kernel for the mass map).
An overlay
of the mass contour ($\kappa > 0.1$) on top of this light distribution created
from the confirmed cluster galaxies
 is presented
in Figure~\ref{fig_massoverlum}. It is apparent that the cluster galaxies trace
the cluster mass fairly well. The mass distribution
over the eastern clump, unlike the other two mass clumps, seems to be somewhat complicated as is also indicated
by the color-coded light distribution. 

\begin{figure}
\plotone{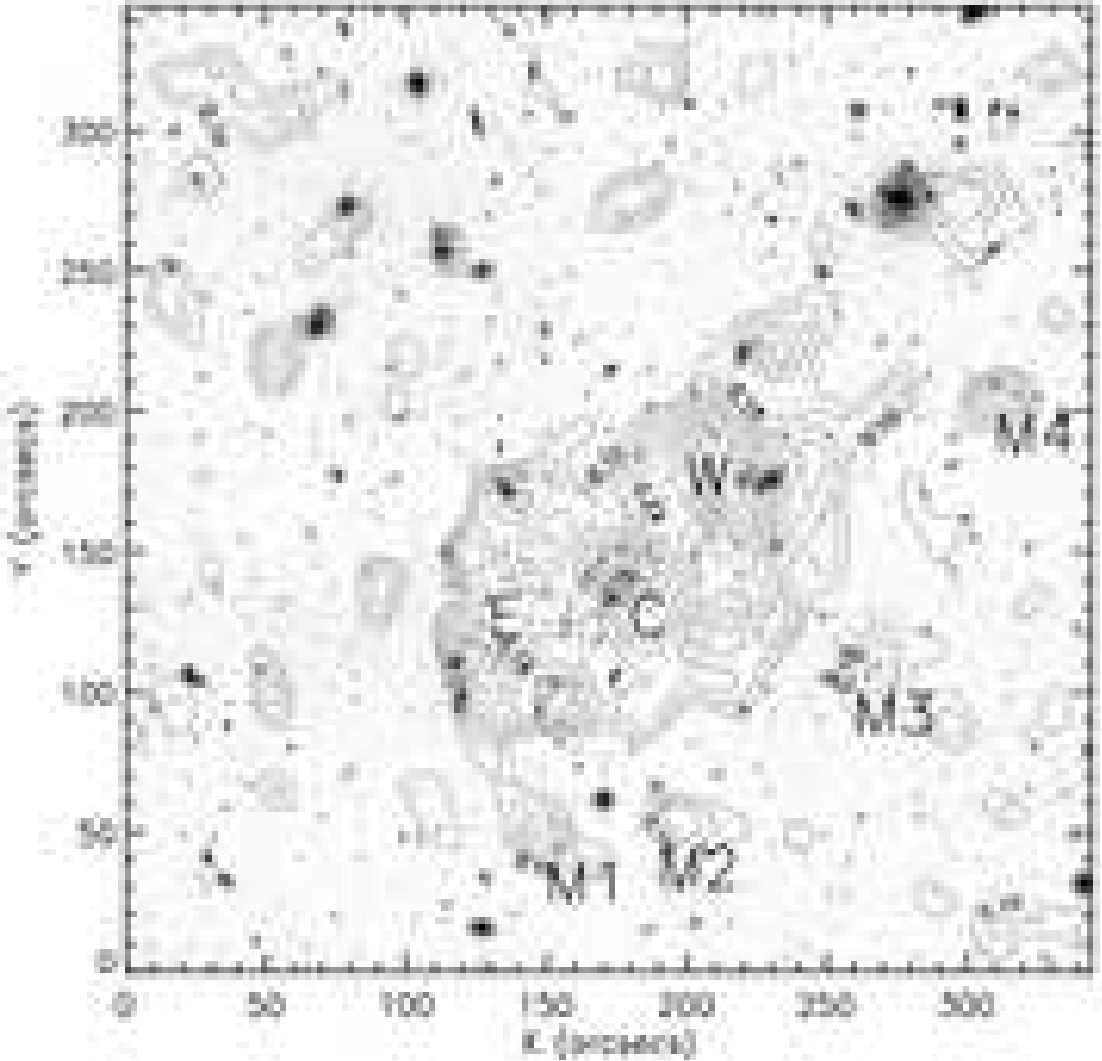}
\caption{Mass contours overlaid on negative ACS image. Same as Figure~\ref{fig_massoverlum} but with the background replaced with
the negative ACS detection image. We identify 7 mass clumps (labeled W, C, E, M1, M2, M3 and M4) whose significance is high ($>3\sigma$) and the galaxy
counterpart is unambiguous. Some of the unlabeled mass clumps seem to be associated with foreground galaxies. 
\label{fig_massoverplot}}
\end{figure}

We conservatively select four mass clumps outside the main body 
where the significance is above the $\sim3\sigma$ level and
the galaxy counterparts are not ambiguous (Figure~\ref{fig_massoverplot}). 
These four minor clumps (M1-4) are located south of the main clumps (C, W, and E) and they
appear to be infalling groups onto the main cluster. Each of these minor clumps contain
some bright early-type galaxies and its luminosity peak is
close to the corresponding mass clump with an average offset of $\sim6\arcsec$.
Because of the relatively weak shear signal and thus
low significance of the mass centroid of each mass peak, it is hard
to infer the dynamics of the clump based on its mass-to-galaxy offset.
It is interesting to note that in our previous investigation 
we also identified four minor mass clumps spatially correlated
with the cluster member galaxies outside the main body of CL 0152-1357 at z=0.84 (J05).
If these clusters are found to represent the typical clusters at this redshift,
the presence of these hypothesized infalling groups can provide direct
support of hierarchical structure formation.

Some of the rest of the minor mass peaks in our MS 1054-0321 mass reconstruction 
seem to be associated with foreground
objects in the field, but here we confine our analysis only to mass clumps obviously related to
the substructure of
MS 1054-0321 at $z=0.83$.

Table 1 summarizes the properties of these mass clumps. The mass
within $30\arcsec$ ($\sim223$ kpc) aperture radius is measured from the rescaled mass map, and
the mass uncertainty is calculated using the rms map (Figure~\ref{fig_err_map}) under the assumption 
that the mass pixels within the aperture are entirely correlated.
The mass-to-light ratio is computed only counting the spectroscopically confirmed members.
Therefore, the $M/L_B$ values quoted here correspond to the upper limits and 
in particular, for the minor clumps we suspect these $M/L_B$ values are largely
overestimated because the incompleteness of the spectroscopic sample is higher for
bluer cluster galaxies (Goto et al. 2005).

\section{$CHANDRA$ X-RAY ANALYSIS}

\subsection{X-ray Substructure}
\begin{figure}
\plotone{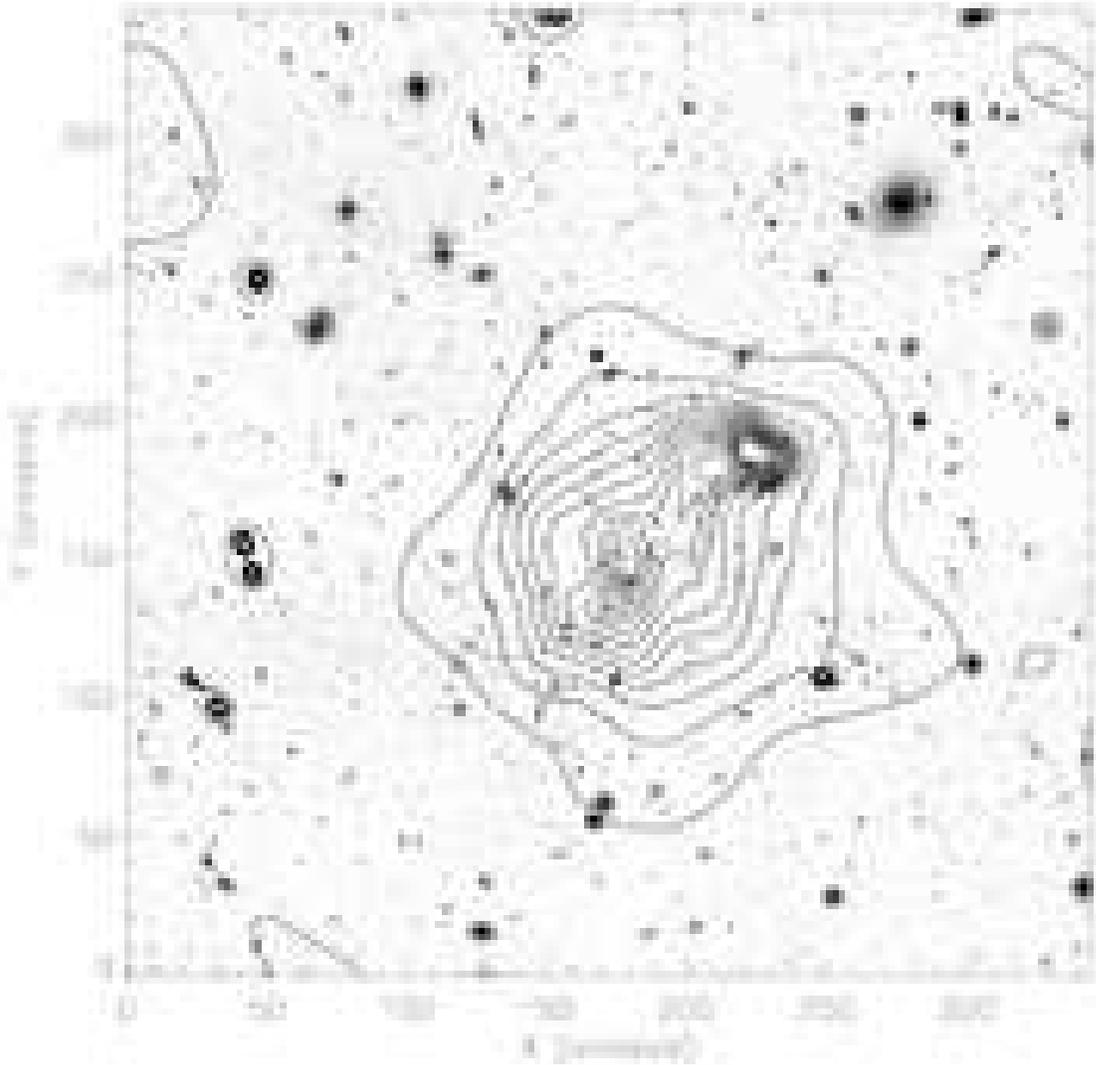}
\caption{Chandra X-ray contours on top of ACS image. The exposure corrected X-ray image (0.8-7 keV photons) is
adaptively smoothed with a minimum significance of 3 $\sigma$. The X-ray peaks are correlated with, 
but slightly offset from the cluster galaxies.
The excellent agreements in point sources verifies the accurate
image alignments between the two images.
\label{fig_xrayoverimage}
}
\end{figure}

We created an exposure corrected X-ray image by extracting 0.8-7 keV photons without excluding point sources. The
image was then adaptively smoothed with the CIAO ``csmooth'' with a minimum significance of 3 $\sigma$.
We carefully aligned the Chandra image with the ACS and the excellent
coincidence of the X-ray point sources with the corresponding galaxies verifies the precise alignment between these
two images (Figure~\ref{fig_xrayoverimage}).  
The resulting X-ray contours clearly show the substructure of the cluster reminiscent of the distribution of
the member galaxies and dark matter.
However, as noted by Jeltema et al. (2001), the $Chandra$ X-ray
reveals only two clumps, which seem to be associated with the central and the western weak-lensing mass peaks. 
The absence of the eastern weak-lensing substructure in X-ray emission is also confirmed by
the XMM-$Newton$ observations (Gioia et al. 2004).
Interestingly, the locations of the two X-ray peaks are not in precise agreement with those of the BCGs as
first indicated by Jeltema et al. (2001).  
Further discussion on this centroid offsets in addition to
the absence of the eastern peak in the X-ray emission 
is deferred to \textsection~\ref{lensing_vs_xray_substructure}. 

Although
the X-ray peak close to the central dark matter clump was referred to as the 
``eastern'' X-ray peak in Jeltema et al. (2001), hereafter we refer to
the clump as the ``central'' peak to maintain a consistency with the previous weak-lensing substructure labeling. 
The central X-ray peak seems to be somewhat less concentrated
than the western X-ray peak, possibly further resolved into two minor peaks. The X-ray contours obtained
from the XMM-$Newton$ observations (Gioia et al. 2004) also indicate this possibility.

\subsection{Cluster X-ray Temperature and Luminosity \label{xraytemp}}
The understanding of the X-ray temperature of MS 1054-0321 based on the $Chandra$ observatory evolves as new calibration data 
on the instrument becomes available. It has been realized that the degradation of the low energy quantum efficiency (QE) of the instrument
is significant and possibly due to the molecular contamination of the ACIS optical blocking filters. This decline of the
low energy sensitivity biases the temperature measurement upward unless accounted for. Based on the observed 
decay rate determined from the external calibration source, Chartas \& Getman (2002)
introduced a time dependent ACIS absorption model (ACISABS), which describes the degradation of the QE
as a function of the number of days since the launch. Gioia et al. (2004) argued that the temperature measurement of MS 1054-0321
by Jeltema et al. (2001) suffered from the absence of this low-energy correction, being
biased toward high temperature.
However, the use of the ACISABS model 
was deprecated by CIAO as of CALDB 2.26 (released on 2 February 2004), and it is recommended
that instead an area response file (ARF) be created with the calibration database that now 
properly reflects the time- and area-dependent QE degradation 
model\footnote{Also, see the note at http://cxc.haravrd.edu/ciao/why/acisqedeg.html.}.
We observe that following this new procedure tends to yield a higher temperature than the ACISABS prescription.   

Another important preprocessing for the spectral analysis includes careful background flare removal. 
The omission of the procedure also results in an overestimation of the cluster
temperature. Although Jeltema
et al. (2001) attempted to remove high background time intervals, their somewhat long net exposure time of $\sim88$ ks
compared to our result of $\sim74$ ks 
indicates that flare identification was insufficient. Tozzi et al. (2003) and Gioia et al. (2004)
also obtained relatively short net exposures of $\sim80$ks and $\sim67$ks, respectively.

\begin{figure}
\plotone{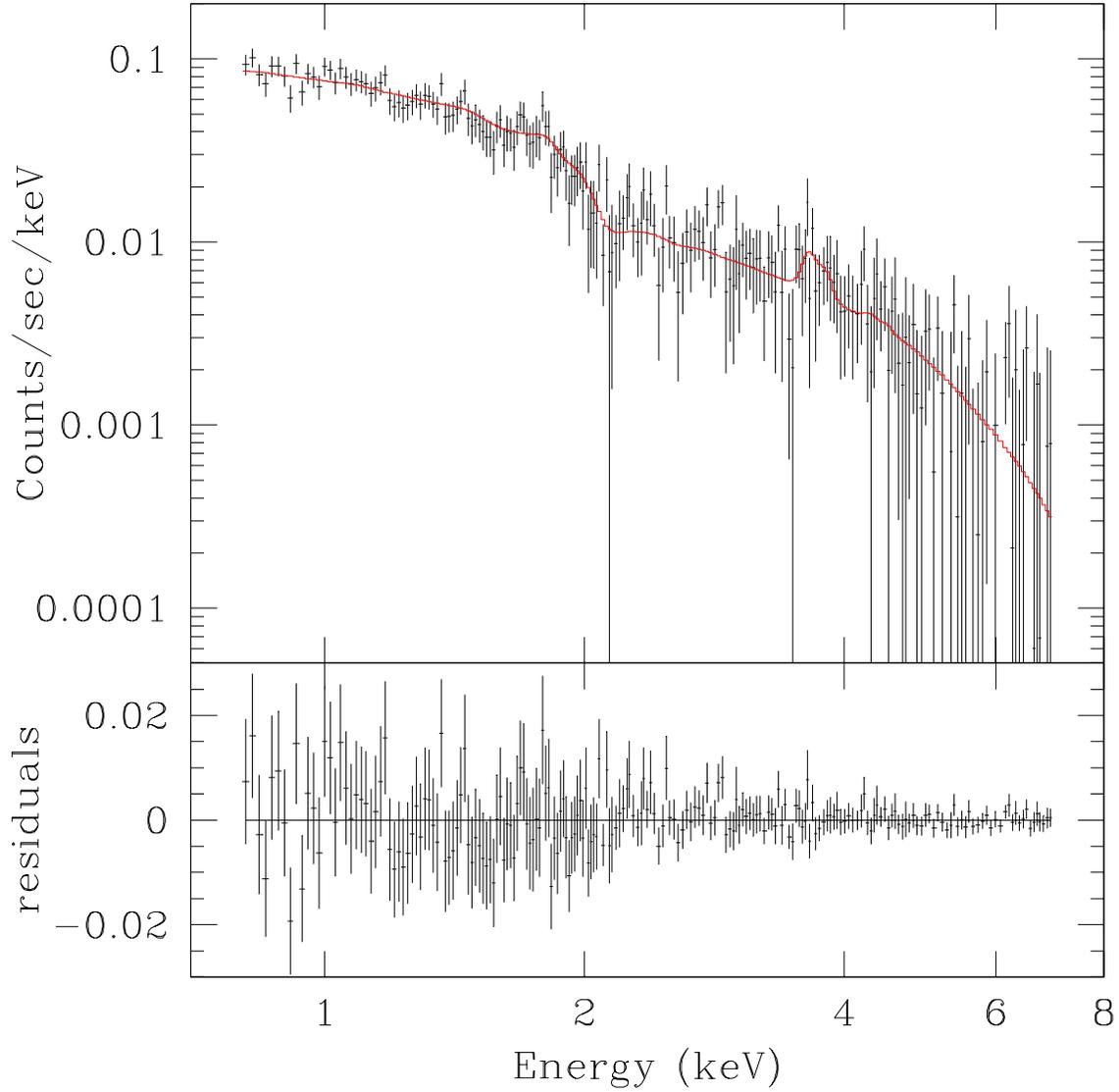}
\caption{X-ray spectrum of MS 1054-0321 as a whole. The spectrum is taken from $90\arcsec$ aperture radius with the local
background subtraction. Each spectral bin contains at least 20 counts. The best-fit
MEKAL plasma model (solid) yields a cluster temperature of $T=8.9_{-0.9}^{+1.0}$ keV with a metal abundance of 
$Z/Z_{\sun}=0.30_{-0.12}^{+0.12}$.
\label{fig_temp_fit_large}}
\end{figure}

We created the ARF 
and the redistribution matrix file (RMF) 
to account for the instrument
response variation. The ARF file was generated using the new CIAO 3.2 tool $mkacisrmf$, which
provides more accurate calibration at low energies ($<1 keV$) than the previous CIAO script $mkarf$.
However, because our spectral analysis was limited 
to the 0.8-7 keV range, we only observed marginal change ($\sim0.2$ keV) due to
this new prescription.
We adopt the cluster center 
($\alpha=10^h56^m58^s.6$ and $\delta=-03\degr37\arcmin36\arcsec.7$)  
defined
by Neumann \& Arnaud (2000). 
The spectrum extracted from a circular region with a radius of $90\arcsec$ 
was fitted to the MEKAL plasma model (Kaastra \& Mewe 1993; Liedahl, Osterheld, \& Goldstein 1995). 
We froze the Galactic hydrogen column density 
and the redshift of the cluster
at $N_H=3.6\times10^{22} \mbox{cm}^{-2}$ (Dickey \& Lockman 1990)
and $z=0.83$, respectively. Figure~\ref{fig_temp_fit_large} shows the best-fit thermal plasma spectrum, which yields a temperature of
$T=8.9_{-0.8}^{+1.0}$ keV and a metal abundance of $Z/Z_{\sun}=0.30\pm0.12$ with a reduced $\chi^2=0.63$ (206 degrees of freedom).
Using the similar parameter constraints, Jeltema et al. (2001) found a temperature of $T=10.4^{+1.7}_{-1.5}$ keV whereas
Gioia et al. (2004) reported that a temperature of $T=7.4^{+1.4}_{-0.9}$ keV was obtained when the two aforementioned corrections
were applied to the $Chandra$ data.
We verified
that, if the ACISABS model was adopted instead, our temperature measurement decreased to $\sim7.8$ keV, a result closer
to the measurement of Gioia et al. (2004).
In Table 2 we summarize the previous temperature measurements of MS 1054-0321 found in the literature.

From the best-fit spectra, we obtain the
observed-frame flux of 
$F_{0.8-7.0~keV}=(5.9\pm0.2)\times10^{-13} \mbox{erg}~\mbox{cm}^{-2}~\mbox{s}^{-1}$ within a $90\arcsec$ radius.
This becomes a rest-frame bolometric (0.01-40 keV) luminosity of $L_X=(3.1\pm0.1)\times10^{45} \mbox{erg}~\mbox{s}^{-1}$
after the k-correction to the rest-frame and the aperture correction to total are applied.

\begin{figure}
\plottwo{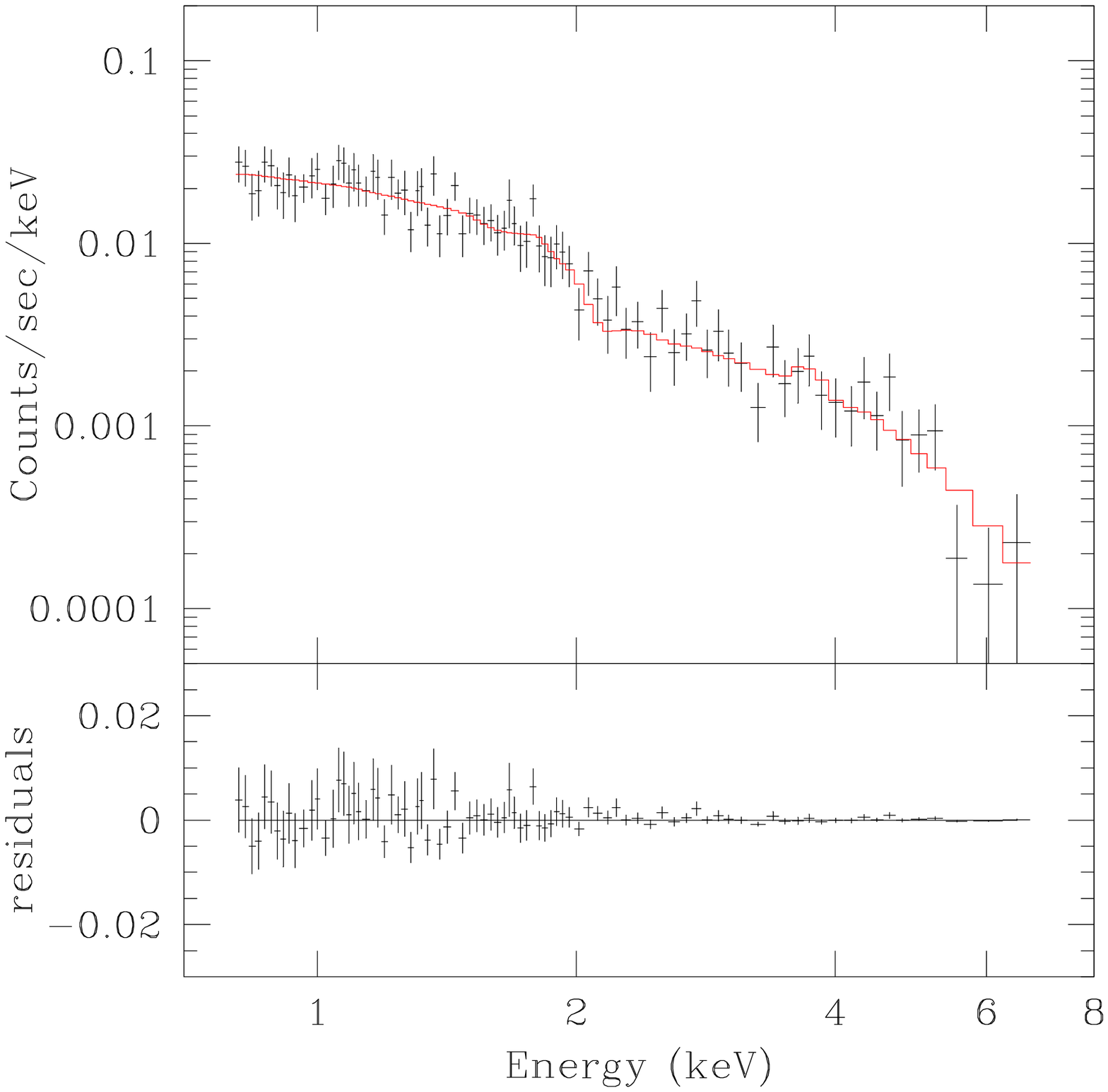}{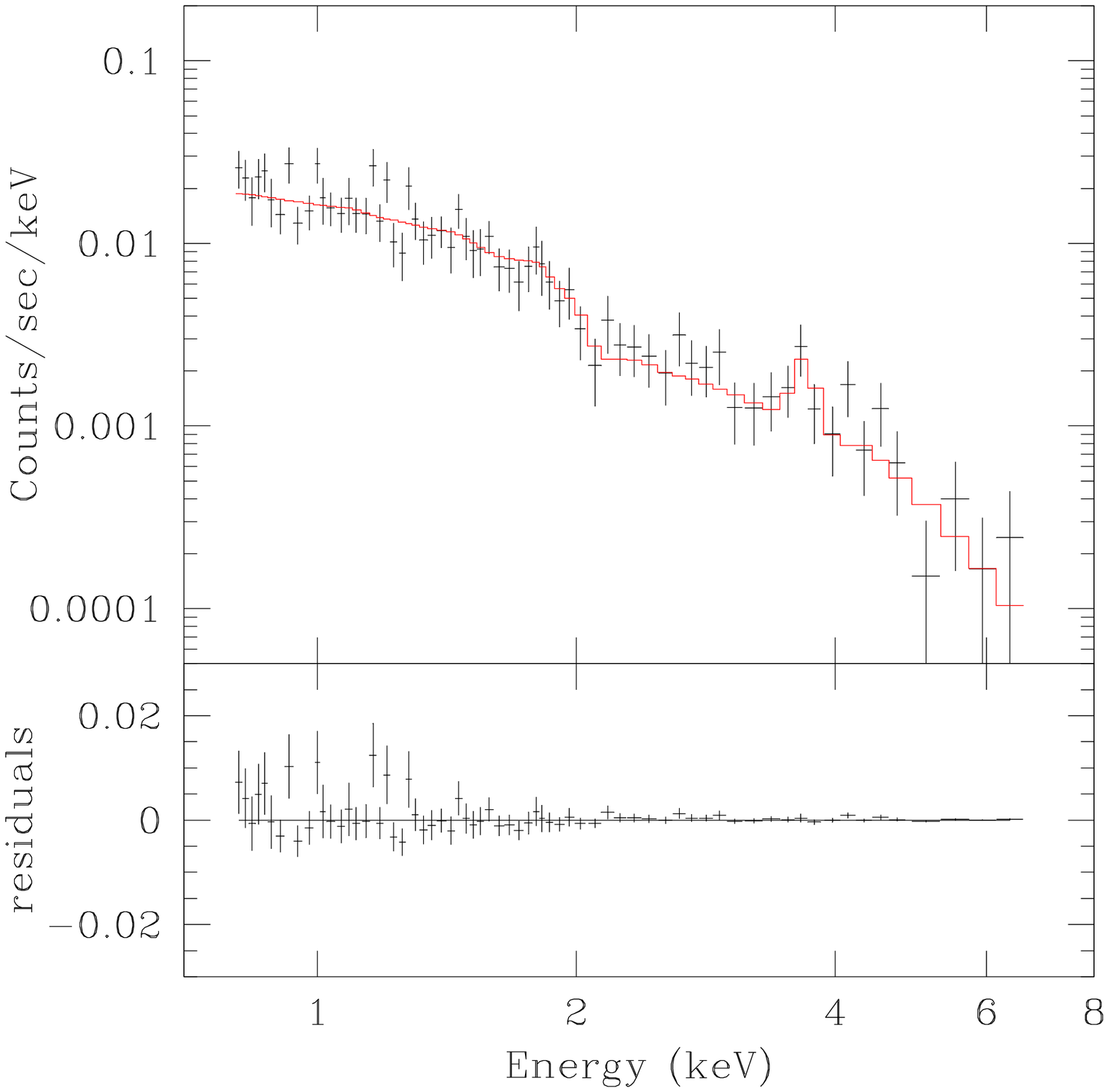}
\caption{X-ray spectra of the individual X-ray clumps. The central clump (left panel) has a higher temperature ($10.7_{-1.7}^{+2.1}$ keV)
than the western X-ray clump (right panel; $7.5_{-1.2}^{+1.4}$ keV). The best-fit MEKAL plasma model is represented
by solid line. The iron emission line is more pronounced in the spectrum of the western clump ($Z=0.47_{-0.23}^{+0.24} Z_{\sun}$) while
the feature is weaker for the central clump ($Z=0.16_{-0.16}^{+0.19} Z_{\sun}$).
\label{fig_temp_fit_small}}
\end{figure}

We find that the temperature
of the central peak is higher than that of the western peak (Figure~\ref{fig_temp_fit_small};Table 3)
as was also noticed
by Jeltema et al. (2001). 
Our temperature measurements of $T=10.7_{-1.7}^{+2.1}$ and $7.5_{-1.2}^{+1.4}$ keV
for the central and western X-ray peaks, respectively, are consistent with
the results presented by Jeltema et al. (2001), who quoted 
$T=10.5^{+3.4}_{-2.1}$ and $6.7_{-1.2}^{+1.7}$ keV. Gioia et al. (2004) also
confirmed that the central substructure has a higher temperature, but
their measurements ($T=8.1^{+1.3}_{-1.2}$ and $5.6_{-0.6}^{+0.8}$ keV) are 
slightly lower. We obtain a metal abundance of $Z=0.16_{-0.16}^{+0.19} Z_{\sun}$ for
the central X-ray peak and a somewhat higher value of $Z=0.47_{-0.23}^{+0.24} Z_{\sun}$
for the western substructure. Despite the large uncertainties, Gioia et al. (2004)
found similar values ($Z=0.12^{+0.35}_{-0.12}$ and $0.51^{+0.36}_{-0.32} Z_{\sun}$).

\begin{figure}
\plotone{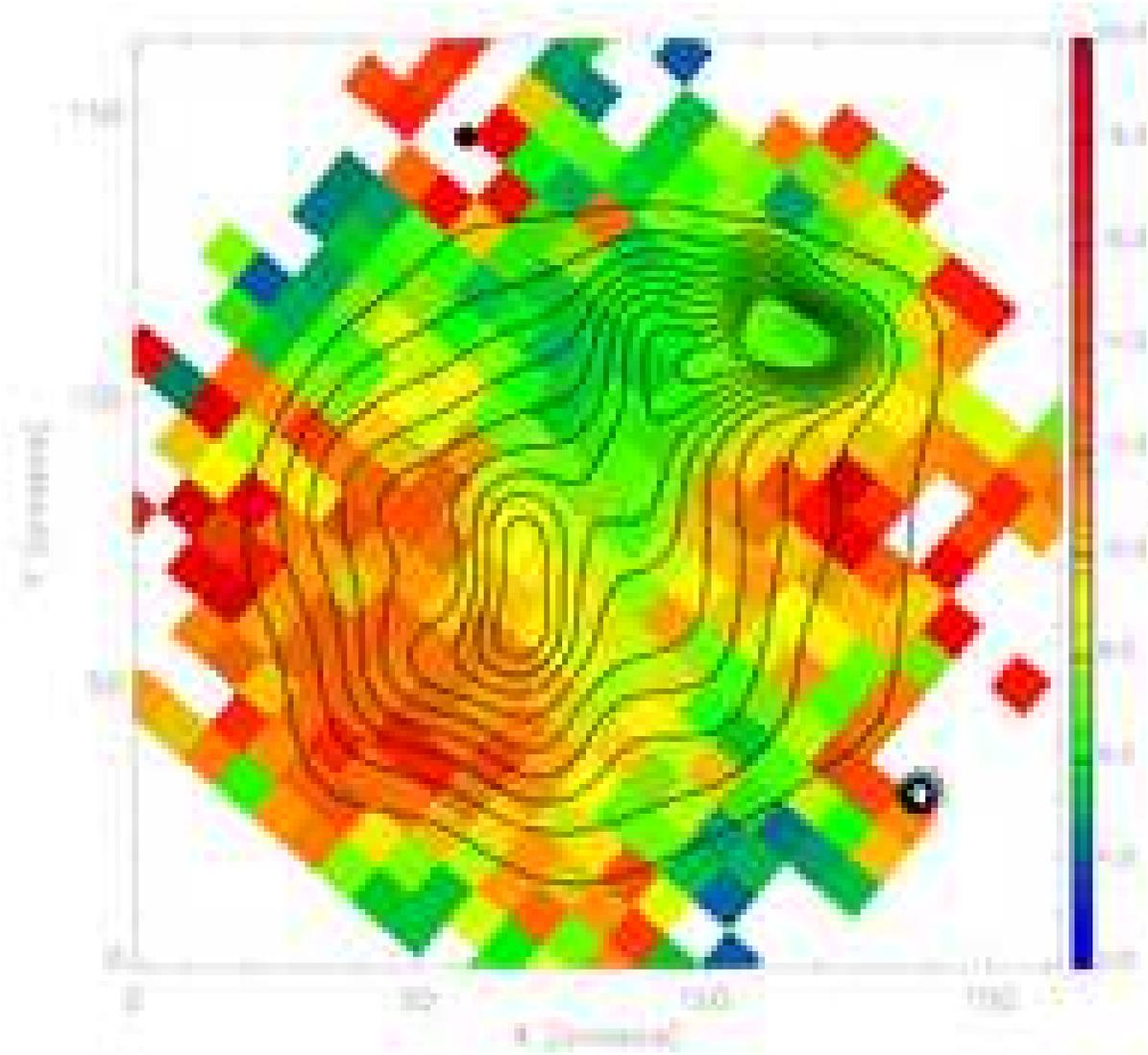}
\caption{Temperature map of MS 1054-0321. The cluster field is divided into $21\times21$ subregions and
the MEKAL plasma model is fit to the spectrum extracted from each cell with a fixed abundance of $Z=0.30$.
The blank (white) cells represent the region where the model fit is judged to be insignificant.
X-ray flux contours are overlaid for reference.
The eastern part of the cluster is estimated to be of higher temperature
than the western part. There appears to be no shock-heated region between the two X-ray peaks.
\label{fig_temp_map}}
\end{figure}

This result indicates that there may exist 
severe temperature variation within the cluster. In
order to address the issue further, we constructed the temperature map of the cluster as follows. We divided the cluster field
$160\arcsec\times160\arcsec$ into $21\times21$ subregions and extracted
a spectrum for each subregion after applying the RMF and the ARF corrections as above. To prevent the poor photon statistics
from leading to a spurious temperature, we not only allowed an ample overlapping between subregions, but also the resulting
spectrum is ensured to contain at least 50 counts per spectral bin. Then, the spectrum is fit to the MEKAL plasma model
with a fixed abundance of $Z=0.30$. We present this temperature map in Figure~\ref{fig_temp_map}. 
We examined the fitting result for each grid and rejected the resulting temperature if the fit appears to be insignificant
because of the poor photon statistics.
The overall temperature structure is
somewhat similar to the hardness ratio map (Jeltema et al. 2001), but this temperature map derived from the direct spectral
fitting enables a quantitative comparison between different regions. We verified that the temperature estimation
based on this map is consistent with the result summarized in Table 3. 

Cosmological simulations
of the intracluster medium (e.g., Schindler \& Mueller 1993; Ricker 1998; Takizawa 1999; Ricker \& Sarazin 2001) 
have demonstrated that there develop shock-heated regions between
sub-clusters approaching each other, stretched perpendicular to the merging axis. However, the temperature structure of MS 1054-0321
does not indicate such a feature between the centers of the two X-ray peaks. Instead, we note that the large scale
temperature variation is reminiscent of the quadrupole temperature structure 
(i.e., two high-temperature regions propagating in opposite directions
along the merger axis) observed at the post-merger stage in numerical simulations. The
disrupted appearance and the flat surface brightness profile (see \textsection\ref{isothermalbeta}) of the central X-ray peak
are also consistent with this conjecture that the central X-ray peak might have gone through a recent merger.
We discuss a possible scenario in association with the
absence of the eastern X-ray peak in \textsection\ref{section_absence}.

\subsection{Isothermal Beta Description of the Cluster \label{isothermalbeta}}
The complicated substructure of the cluster makes it challenging to fit 
the surface brightness profile to an isothermal beta model, which assumes a spherical symmetry, and the attempts
often lead to physically unreasonable parameters. 

We created an 
exposure-corrected image by binning 0.8-7 keV photons with $\sim2\arcsec\times2\arcsec$ pixels.  
Initially, we attempted to model the surface brightness profile of the cluster with a 
superposition of two isothermal beta profiles. However, as noted by Jeltema et al. (2001), 
the central X-ray peak requires a large beta index ($\beta\sim1.3$) whereas the
western X-ray peak profile is nicely fit with a beta index of $\beta=0.58\pm0.15$ and a 
core radius of $r_c=17\arcsec\pm7\arcsec$.
We suspect that the large beta index for the central X-ray peak originates from the flat
core, most probably due to recent merger activity. Therefore, the mass calculation of the cluster
from this approach is flawed and subject to overestimation.

Instead, we decided to fit a single isothermal beta profile to the entire cluster, excluding
the complicated inner substructure. Numerical simulations (e.g., Rowley et al. 2004)
show that the merger of the global ICM precedes that of the core merger, which tends to maintain
the bulk motion up to the late stage of the event. If we assume MS 1054-0321 is
undergoing a major merger between the western and central clumps, and also
the central clump has already suffered a recent merger, we argue that
our fitting a isothermal beta profile to the outer region of the X-ray profile
minimizes the artifact arising from this hypothesized merger activity.

To this end, the binned X-ray image was first fitted 
to a single two-dimensional isothermal beta model to determine the
centroid that maximizes the azimuthal symmetry.
After the optimal centroid was determined at $\alpha \simeq 10^h56^m58^s.7$ and 
$\delta \simeq -03\degr37\arcmin36\arcsec.1$, we constructed an azimuthally averaged 
surface brightness profile, which was then fit to one-dimensional isothermal beta model.
Because the centroid was placed $\sim40\arcsec$ apart from both X-ray peaks,
we excluded the inner region ($r<45\arcsec$) from the fit, which
yielded a core radius of $16\pm15\arcsec$ and a $\beta$ index of $0.78\pm0.08$.
The uncertainties are computed with both parameters allowed to vary.

The $\beta_{spec}$ index or specific energy ratio of the dark matter and ICM is defined as
\begin{equation}
\beta_{spec}=\frac{\mu m_p \sigma^2}{k_B T}, \label{eqn_specific_ratio}
\end{equation}
\noindent
and we can predict a value of $\beta_{spec}\simeq0.9$ using $T=8.9$ keV and $\sigma=1150 \mbox{km}~\mbox{s}^{-1}$
for the entire cluster. If we adopt the correction factor ($\beta_{spec}=(1.25\pm0.1)\beta_{fit}$) from Bahcall \& Lubin (1994),
the $\beta_{fit}$ is estimated to be $\sim0.73$, which nicely matches the measured $\beta$ index
of $0.78\pm0.08$ above.

\begin{figure}
\plotone{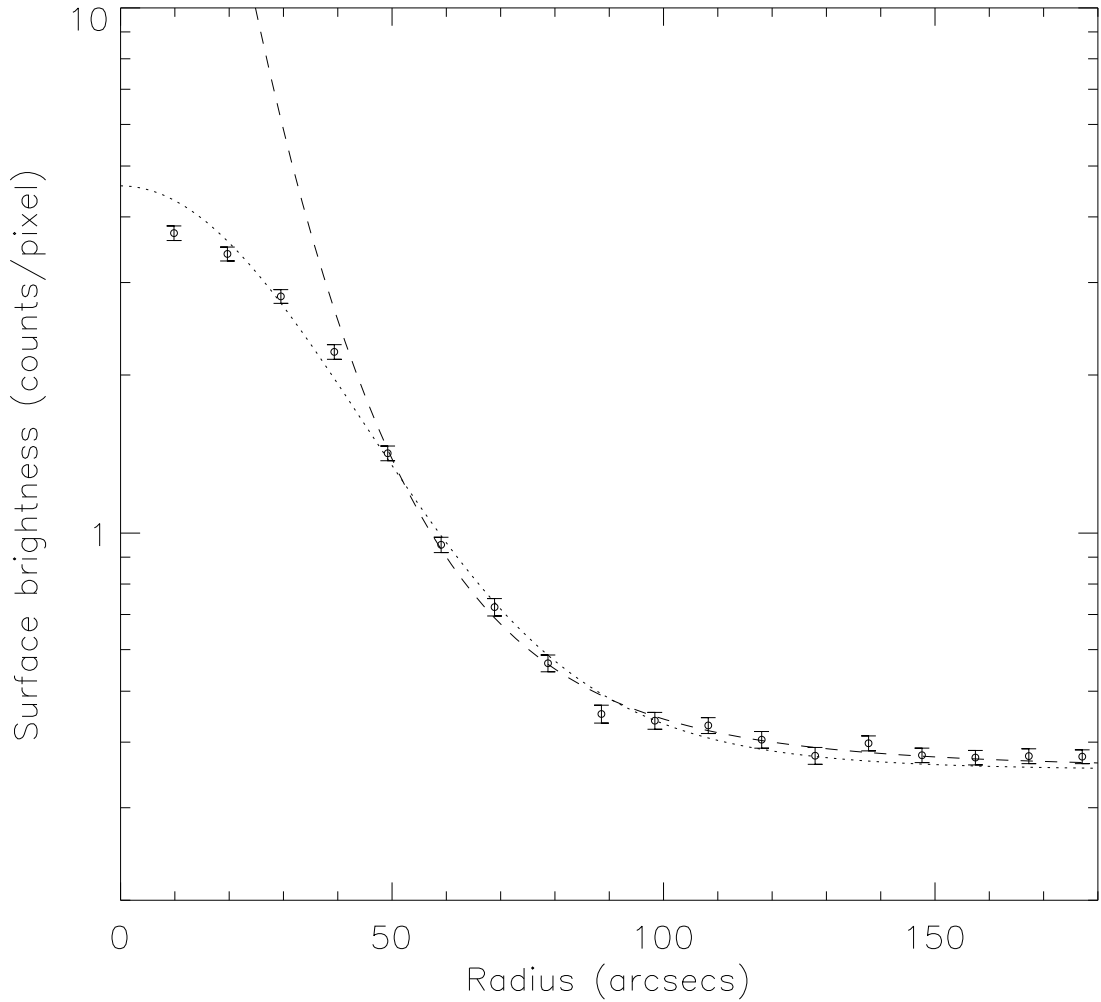}
\caption{X-ray surface brightness profile.
The open circles represent the
azimuthally averaged surface brightness. Because of the 
substructure, the central region ($r<45\arcsec$) was excluded in
isothermal beta fitting, yielding $r_c=16\arcsec\pm15\arcsec$ and
$\beta=0.78\pm0.08$ (dashed). 
If we use all the data points,
the fitting requires a physically unrealistic value of $\beta\simeq1.66$ and
$r_c\simeq 81.7\arcsec$ (dotted).
\label{fig_xray_beta}}
\end{figure}

We show the radial surface brightness profile from the X-ray emission in
Figure\ref{fig_xray_beta} with the best-fit isothermal beta profile (dashed line). 
When the inner region is not excluded, 
the resulting isothermal beta profile
(dotted line) must be described with a physically
unrealistic value of $\beta\simeq1.66$ (i.e., 
much hotter dark matter than the ICM) and $r_c \simeq81.7\arcsec$.
 
\subsection{X-ray Mass Estimation \label{xray_mass_estimation}}
One can estimate the total gravitating mass of the cluster once the core radius, the $\beta$ index, and
the temperature are determined (Evrard et al. 1996):
\begin{equation}
M(r)=1.13\times10^{14} \beta \left ( \frac{T}{\mbox{keV}} \right ) \left ( \frac{r}{\mbox{Mpc}} \right ) \frac { (r/r_c)^2 } { 1+(r/r_c)^2} M_{\sun}, \label{eqn_xray_mass_3d}
\end{equation}
\noindent
where $r_c$ is the core radius and the mean molecular weight of the gas is assumed to be 0.59 times the proton mass.

Equation~\ref{eqn_xray_mass_3d} yields a total mass of the cluster within
a spherical volume at $r$ and therefore it is not straightforward to compare this X-ray mass with
the weak-lensing estimation, which typically gives an aperture mass
inside a cylinder. However, because equation~\ref{eqn_xray_mass_3d} already assumes a spherical symmetry,
one can also estimate an analogous projected X-ray mass without any loss of generality. 
We present a brief derivation of an analytic expression for the X-ray aperture mass as follows.

After converting 
equation~\ref{eqn_xray_mass_3d} to an expression for a mass density $\rho(r)$, we can integrate the result
along the line of sight to obtain the following surface mass density:
\begin{equation}
\Sigma (r) = 2.83\times10^{13} \frac{\beta T}{r_c} \left ( 1+ \left (\frac{r}{r_c} \right )^2 \right)^{-3/2} \label{eqn_sigma}
\left (2 + \left ( \frac{r}{r_c} \right )^2 \right ) M_{\sun} \label{clowe_equation}
\end{equation}
\noindent
This result was also derived by Clowe, Luppino, \& Kaiser (2003) in their comparison of weak-lensing 
surface mass density with X-ray measurements.
The surface mass density can be now integrated from the cluster center to the aperture radius $r$:
\begin{equation}  
M_{ap}(r)= 2 \pi \int_0^r r^{\prime} \Sigma (r^{\prime}) dr^{\prime} =    
      1.78  \times 10^{14} \beta \left ( \frac{T}{\mbox{keV}} \right ) \left (  \frac{r}{\mbox{Mpc}}  \right )  \frac{r/r_c}{\sqrt{1+(r/r_c)^2}} M_{\sun}, \label{eqn_xray_mass_2d}
\end{equation}
\noindent
where again $r$ and $T$ must be given in units of Mpc and keV, respectively.

Using the cluster temperature and the surface brightness profile determined in \textsection\ref{xraytemp} and~\ref{isothermalbeta},
we estimate the projected mass of the cluster within 1 Mpc aperture radius to be $(1.2\pm0.2)\times10^{15} M_{\sun}$.
This mass is consistent with our parameter-free weak-lensing estimation. In \textsection\ref{xrayvslensing} we
present a detailed comparison of the X-ray mass profile with the weak-lensing result.

\section{DISCUSSION}

\subsection{X-ray Mass versus Lensing Mass \label{xrayvslensing}}

In weak-lensing analysis, the largest uncertainty is introduced when the
tangential shear is fit to a particular mass profile in order to constrain $\kappa$ in
a certain annulus. The accuracy in determining $\kappa$ for a specified area depends on the conformity
of the cluster mass profile to the parameterized model as well as the extent of azimuthal symmetry.
Nevertheless, we consider the uncertainty arising from these assumptions to be small for the current
analysis because, in spite of the apparent substructure, the mass profile of MS 1054-0321 seems
to be well described by the isothermal profile out to the field boundary 
at least in an azimuthally averaged sense. In addition, $\kappa$ is expected to be small
at the annulus near the field boundary and to approach $\kappa\approx\gamma$ with a minor
dependence on the assumed model.

On the other hand, the X-ray mass determination potentially harbors many ambiguities due to the indefiniteness
of the assumptions involved. The hypothesis that the system is in hydrodynamic equilibrium is often
questionable especially in high-redshift clusters where the clusters appear to be still in their forming stage.
Besides, many of these high-redshift clusters possess filamentary structure that makes it difficult to
use equation~\ref{eqn_xray_mass_3d} or~\ref{eqn_xray_mass_2d}, which necessitates an assumption of a spherically symmetric
mass distribution. In addition, the accurate measurement of the X-ray structural parameters (i.e., the determination of
the core radius and $\beta$ index) is a challenging task, especially for high-redshift clusters where
the cosmological surface brightness dimming reduces the photon statistics substantially. 
 
\begin{figure}
\plotone{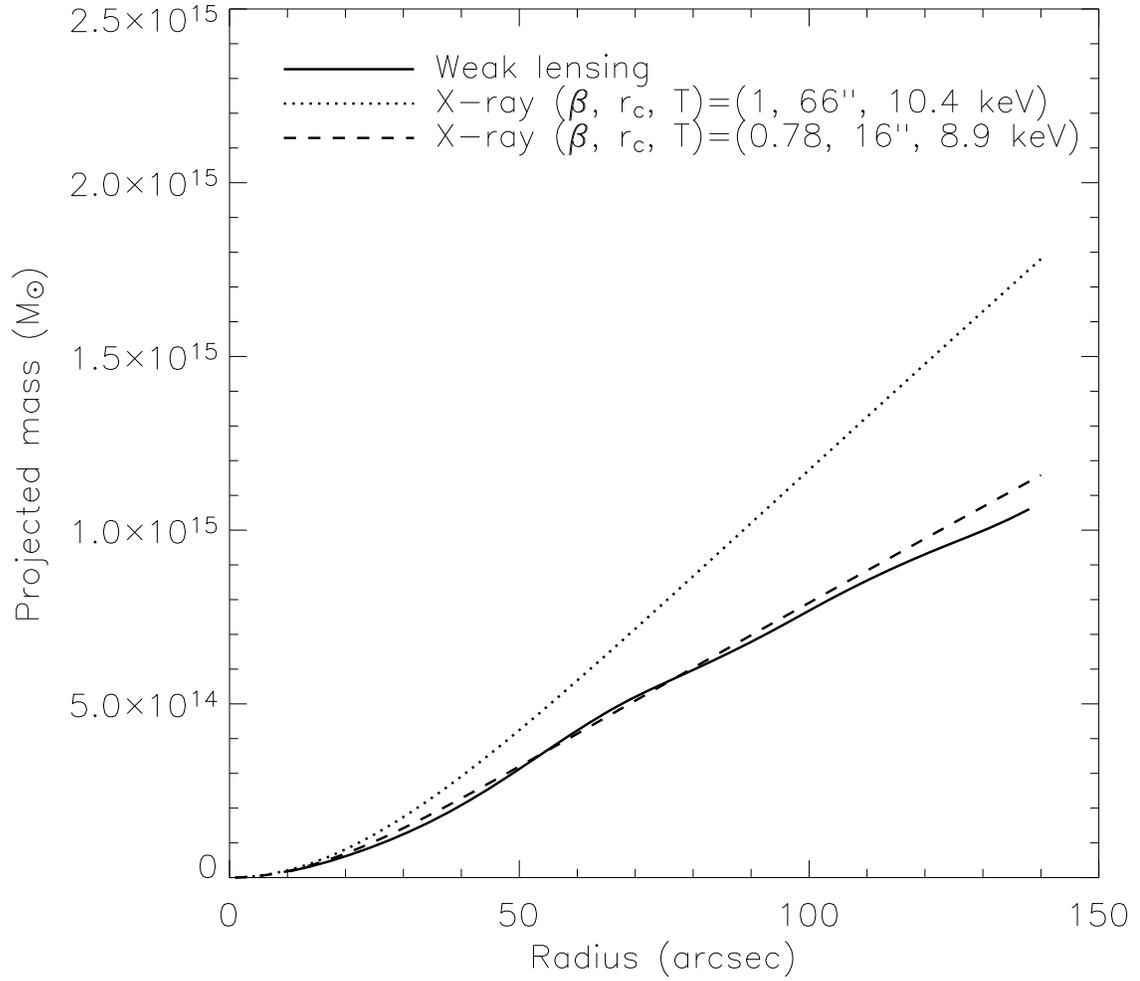}
\caption{Comparison of weak-lensing mass with X-ray results for different
isothermal beta parameters. Our adopted X-ray structual parameters ($\beta=0.78$ and $r_c=16\arcsec$)
provide the mass profile consistent with the weak-lensing result. 
\label{fig_many_profiles}}
\end{figure}

We compare a variety of the mass profiles of MS 1054-0321 in Figure~\ref{fig_many_profiles}. The X-ray mass profiles are
computed using equation~\ref{eqn_xray_mass_2d}
for different isothermal beta model parameters. First, note that the X-ray mass profile for the parameters
determined in \textsection~\ref{isothermalbeta} ($\beta=0.78$ and $r_c = 16\arcsec$) 
maintains a consistency with the weak-lensing mass profile.

A blind isothermal beta profile fitting gives unrealistically large values for the beta ($\sim1.66$) and the
core radius ($\sim81.7\arcsec$) of MS 1054-0321 as already mentioned in \textsection\ref{isothermalbeta}. The
resulting aperture mass at $r=1$ Mpc radius would also become unacceptably
huge ($\sim2.3\times10^{15} M_{\sun}$). Jeltema et al. (2001) obtained a core radius
of $\sim1.1\arcmin$ when the beta is frozen to unity. These two parameters along with
their temperature measurement of $\sim10.4$ keV suggest a total projected mass of 
$M(r<1~\mbox{Mpc})=\sim1.7\times10^{15}M_{\sun}$, which is still larger than our X-ray estimate of 
$M(r<1~\mbox{Mpc})=(1.2\pm0.2)\times10^{15} M_{\sun}$.
However, we note that the virial mass that Jeltema et al. (2001) computed from
the mass-temperature scaling relation (Evrard et al. 1996; Arnaud \& Evrard 1999) is similar to our
result and the comparison is presented in \textsection\ref{section_virialradius}.

Although it is encouraging to observe a nice concordance between the X-ray and weak-lensing masses in this particular study,
we do not expect that there exists a general agreement for other clusters. Mainly, we attribute the reconciliation of the mass
properties in the current study to the excellent conformity of the large scale gas and mass profiles of MS 1054-0321
to that of the isothermal model (i.e., $\rho_{DM}=(1+(r/r_c))^{-3/2}$ and $\rho_{ICM}=\rho_{DM}^{\beta}$) out to a large distance from the
cluster center despite the apparent temperature variation and cluster substructure.
The current method of X-ray mass estimation is limited by the assumption of
the isothermality of the cluster profile. Accordingly, the X-ray mass estimation is not
always guaranteed to lead to an unbiased result at large radii where the extrapolation
of the X-ray profile determined within a small aperture is often problematic.
For example, although J05 obtained a weak-lensing mass similar to the X-ray result
at small radii for CL 0152-1357 (a difference of $\sim10$\% at $\sim50\arcsec$), 
the discrepancy increases continuously toward the field edge
because the actual cluster mass profile determined from weak-lensing rises more slowly 
than the isothermal sphere at large radii, favoring an NFW description. 

\subsection{Uncertainties of Gravitational Lensing Mass Due to the Cosmic Shear \label{section_cosmic_shear}}

The gravitational lensing signal is sensitive to all matter along the line of sight.
As a result, the shape of a background galaxy is distorted not only by the
cluster mass, but also by the background large scale structure in front of and behind
the cluster. In the general application of weak-lensing to cluster mass estimation, the cluster is
assumed to be the dominant source of signal and the contribution from
this cosmic structure is often neglected. Such lack of concern for the background structure
can be justified in many situations
where the statistical errors resulting from the discrete sampling of the signal overwhelms
the uncertainties caused by the large scale structure.
However, in our current ACS analysis, 
the unprecedentedly high number density of source galaxies reduces the measurement
errors substantially, making other sources of errors more important.

In this subsection, we assess the effect of the large scale density field
in our mass estimation of MS 1054-0321 following the formalism
of Schneider et al. (1998) and Hoekstra (2001, H01 hereafter) (see also Kaiser 1992;1998 for
the original derivations of many useful equations). 
The variance of the aperture mass due to the large scale density fluctuation is given by:

\begin{equation}
\left<M_{AP}(\theta)^2\right> = 2 \pi \int_0^{\infty} \mbox{d}s~ P_{\kappa}(s) I(s,\theta)^2,
\end{equation}
\noindent
where $P_{\kappa}$ is the projected power spectrum defined as:
\begin{equation}
P_{\kappa}(s) = \frac{9}{4} \left ( \frac{H_0}{c} \right )^4 \Omega_M^2
\int_0^{w_H} \mbox{d} w \frac {g^2(w)}{a^2(w)} P \left( \frac{s}{f_K(w)} ; w \right) \label{eqn_projected_p}
\end{equation}
\noindent
and $I(s,\theta)$ is a filter function dependent on the mass estimation method. The form
of the functions for the SIS model and the aperture mass densitometry ($\zeta$-statistic)
can be found in H01. In equation~\ref{eqn_projected_p}, $g(w)$ is
a source-averaged angular diameter distance ratio $D_{ls}/D_s$, 
$a(w)$ is the scale factor, and $f_K(w)$ is the comoving angular diameter distance.
We computed $g(w)$ using the redshift distribution obtained from the UDF after
applying the same source selection criteria to the UDF catalog.
In the evaluation of the projected power spectrum (eqn.~\ref{eqn_projected_p}), we first
constructed the non-linear power spectrum (Peacock \& Dodds 1996) for the current WMAP
cosmology and then integrated it out to
the comoving distance to horizon $w_H$.

\begin{figure}
\plottwo{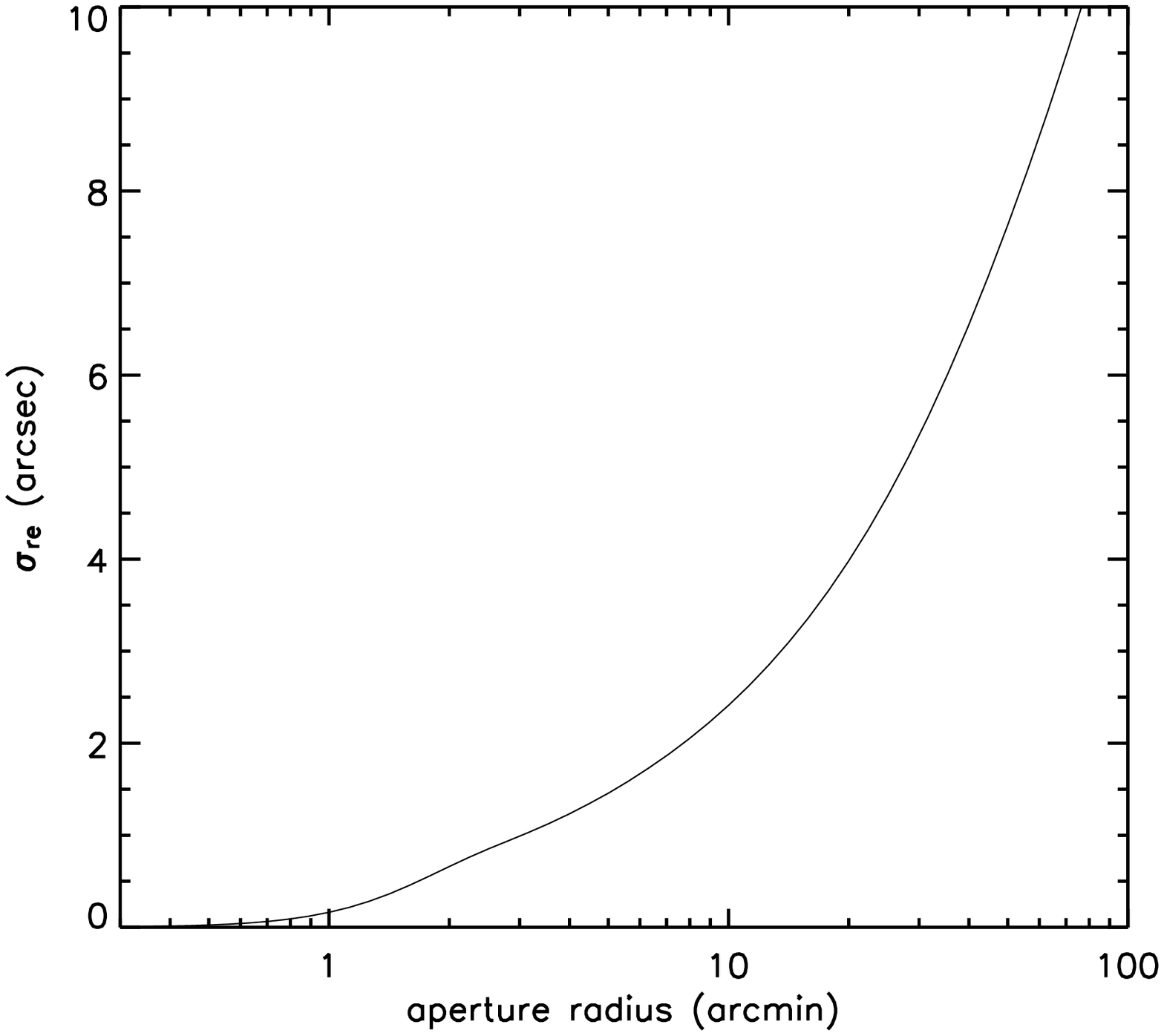}{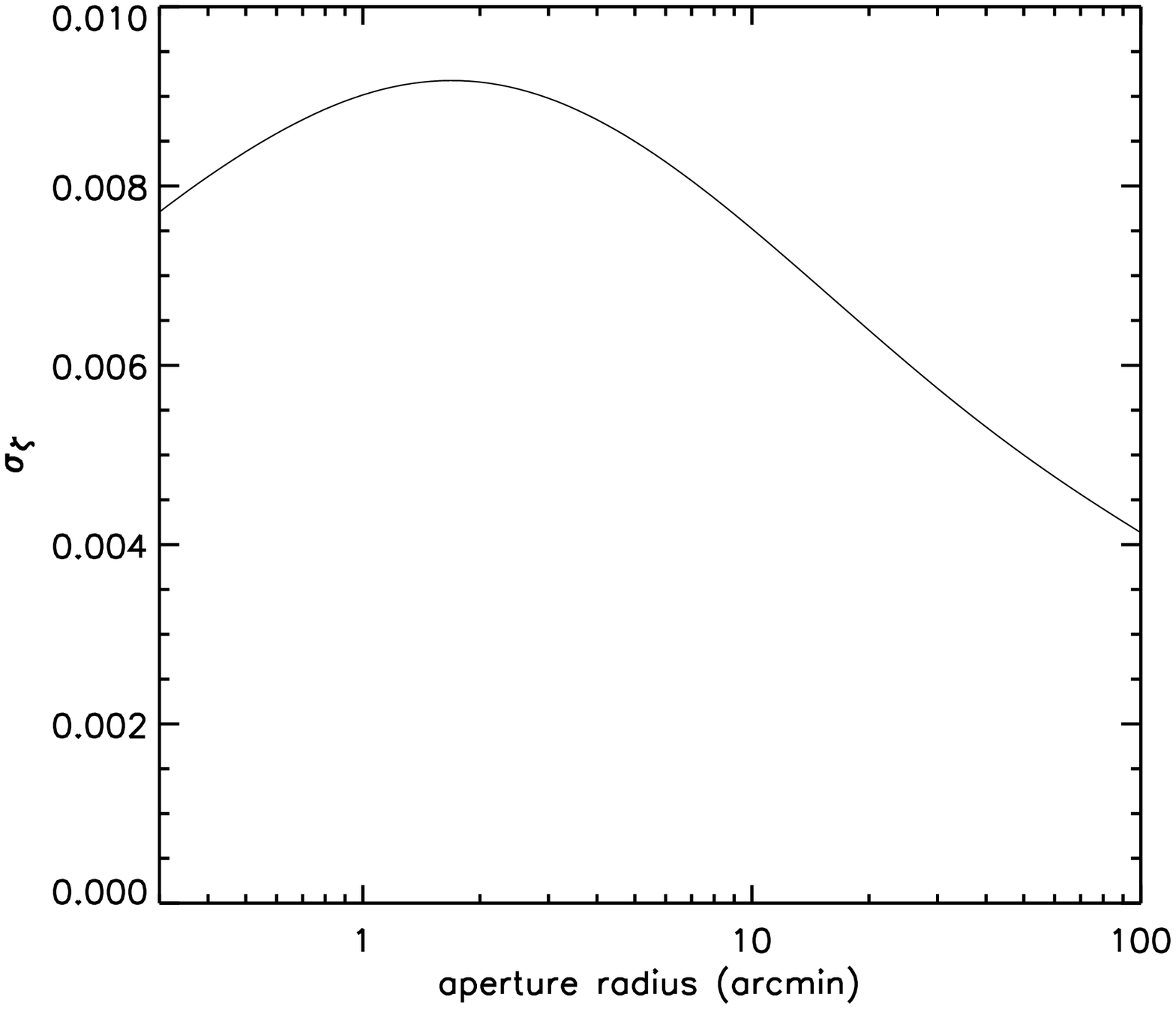}
\caption{Predicted uncertainties in the measurement of the Einstein radius from the SIS fit (left panel)
and the aperture mass densitometry (right panel).
\label{fig_uncertainty}}
\end{figure}

We show the predicted uncertainties in the estimate of the Einstein radius from the SIS fit and
the aperture mass densitometry for our source redshift distribution
in Figure~\ref{fig_uncertainty}. The overall shape of the aperture-dependent variation
and the magnitude of the predicted uncertainties are
similar to those estimated by Schneider et al. (1998) and H01.
At 1 Mpc ($\sim132\arcsec$), we predict $\sigma_{\theta_E}\simeq0.74\arcsec$ and $\sigma_{\zeta}\simeq9\times10^{-3}$
for the SIS fit and the aperture mass densitometry, respectively, due
to the cosmic shear effect. The uncertainty of the best-fit SIS Einstein radius is only marginally increased
from $\theta_E = 11\arcsec.1\pm1.\arcsec0$ to $\theta_E = 11\arcsec.1\pm1.\arcsec2$.
On the other hand, the effect is relatively large for the aperture mass densitometry.
Assuming that $\sigma_{\theta_E}$ and $\sigma_{\zeta}$ are correlated, we
can estimate the upper limit on the uncertainty for the aperture mass densitometry
to be $\sigma_{\bar{\kappa}}\simeq0.012$. This value of $\sigma_{\bar{\kappa}}$ due to the 
cosmic shear introduces an additional $\sim14$\% error to the total cluster mass estimation
at 1 Mpc, yielding $M(r<1\mbox{Mpc})=(1.02\pm0.15)\times10^{15}M_{\sun}$ (adding the statistical
error and the cosmic shear effect in quadrature).

The brief analysis above demonstrates that the cosmic shear effect is a significant
factor in the total error budget of the cluster mass estimate. In general, the uncertainty due
to the effect increases as we probe into higher redshift regimes. However, as
indicated by H01, the fractional uncertainty ($\left <\kappa_{cs} \right >^{1/2} / \left <
\kappa \right >$ increases only moderately with cluster redshift 
for rich clusters (e.g., $\sigma=1000 \mbox{km}~\mbox{s}^{-1}$
) if we choose faint (e.g., $20<R<26$) source population. In the current analysis, the high-resolution
of the ACS images as well as the thorough knowledge of the instrumental PSF
allow us to select the unprecedentedly faint ($24<z_{850}<28$) galaxies as our sources, minimizing
the cosmic shear effect. H01 shows that if the $20<R<24$ (a typical magnitude range
for ground-based weak-lensing analysis) galaxies are used instead, the cosmic
shear can contribute a nearly 50\% uncertainty in the total mass even for a rich cluster at $z\simeq0.8$.

A promising way to improve the accuracy dramatically in cluster mass estimate yet not attempted
in the current analysis is to perform a so-called weak-lensing tomography. When
good photometric redshift information for individual source galaxies becomes
available in the future, it is possible to separate the cluster lensing signal from the cosmic shear 
utilizing the differential lensing efficiency.

\subsection{Comparison of the Cluster Substructure with HFK00 \label{section_hfk00_compare}}
\begin{figure}
\plotone{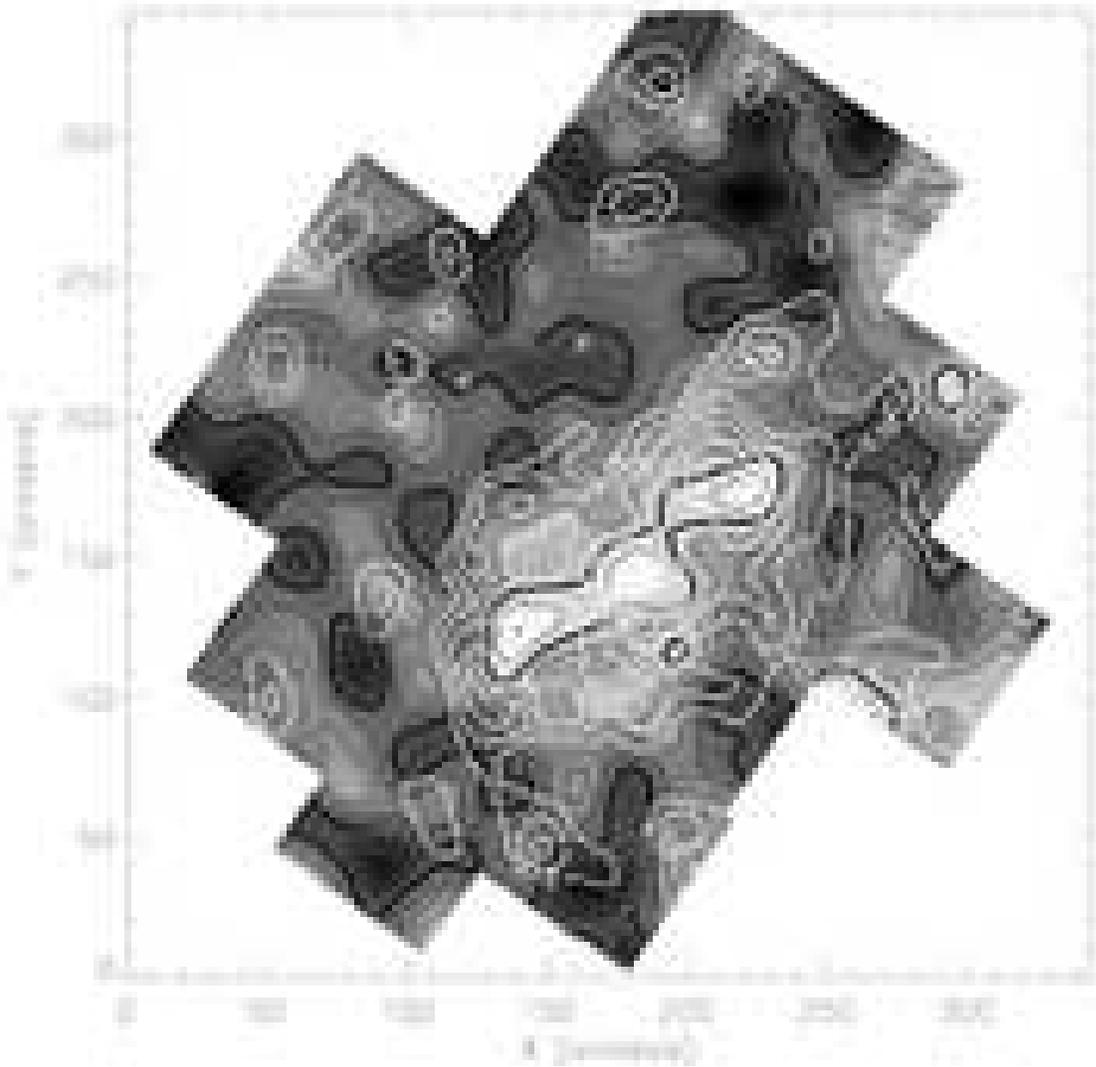}
\caption{Comparison of the mass reconstruction with the WFPC2 observations (HFK00). The background gray-scale
is the mass map presented in HFK00 and the white contours are the current mass reconstruction
based on the ACS observations. The alignment between the two sets of contours is only approximate ($\sim6\arcsec$). 
\label{fig_hfk00}}
\end{figure}

Figure~\ref{fig_hfk00} shows our mass reconstruction contours on top of the 
HFK00 result. The alignment between the two sets of the contours is only approximate (
$\sim6\arcsec$). 
The main structure of MS 1054-0321 characterized by the three major clumps is obvious
in both results. The central and western mass clumps in HFK00 are in good spatial
agreement with our results. However, we note that their eastern clump is displaced to 
the west by $\sim20\arcsec$. In our mass reconstruction the eastern
clump is further resolved into two or possibly three smaller mass peaks, in better
spatial agreement with the cluster optical lights (see Figure~\ref{fig_massoverlum}).

There are some indications that the minor peaks  
labeled as M1-4 in Figure~\ref{fig_massoverplot} are also present
in HFK00 mass map, but the significance of their
detections appears to be marginal and the locations have greater offsets
from the corresponding luminosity clumps.

\subsection{Virial Radius \label{section_virialradius}}
There exist different definitions for a virial radius in the literature, mostly due to
the discrepant definitions of a characteristic density of a halo. 
We define the virial radius of the cluster as a radius
where a mean density inside the spherical volume
becomes 200 times the critical density at the redshift of the cluster. Although 
spherical collapse simulations suggest different criteria for
a virialized halo depending on the cosmological parameters and normalization, we adopt this
conventional definition to ensure that a straightforward comparison can be made to previous work.
 
If the isothermal beta model (Eqn.~\ref{eqn_xray_mass_3d}) is used, the X-ray measurements
predict a virial radius of $1.7\pm0.2$ Mpc ($\sim224\arcsec$) for MS 1054-0321. A similar
value of $1.5\pm0.1$ Mpc is
obtained if the weak-lensing mass from the SIS fit result is extrapolated assuming spherical symmetry.
The corresponding virial masses $within$ $the$ $sphere$, then,
become $(1.2\pm0.2)\times 10^{15} M_{\sun}$ and
$(1.1\pm0.1) \times 10^{15} M_{\sun}$ from the X-ray and weak-lensing analyses, respectively.

Jeltema et al. (2001) quoted $M_{200}=6.2_{-1.3}^{+1.6}\times10^{14} M_{\sun}$ within $r=0.76$ Mpc
from the mass-temperature scaling relation (Evrard et al. 1996; Arnaud \& Evrard 1999) in the $(\Omega_M,\Omega_{\Lambda},h)=(1,0,1)$
universe. Using the same cosmological parameters, our X-ray virial mass is transformed to a
similar value of $M (r<0.74$ Mpc)=$(5.8\pm0.7) \times10^{14} M_{\sun}$ 
from equation~\ref{eqn_xray_mass_3d}. 
Although Jeltema et al. (2001) claimed that their mass estimate is lower than the HFK00 weak-lensing
mass by $\sim38$\% (compared at the same radius of r=0.76 $h_{100}$ Mpc in the above $\Omega_M=1$ cosmology),
our analysis shows that the two results are consistent with each other (on the contrary, their
X-ray mass becomes even slightly higher than the weak-lensing mass) when the projected
weak-lensing mass is properly converted to the total cluster mass within a spherical volume.

\subsection{Comparison of the Velocity Dispersion and Sunyaev-Zeldovich Analyses}

The most recent determination of the velocity dispersion of MS 1054-0321 from the spectroscopic catalog
of the cluster members (Tran et al. in prep.) is
$1153\pm80 \mbox{km}~\mbox{s}^{-1}$. The value is very close to our weak-lensing prediction
(\textsection\ref{mass_estimate}) of $1150_{-51}^{+49}$ km$s^{-1}$ under the SIS assumption
The empirical relation between X-ray temperature and velocity dispersion is rather scattered
around the theoretical $\sigma \propto T^{1/2}$ line.  Adopting 
($\sigma_v / \mbox{km~s}^{-1})=10^{2.57\pm0.13}(kT/\mbox{keV})^{0.59\pm0.14}$
(Wu et al. 1998), we obtain $\sigma_v = 1349_{-340}^{+461} \mbox{km} \mbox{s}^{-1}$. If we
use the theoretical relation instead assuming energy equi-partition, the X-ray
temperature of $T=8.9_{-0.8}^{+1.0}$ keV corresponds to $\sigma_v=1202_{-55}^{+66} \mbox{km s}^{-1}$.

From Sunyaev-Zeldovich (SZ) imaging of the cluster, Joy et al. (2001) estimated a temperature 
of $10.4_{-2.0}^{+5.0}$ keV and a projected mass ($r<94\arcsec$) of $(4.6\pm0.8)\times10^{14} M_{\sun}$
(under the $(\Omega_M,\Omega_{\Lambda},h_{100})=(0.3,0.7,1)$ cosmology). The conversion of
our weak-lensing result to their cosmological parameters yields a projected mass of 
$(4.9\pm0.3) \times 10^{14} M_{\sun}$ ($r<94\arcsec$), which is in good agreement with
the SZ estimation.

It may be considered surprising to observe the converging results between weak-lensing
and other various analyses above especially for this high-redshift cluster.
One interpretation of these impressive agreements is that the dynamical structure of MS 1054-0321 may already have
matured even at $z\sim0.83$ to the extent that at least the azimuthally averaged properties do not depart
greatly from that of a relaxed cluster in spite of the elongated appearance of the cluster galaxy distribution. 
It will be interesting to examine in future investigations how critical the
role of the ICM is in the global virialization of the cluster.

\subsection{Offsets between Galaxies, ICM, and Dark Matter \label{lensing_vs_xray_substructure}}

\begin{figure}
\plottwo{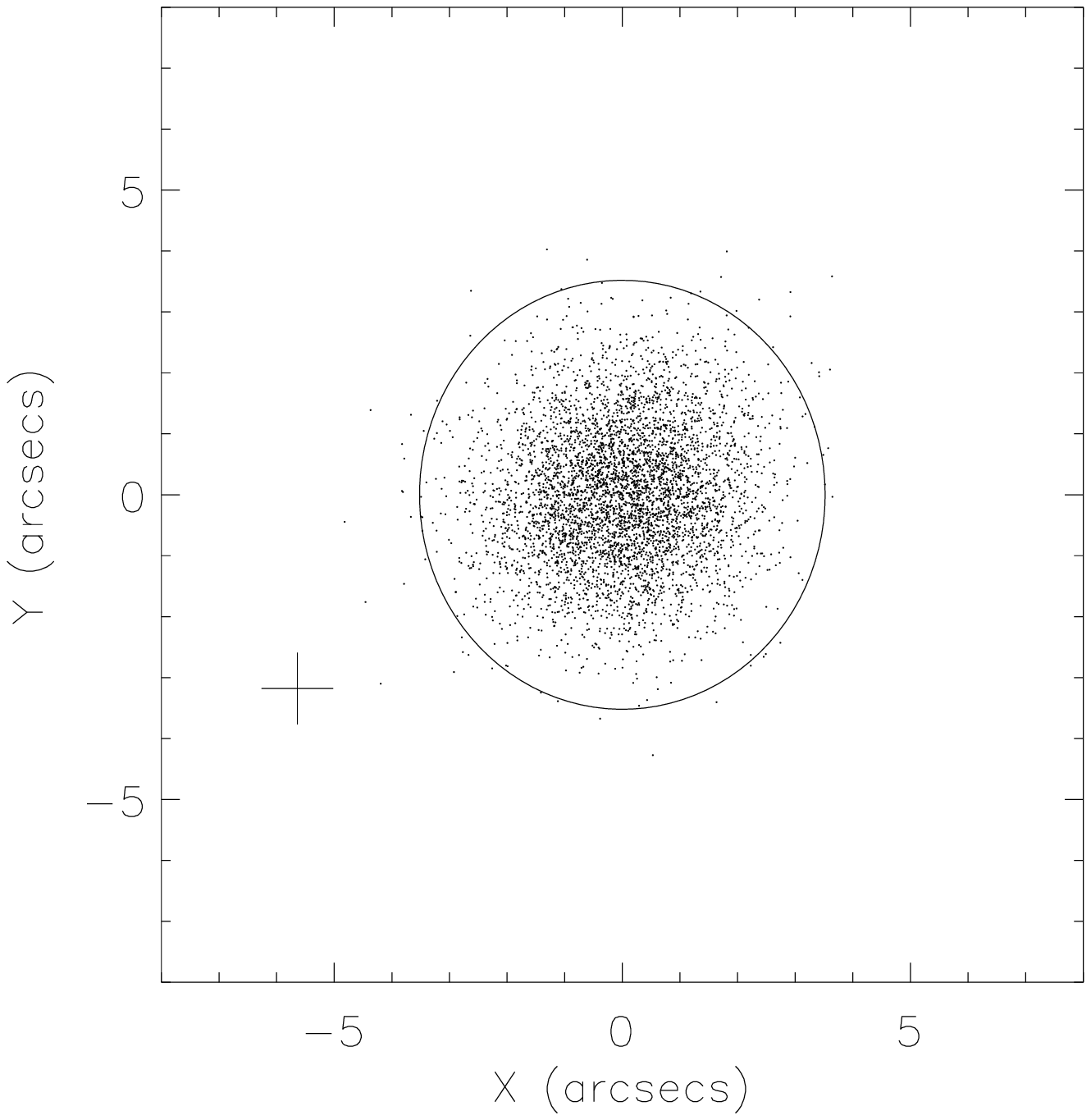}{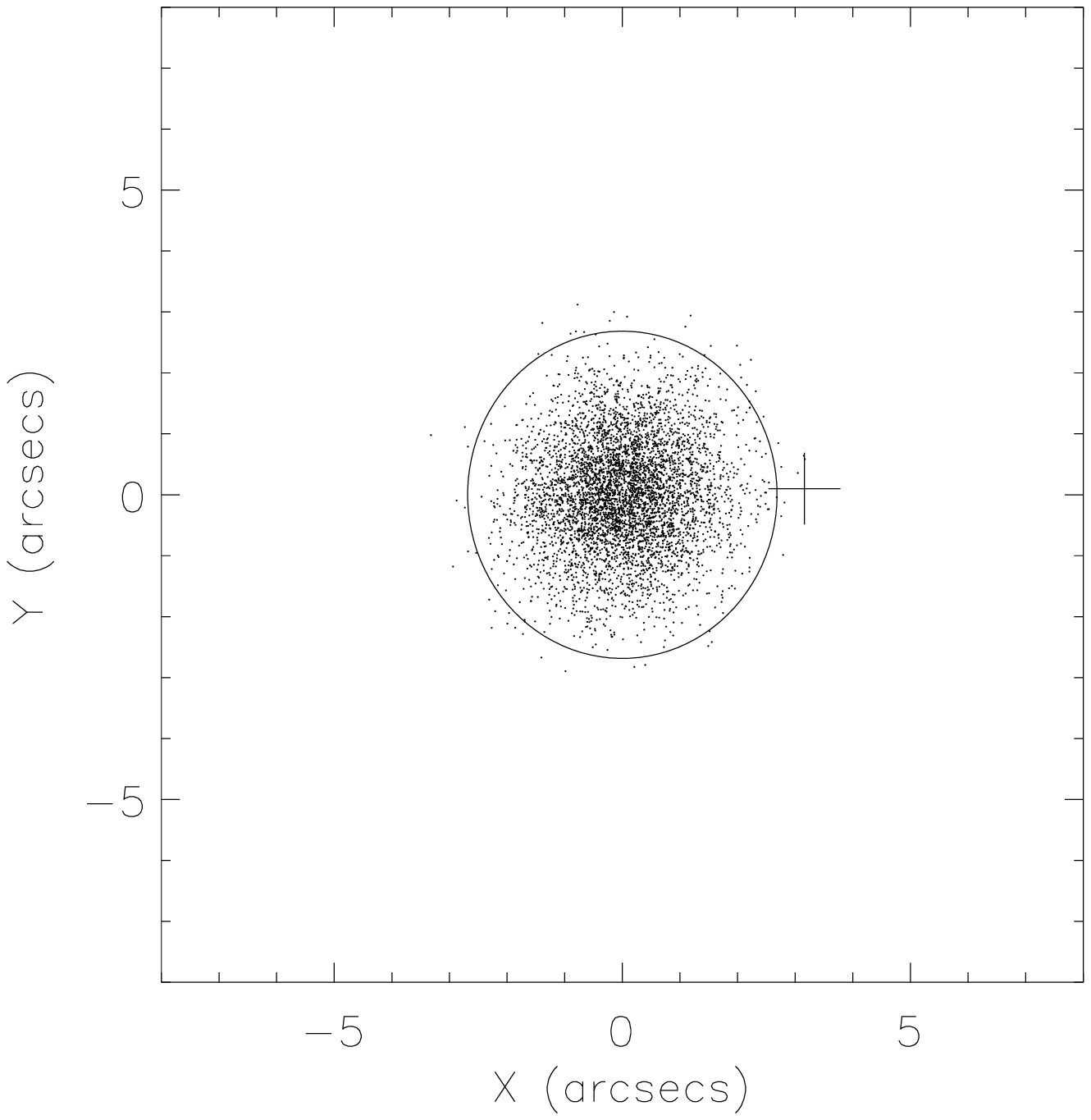}
\caption{Centroid distribution in 5000 bootstrap resampling of source galaxies. The central mass clump (left panel)
has a larger distribution than the western mass clump (right panel), which is consistent with their different masses
(thus with different significance). The corresponding luminosity centers are outside the boundary (solid), which
encloses 99\% of the centroids.
\label{fig_centroid_test}}
\end{figure}

In \textsection\ref{substructure}, we discussed the
cluster substructure revealed by the weak-lensing mass reconstruction
in comparison with the cluster luminosity distribution. A careful
examination of the mass/light overlay (Figure~\ref{fig_massoverlum}) shows
that there present some offsets between the mass and luminosity centroids.
In order to investigate the statistical significance of these offsets, we
ran a 5000 bootstrap resampling of background galaxies and measured the
centroid of the mass peaks from each realization. We found that only the central
and western mass clumps possess statistically significant offsets
with respect to the corresponding luminosity peaks. Figure~\ref{fig_centroid_test}
demonstrates that both luminosity centers are outside the 99\% circle of the mass peak
centroid distribution. 

\begin{figure}
\plottwo{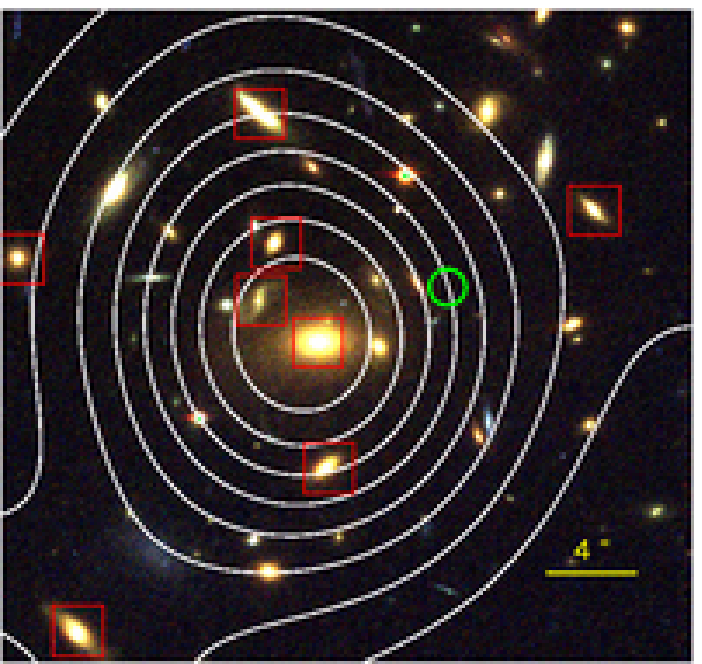}{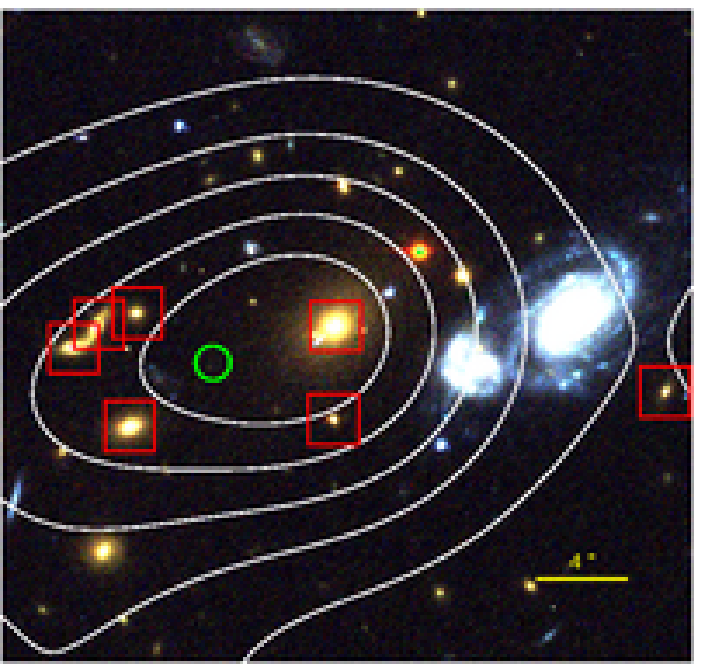}
\caption{The luminosity contours overlaid on the enlarged color images of the central (left) and western (right)
mass clumps. The luminosity contours are drawn in white, and red
squares are placed on the spectroscopically confirmed members. We also mark the mass centroids
with green circles. It is apparent that the luminosity centroids are
mostly influenced by the BCGs. The smoothed central luminosity
peak has its centroid only $\sim 1\arcsec$ apart from the location of the BCG.
The offset is $\sim 2\arcsec$ for the western luminosity clump.
\label{fig_lum_centroids}}
\end{figure}

Another important question in this mass-light centroid
comparison is whether the above luminosity centroids are fair locations to compare
with the mass centroids. If the luminosity centroids are severely
affected by the incompleteness and/or the smoothing kernel size, we may
also need to consider the resulting uncertainties in our analysis. 
We present the overlay of the detailed luminosity contours of the central
and western clumps on top of their color images in
Figure~\ref{fig_lum_centroids}. The luminosity contours are drawn in white, and red
squares are placed on the spectroscopically confirmed members. We also mark the mass centroids
with green circles. It is apparent that the luminosity centroids are
mostly influenced by the BCGs. The smoothed central luminosity
peak has its centroid only $\sim 1\arcsec$ apart from the location of the BCG.
The offset is $\sim 2\arcsec$ for the western luminosity clump. We remember
that the luminosity map is smoothed with a FWHM$\simeq14\arcsec$ Gaussian kernel whereas we
used a FWHM$\sim10\arcsec$ kernel for the smoothing of the mass map. 
Considering an additional smoothing ( $\sim10\arcsec$) implicitly done in averaging source
ellipcities before the mass reconstruction, we estimate that the FWHM of $14\arcsec$ for the
luminosity smoothing is a proper kernel size to ensure a comparable smoothing scale
(dispersion by finite sizes of galaxies can be neglected because the 
FWHM$\simeq14\arcsec$ Gaussian kernel is so dominant).
If one argues that a smaller kernel must have been used in smoothing the light distribution 
in order to secure a fair comparison, it is easy to ascertain that a smaller kernel
will only shift the luminosity centroids toward the BCGs, enlarging
the offset between the luminosity and mass centroids. Alternatively, if one prefers
a larger kernel for a luminosity map, we find that increasing kernel size reduces the
offsets between the luminosity and mass centroids, which, nevertheless,
still remain significant at a 2$\sigma$ level ($\gtrsim 4\arcsec$ and $\gtrsim 2.5\arcsec$ for the central 
and western clumps, respectively)
up to a FWHM$\sim20\arcsec$. Of course, for this large kernel (FWHM$\sim20\arcsec$), the contribution
from the cluster members absent from the spectroscopic catalog may become relatively important.
However, the examination of the distribution of unconfirmed cluster candidates around the BCGs
does not convince us that
this will modify the significance of the mass/luminosity offsets substantially.

Given the above argument that the luminosity centroids stay close to
the BCGs for the reasonable values of the smoothing kernel, we list the following
possibilities as causes of the observed offsets:
\begin{itemize}
\item noises from discrete sampling of background galaxies,
\item systematics shifts due to foreground masses,
and
\item real features reflecting different hydrodynamical properties of individual cluster constituents.
\end{itemize}

1. {\it Noises from discrete sampling of background galaxies.} In general, the finite sampling 
of the background galaxies whose spatial distribution is not uniform scatters
the centroids of mass reconstruction from their true positions. The stability of the centroids
are mainly determined by the strength of the lensing signal as well as the number density
of available source galaxies. As we found that the signal
around the two mass peaks is very strong and the number density of sources is 
unprecedentedly high ($\sim154~\mbox{arcmin}^{-2}$), the uncertainty
of the centroids are expected to be small.
One of the most useful tests to assess the errors is
a bootstrap resampling of background galaxies discussed above. The result is consistent with
our intuition that the more pronounced peak (the western mass clump) has
a smaller dispersion than the weaker one (the central mass clump).  
Because both luminosity centroids are significantly outside the 99\% enclosing circle of the corresponding
mass centroids (Figure~\ref{fig_centroid_test}), it is very unlikely that
the observed mass/light offsets originate from the noises.

2. {\it Systematic shifts due to foreground masses.} As light rays are perturbed by any objects between
the observer and source, we cannot preclude the possibility that there might exist some
foreground/background masses responsible for the mass centroids. 
In \textsection\ref{section_cosmic_shear}, we have estimated the uncertainties in our cluster mass
due to the presence of other large scale structures. In particular, lower redshift foreground objects can
work as more efficient lens and measurably alter the mass map even if their masses
are not as significant as that of MS 1054-3021. One of the most unambiguous tests is
to perform a so-called weak-lensing tomography, which can, in principle, separate the cluster mass from
the dynamically uncorrelated background/foreground structures. The technique, however, requires
us to obtain a good photometric redshift knowledge of individual sources and thus can become
practical only in the future.

Nevertheless, without this sophisticated check, it is still possible to discuss the effects
of these possible interlopers by scrutinizing our mass map. If the foreground masses
are to greatly affect the centroids of the cluster mass clumps, they must have a rather
peaked distribution whose width (e.g., FWHM) is less than or comparable to those of the cluster mass peaks.
(if the foreground mass distributions are much smoother than the characteristic scale
of the mass peaks, they cannot shift the cluster centroids by the large amount as observed in our case).
Besides, the peaked foreground mass must be located very close to the mass centroids of the cluster
perhaps within a smoothing kernel (FWHM$\simeq10\arcsec$) in order to
shift the centroids. Otherwise, we suspect that they should reveal themselves as separate peaks
in our high-resolution mass reconstruction. Finally, we analyze the spectroscopic catalog of the cluster field
to examine if there is any concentration of foreground galaxies. Out of 325 objects whose
redshifts are known, about $150$ galaxies are found to lie foreground. Although the spectroscopic 
survey is incomplete and still in progress (Tran et al. in preparation), we found no significant
foreground groups near the two cluster mass centroids. 

3. {\it Real features reflecting different hydrodynamical properties of individual cluster constituents.}
There is a growing list of X-ray clusters whose intracluster gas might have been swept back
behind the corresponding cluster galaxies possibly due to the ram pressure (e.g., Markevitch et al. 2002;
Maughan et al. 2003; Scharf et al. 2004). Although the discovery is rather recent,
the ICM-galaxy offsets seem to be acquiring increasing observational supports. On the other hand,
the mass-galaxy offsets have not been reported in any other investigations yet except in
our previous weak-lensing analysis of CL 0152-1357 (J05).
Markevitch et al. (2004) suggest that one can use the mass-galaxy displacements to
constrain the collisional cross-section of the dark matter particles because, if
cluster galaxies are to suffer any deceleration from the dark matter collisions,
the cluster galaxies should be hauled by the dominant dark matter potential well
and appear to be displaced away from the merging direction. 

Alternatively, one can also propose a different scenario wherein the dark matter
particles are purely collisionless, but collisional properties of the cluster galaxies
are not negligible. This is against a conventional yet disputable belief that galaxies
are effectively collisionless. Detailed numerical simulations must be followed in
order to estimate whether the effect is measurable and can lead to such an observed offset.
   
\begin{figure}
\plotone{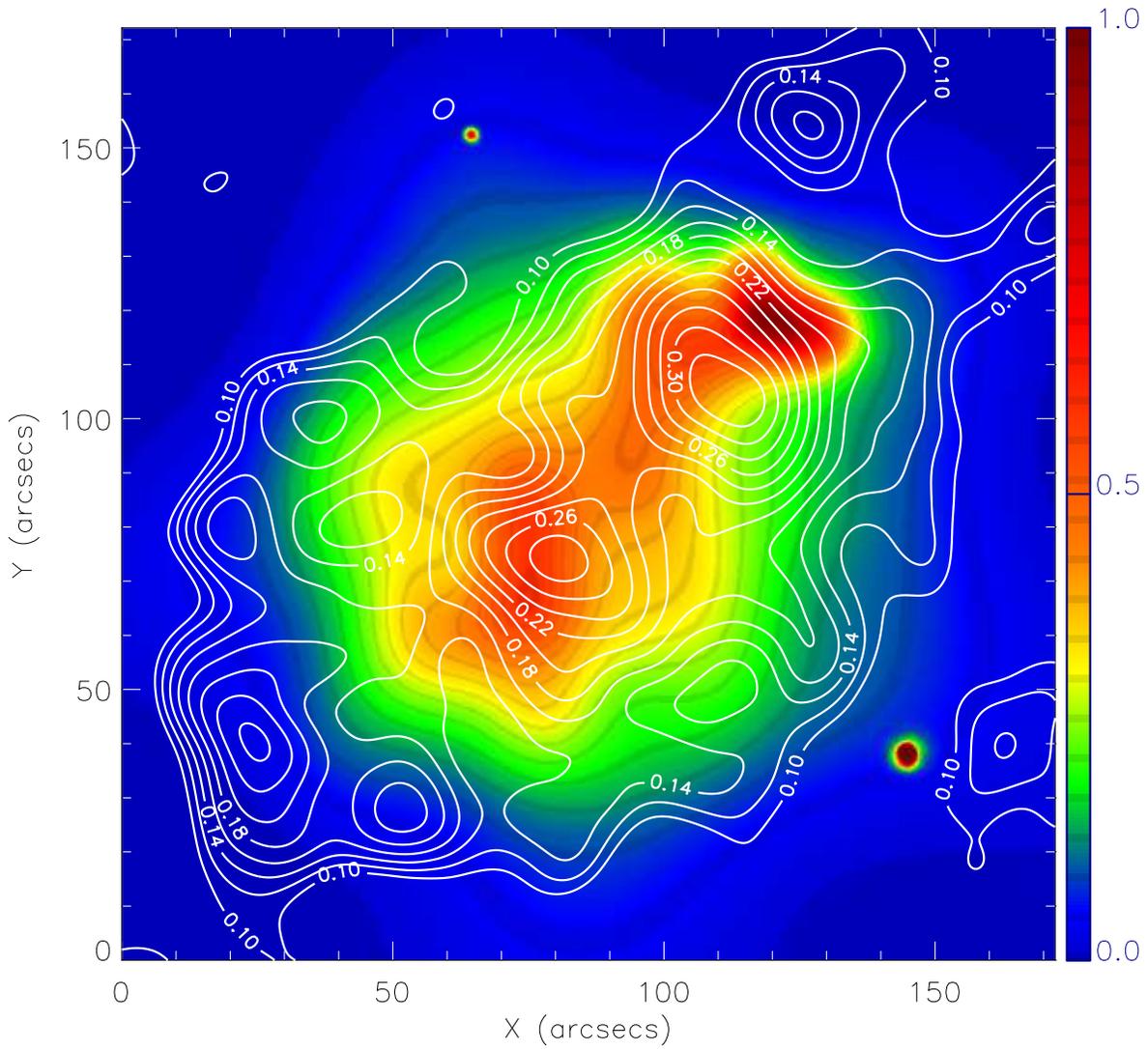}
\caption{Mass reconstruction on top of the X-ray map. The eastern mass clump is not
detected in the $Chandra$ X-ray observation. Under the hypothesis of a
merger, the central and western dark matter clumps seem to be moving ahead
of the corresponding X-ray peaks.
\label{fig_mass_xray}}
\end{figure}

As argued by Clowe et al. (2004), the mass-ICM offset can be used to
support the existence of dark matter.
There have been some attempts to explain
the ``missing matter'', particularly in clusters of galaxies, in the context of the Modified Newtonian
dynamics (MOND; Milgrom 1983). The MOND theory has been
partly successful in describing the observed velocity dispersions of galaxy
clusters as well as rotation curves of spiral galaxies 
without inclusion of dark matter by modifying the Newtonian
gravity so that the inertia mass of a particle is decreased in the limit of low accelerations
(e.g., McGaugh \& de Blok 1998; Sanders 2003; Milgrom \& Sanders 2003).
Recently, Bekenstein (2004) has proposed a relativistic extension of MOND, which now
predicts definite gravitational lensing. 
If we view the cluster in this MOND paradigm, the centroids of the mass clumps
are expected to lie
on top of the X-ray peaks, which trace the dominant baryonic component of the cluster.
However, our lensing analysis (Figure~\ref{fig_mass_xray}) shows with high significance that the locations of these X-ray peaks are
not where the dominant cluster mass is concentrated. 
A similar trend is observed in the weak-lensing analysis of the interacting cluster 1E 0657-558 (Clowe et al. 2004).
The 1E 0657-558 cluster shows not only the dramatic bow-shock
from the X-ray emission, but also the two dark matter clumps conspicuously offset from the
corresponding X-ray peaks.

\subsection{Absence of the Eastern X-ray Peak \label{section_absence}}

As is clear in Figure~\ref{fig_mass_xray}, the Chandra observation does not detect
any X-ray excess over the eastern region of MS 1054-0321 where the weak-lensing and
luminosity map reveal distinct mass concentrations. The absence of this third weak
lensing peak in the X-ray observation was first observed by Clowe et al. (2000)
in the comparison of the ROSAT/HRI X-ray analysis with their Keck weak-lensing result.

Jeltema et al. (2001) proposed two possibilities regarding the discrepancy: immaturity
of the eastern clump or an artifact of noisy mass reconstruction caused by sub-optimal smoothing (Marshall 2002).
The high significance of the substructure detected by the unprecedentedly high number density
of source galaxies using $HST$/ACS images immediately rules out the second possibility. 
Besides, the consistency of our weak-lensing substructure with the result of HFK00 despite the different
selection and measurement of background galaxies, as well as 
the overdensity of the spectroscopically confirmed cluster galaxies at the eastern mass clump
evidence that this eastern substructure is real.

The first scenario that the third clump has not fully developed into an X-ray
system is an interesting possibility, considering the extremely high temperature ($10^7-10^8 \mbox{K}{\degr}$) required
for thermal Bremsstrahlung radiation. As seen by the cluster light distribution, the cluster galaxies
in this region appear to be less concentrated and luminous than in the other two clumps. The weak-lensing
mass in the eastern clump is also somewhat lower ($\sim70$\%
of the western mass clump). 

However, the predicted gas temperature derived from 
an empirical $\sigma-T$ relation (e.g., Wu et al. 1998) is $3.3\pm0.6$ keV after the projected mass 
($(9.3\pm1.5)\times 10^{13}M_{\sun}$
within $30\arcsec$) is converted to the velocity dispersion ($747\pm65 \mbox{km}~\mbox{s}^{-1}$) under SIS assumption.
Hence, even considering
the scattered $\sigma-T$ relation and the surface brightness dimming at $z\sim0.83$, it is
difficult to justify the lack of the X-ray emission exclusively by the low temperature of the subcluster.
The western Lynx cluster RXJ 0848+4453 at z=1.27 (Stanford et al. 2001) has
even a lower X-ray temperature of $T<2$ keV with a weak-lensing mass of $M(<0.5$ Mpc$)\sim2\times10^{14}$
(M. Jee et al., in preparation), but the $Chandra$ X-ray image confirms the presence of the
X-ray emission associated with the cluster.

An alternative scenario can be envisaged, in which most of the intracluster gas of the eastern clump has been stripped
while passing through the dense region of the central substructure.
If we imagine that clump E traveled from the southwest of clump C and passed through it
possibly with an off-center collision, clump E would then lose a significant fraction
of its ICM, or most of its ICM would become bound to that of clump C, leaving a double-peaked
structure as is hinted in our current X-ray contours.
This post-merger picture is partly supported by the temperature structure of the cluster (Figure~\ref{fig_temp_map}) showing
that the cluster temperature is rapidly rising in opposite directions from the central
X-ray peak to the east and the west as mentioned in \textsection\ref{xraytemp}. 
A similar temperature structure is observed in numerical simulations of cluster ICM and particularly
Takizawa (2000) demonstrated in his off-center merger simulation that double-peak features in X-ray emission
would survive even after the most violent epoch.

The flat inner region (possibly with double core) of the central X-ray peak yields a $\beta$ index that
is greater than unity (\textsection~\ref{isothermalbeta}) when the azimuthally averaged radial
profile is fit to the isothermal beta profile. The large $\beta$ value indicates
that the ICM of the region might have been distorted due to the recent merger. The numerical
simulation by Takizawa (1999) showed that the specific energy ratio of dark matter and the ICM (eqn.~\ref{eqn_specific_ratio})
remains greater than unity even in the late stage of the merger event. In addition, the rather scattered distribution of cluster galaxies
and dark matter in the eastern region might be associated with
the gravitational disruption during this hypothesized pass-through. 

Because the current resolution of the temperature map is not limited by the optical resolution, but
by the poor photon statistics,
a longer exposure X-ray observation of MS 1054-0321 
will provide more significant details of the thermodynamical structure out to a larger distance from the
cluster center and enable us to test this merger hypothesis further.

\subsection{Formation Sequence of MS 1054-0321}

Since more massive clusters are believed to have collapsed earlier within a hierarchical structure
formation paradigm, it is natural to conjecture that these clusters are also more likely to
possess higher $M/L_B$ values resulting from their older stellar population (Bahcall \& Comerford 2002).
Compared with CL 0152-1357 (J05) at a similar redshift of z=0.84, MS 1054-0321 has both higher
mass and $M/L_B$ within 1 Mpc, consistent with the hypothesis that $M/L_B$ values increase with cluster richness.

If MS 1054-03 is indeed being formed by the hierarchical infall of the smaller mass clumps, which
have collapsed at different times, we can also extend and apply the above argument to the wide range of 
$M/L_B$ ratios of the substructure. Table 1 shows that the $M/L_B$ ratios of the cluster main clumps
(E, C, and W) increase with their masses (we do not include the minor clumps (M1-4) in the discussion
because of the reasons explained in \textsection\ref{substructure}).
If this trend is interpreted as an indication of the formation time sequence,
the highest $M/L_B$ value of the western clump may signify the earliest formation epoch of the substructure.
This speculation is further supported by the cluster metallicty distribution (Table 3;\textsection\ref{xraytemp}). 
The western X-ray peak has higher metal abundance than the other X-ray peak, which
suggests that the stellar contents of the galaxies
associated with the western X-ray peak might be older.

\section{SUMMARY AND CONCLUSIONS}
We have presented our $HST$/ACS weak-lensing and $Chandra$ X-ray analysis of MS 1054, guided with
an extensive spectroscopic survey data of the cluster field. The weak-lensing mass reconstruction,
empowered by the superb resolution and sensitivity of ACS, not only confirms the existence
of the three dominant mass clumps reported by HFK00, but also further resolves
the cluster substructure in unprecedented detail. 

In the $Chandra$ X-ray analysis, we took care to properly
compensate for the low-energy DQE degradation of the instrument. As demonstrated
by a variety of different results in the literature, the temperature measurement
is sensitive to the applied correction. In addition to the low-energy
DQE correction, it has been realized that even a mild background flare can
bias the temperature measurement by a few keV. Utilizing the newly
available calibration data and recommendation from CIAO, we obtained $T=8.9_{-0.8}^{+1.0}$ keV
for the cluster temperature as a whole. As first noted by Jeltema et al. (2001), the
western X-ray peak has a lower temperature than the central X-ray peak. 
In our temperature map, the cluster seems to have a severe temperature variation
across the field. 

We excluded the central region ($r<45\arcsec$) in our
determination of the structural parameter because otherwise the substructure
leads to a physically unacceptable condition for a relaxed isothermal beta sphere, where the cluster dark matter
possesses a significantly higher temperature than the ICM particles.
We found that the isothermal beta description of the cluster with $\beta=0.78\pm0.08$ and
$r_c=16\arcsec\pm15\arcsec$ yields a consistent mass profile with
that from weak-lensing.

The comparison of the weak-lensing mass map with the X-ray contours provides an invaluable 
opportunity to examine how differently these two cluster constituents are distributed within the cluster.
The relatively circular distribution of the ICM 
in contrast to the rather elongated arrangement of cluster galaxies
is consistent with the expectation that the ICM has reached a higher degree
of virialization than the cluster galaxies and dark matter. On the other hand,
the distribution of the cluster dark matter looks somewhat similar to that
of the cluster's optical luminosity. 

Apart from the offsets between the two X-ray peaks and the corresponding dark matter clumps,
as first observed in our previous study of CL 0152-1357, we note that
there present offsets between the dark matter and
the cluster galaxy centroids in MS 1054-0321.
Through bootstrap resampling experiments,we demonstrate that the
offsets between the dark matter and cluster galaxy centroids are statistically
significant. We argue that it is unlikely that the observed offsets are mainly
caused either by the different smoothing scale of the light map or the incompleteness
of the spectroscopic survey catalog. It is possible that any significant foreground
mass lying close to the mass peaks can potentially affect the mass centroids.
Although we cannot prove with current data that this possibility is completely ruled out 
using the so-called weak-lensing tomography, the analysis of the current spectroscopic catalog
does not support any presence of such foreground groups nearby the mass centroids.
In addition, it is hard to imagine that all the four mass peaks in MS 1054-0321 and CL 0152-1357
coincidently possess close neighbors in projection, which all cause the centroids
to look shifted toward the hypothesized merging direction.
Therefore, we suggest that the observed offsets between the mass and galaxy centroids might
reflects the real features, indicative of a merger between the substructures.

The eastern weak-lensing clump is not detected in X-rays, and possible
scenarios have been discussed. Because the 
eastern mass peak was also clearly detected by HFK00 even with a different selection of
background galaxies and the overdensity of the cluster galaxy is present at the location,
we rule out the possibility that it is an
artifact of the mass reconstruction.
As the estimated X-ray temperature ($\sim3.5$ keV) of the unseen eastern clump
based on the projected mass is sufficient for thermal bremsstrahlung radiation,
it is not clear whether the lack of the detection is entirely due to
the immaturity of the subcluster. Alternatively,
we suspect that the ICM of the eastern clump might
have been substantially stripped off during its merger with the cluster main body.
The first indication of this scenario comes from the temperature structure
resembling the propagation of the shock-heated region
along the merger axis observed in numerical simulations.

If the subcluster had passed through the dense region of the central clump
from the southwest, we suspect that it is possible to observe this kind of
temperature gradient increasing in opposite directions. 
Although this interpretation of the temperature map is not unique, the
scenario is further supported by the rather anomalous X-ray profile
of the central X-ray peak. Even after the western substructure
is carefully removed, the remaining structure cannot be well described
by an isothermal beta model because the core region is too diffuse. This
may suggest that the ICM structure might have been recently disrupted and
we are just observing the transient state prior to
its returning to relaxed distribution.

ACS was developed under NASA contract NAS5-32865, and this research was supported
by NASA grant NAG5-7697. We are grateful for an equipment
grant from Sun Microsystems, Inc. We thank the anonymous referee
for constructive criticisms and a careful reading of the manuscript.

\clearpage

\clearpage

\clearpage

\clearpage

\clearpage

\clearpage

\clearpage

\clearpage

\clearpage

\clearpage

\clearpage

\clearpage

\clearpage

\clearpage

\clearpage

\clearpage

\clearpage

\clearpage

\clearpage

\clearpage

\clearpage

\clearpage

\clearpage 

\begin{figure}
\plotone{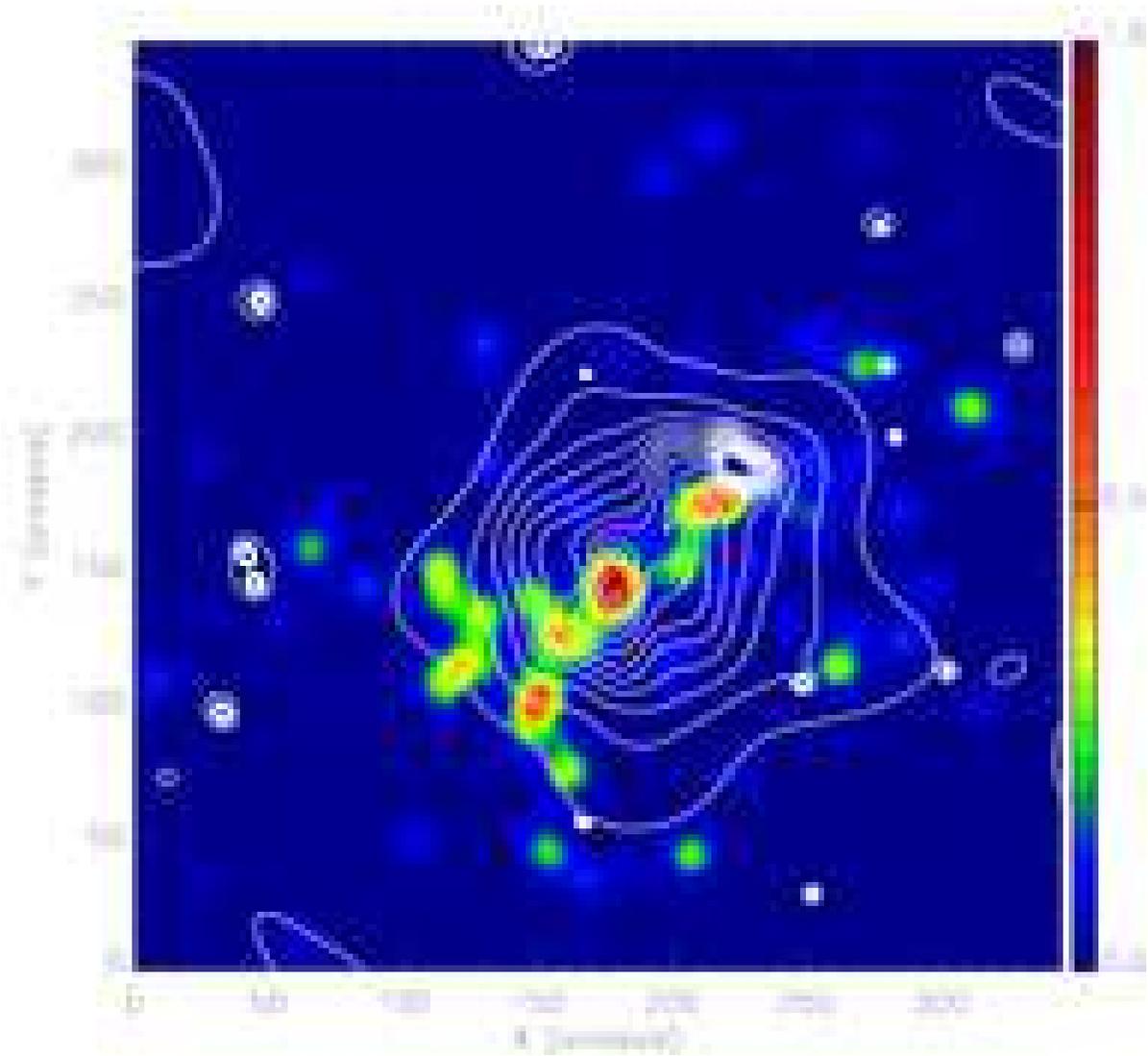}
\caption{X-ray contours overlaid on the luminosity map. The alignment is precise within $1\arcsec$.
The two X-ray peaks are separated further than the corresponding cluster galaxies. The eastern
substructure (here shown by the cluster galaxies) is not visible in the X-ray image.
\label{fig_xray_opt}}
\end{figure}
\clearpage

\clearpage

\clearpage

\clearpage

\begin{deluxetable}{cccc}

\tablecaption{PROPERTIES OF THE SUBSTRUCTURES}

\tablenum{1}

\tablehead{\colhead{Clump} & \colhead{Coordinate} & \colhead{Mass ($r<30\arcsec$)}\tablenotemark{a}
\tablenotetext{a}{Mass uncertainty is calculated using the rms map (Figure~\ref{fig_err_map}) under the assumption 
that the mass pixels within the aperture are entirely correlated. We do not include
the cosmic shear effect discussed in \textsection~\ref{section_cosmic_shear}.}
 & \colhead{$M/L_B$} \\
\colhead{} & \colhead{(RA,DEC)} & \colhead{($10^{14} M_{\sun}$)} & \colhead{($M_{\sun}/L_{B\sun}$)} \\ }
\tablewidth{0pt}

\startdata
W  & $10^h56^m56^s.7$,$-03\degr37\arcmin43\arcsec.0$ & $1.31\pm0.14$ & $219\pm23$ \\
C  & $10^h56^m59^s.7$,$-03\degr37\arcmin38\arcsec.4$ & $1.23\pm0.14$ & $144\pm16$ \\
E  & $10^h57^m03^s.8$,$-03\degr37\arcmin16\arcsec.3$ & $0.93\pm0.15$ & $79\pm13$  \\
M1 & $10^h57^m06^s.2$,$-03\degr38\arcmin12\arcsec.3$ & $0.41\pm0.15$ & $137\pm50$  \\
M2 & $10^h57^m03^s.6$,$-03\degr38\arcmin46\arcsec.0$ & $0.44\pm0.15$ & $290\pm99$  \\
M3 & $10^h56^m57^s.9$,$-03\degr39\arcmin03\arcsec.9$ & $0.48\pm0.15$ & $219\pm67$  \\
M4 & $10^h56^m51^s.1$,$-03\degr38\arcmin37\arcsec.7$ & $0.36\pm0.16$ & $202\pm89$ \\
\enddata

\end{deluxetable}

\begin{deluxetable}{ccc}

\tablecaption{STUDIES OF MS 1054-0321 TEMPERATURE}

\tablenum{2}

\tablehead{\colhead{Study} & \colhead{Temperature (keV)} & \colhead{Instrument} }
\tablewidth{0pt}

\startdata
Donahue et al. (1998)  & $12.3_{-2.2}^{+3.1}$  & $ASCA$ and $ROSAT$  \\
Jeltema et al. (2001) & $10.4_{-1.5}^{+1.7}$ & $Chandra$  \\
Joy et al. (2001)  & $10.4_{-2}^{+5}$  & SZ analysis  \\
Vikhlinin et al. (2002) & $7.8\pm0.6$  & $Chandra$ \\
Tozzi et al. (2003) & $8.0\pm0.5$  & $Chandra$ \\
Gioia et al. (2004) & $7.2_{-0.6}^{+0.7}$ $(7.4_{-0.9}^{+1.4})$  & XMM-$Newton$ ($Chandra$)   \\
This study & $ 8.9_{-0.8}^{+1.0}$  & $Chandra$  \\
\enddata
\end{deluxetable}

\begin{deluxetable}{ccccccc}
\tabletypesize{\scriptsize}

\tablecaption{X-RAY TEMPERATURES OF MS 1054-0321}

\tablenum{3}

\tablehead{\colhead{Region} & \colhead{Center} & \colhead{Radius} &
\colhead{Temperature} & \colhead{Abundance} & \colhead{Reduced $\chi^2$} & \colhead{DOF} \\
\colhead{} & \colhead{(RA,DEC)} & \colhead{(\arcsec)}
 & \colhead{(KeV)} & \colhead{($Z/Z_{\sun}$)} & \colhead{} & \colhead{} \\  }
\tablewidth{0pt}

\startdata
Cluster as a whole & $10^h56^m58^s.6$,$-03\degr37\arcmin36\arcsec.7$ & $90$ & $8.9_{-0.8}^{+1.0}$ & $0.30_{-0.12}^{+0.12}$ & 0.63 & 206 \\
Central peak       & $10^h57^m00^s.2$,$-03\degr37\arcmin39\arcsec.6$ & $24$ & $10.7_{-1.7}^{+2.1}$ & $0.16_{-0.16}^{+0.19}$ & 0.72 & 81  \\
Western peak       & $10^h56^m55^s.6$,$-03\degr37\arcmin42\arcsec.5$ & $24$ & $7.5_{-1.2}^{+1.4}$ & $0.47_{-0.23}^{+0.24}$ & 0.75 & 68  \\
\enddata
\end{deluxetable}


\begin{thebibliography}{}
\bibitem[Anderson(2002)]{anderson2002} Anderson, J. 2002 HST Calibration Workshop, Eds. S. Arribas,
  A. M. Koekemoer, B. Whitmore (STScI: Baltimore), p. 13
\bibitem[Arnaud \& Evrard(1999)]{1999MNRAS.305..631A} Arnaud, M., \& 
Evrard, A.~E.\ 1999, \mnras, 305, 631 
\bibitem[Bahcall \& Comerford(2002)]{2002ApJ...565L...5B} Bahcall, N.~A., 
\& Comerford, J.~M.\ 2002, \apjl, 565, L5 
\bibitem[Bahcall \& Lubin(1994)]{1994ApJ...426..513B} Bahcall, N.~A., \& 
Lubin, L.~M.\ 1994, \apj, 426, 513 
\bibitem[Bahcall \& Fan(1998)]{bf} Bahcall, N.~A.~\& Fan, X.\ 1998, \apj, 504, 1
\bibitem[Bartelmann(1995)]{1995A&A...303..643B} Bartelmann, M.\ 1995, \aap, 
303, 643 
\bibitem[Beckwith, Somerville, \& Stiavelli(2003)]{beckwith03} Beckwidth, S., Somerville, R., Stiavelli M., 2003, STScI Newsletter vol 20 issue 04
\bibitem[Ben{\'{\i}}tez et al.(2004)]{benitez04} Ben{\'{\i}}tez, N., et al.\ 2004, \apjs, 150,
\bibitem[Bekenstein(2004)]{2004PhRvD..70h3509B} Bekenstein, J.~D.\ 2004, 
\prd, 70, 083509 
\bibitem[Blakeslee et al.(2003)]{blakeslee03} Blakeslee, J.~P., Anderson, K.~R., Meurer, G.~R., Ben{\'{\i}}tez, N., \& Magee, D.\ 2003, ASP Conf.~Ser.~295: Astronomical Data Analysis Software and Systems XII, 12, 257
\bibitem[Bernstein \& Jarvis(2002)]{bj02} Bernstein, G.~M.~\& Jarvis, M.\ 2002, \aj, 123, 583 (BJ02)
\bibitem[Bertin \& Arnouts(1996)]{1996A&AS..117..393B} Bertin, E., \& 
Arnouts, S.\ 1996, \aaps, 117, 393 
\bibitem[Bruzual \& Charlot(2003)]{2003MNRAS.344.1000B} Bruzual, G.~\& 
Charlot, S.\ 2003, \mnras, 344, 1000 
\bibitem[Carlberg, Yee, \& Ellingson(1997)]{cye97} Carlberg, R.~G., Yee, H.~K.~C., \& Ellingson, E.\ 1997, \apj, 478, 462
\bibitem[Chartas \& Getman(2002)]{cg02} Chartas, G. \& Getman, K. 2002, http://www.astro.psu.edu/users/chartas/xcontdir/xcont.html 
\bibitem[Clowe, Luppino, Kaiser, \& Gioia(2000)]{2000ApJ...539..540C} 
Clowe, D., Luppino, G.~A., Kaiser, N., \& Gioia, I.~M.\ 2000, \apj, 539, 
540 

\bibitem[Clowe et al.(2003)]{2003A&A...409..851C} Clowe, D., Luppino, 
G.~A., \& Kaiser, N.\ 2003, \aap, 409, 851 
\bibitem[Clowe et al.(2004)]{2004ApJ...604..596C} Clowe, D., Gonzalez, A., 
\& Markevitch, M.\ 2004, \apj, 604, 596 
\bibitem[Coleman, Wu, \& Weedman(1980)]{cww80} Coleman, G.~D., Wu, C.-C., \& Weedman, D.~W.\ 1980, \apjs, 43, 393
\bibitem[Dickey \& Lockman(1990)]{1990ARA&A..28..215D} Dickey, J.~M.~\& 
Lockman, F.~J.\ 1990, \araa, 28, 215 
\bibitem[Donahue et al.(1998)]{1998ApJ...502..550D} Donahue, M., Voit, 
G.~M., Gioia, I., Lupino, G., Hughes, J.~P., \& Stocke, J.~T.\ 1998, \apj, 
502, 550
\bibitem[Evrard et al.(1996)]{1996ApJ...469..494E} Evrard, A.~E., Metzler, 
C.~A., \& Navarro, J.~F.\ 1996, \apj, 469, 494  
\bibitem[Fischer \& Tyson(1997)]{ft97} Fischer, P.~\& Tyson, J.~A.\ 1997, \aj, 114, 14
\bibitem[Gioia et al.(1990)]{1990ApJ...356L..35G} Gioia, I.~M., Henry, 
J.~P., Maccacaro, T., Morris, S.~L., Stocke, J.~T., \& Wolter, A.\ 1990, 
\apjl, 356, L35 
\bibitem[Gioia \& Luppino(1994)]{1994ApJS...94..583G} Gioia, I.~M.~\& 
Luppino, G.~A.\ 1994, \apjs, 94, 583 
\bibitem[Gioia et al.(2004)]{2004A&A...419..517G} Gioia, I.~M., Braito, V., 
Branchesi, M., Della Ceca, R., Maccacaro, T., \& Tran, K.-V.\ 2004, \aap, 
419, 517 
\bibitem[Goto et al.(2005)]{2005ApJ...621..188G} Goto, T., et al.\ 2005, 
\apj, 621, 188 
\bibitem[Henry et al.(1992)]{1992ApJ...386..408H} Henry, J.~P., Gioia, 
I.~M., Maccacaro, T., Morris, S.~L., Stocke, J.~T., \& Wolter, A.\ 1992, 
\apj, 386, 408 
\bibitem[Hoekstra, Franx, \& Kuijken(2000)]{2000ApJ...532...88H} Hoekstra, 
H., Franx, M., \& Kuijken, K.\ 2000, \apj, 532, 88 
\bibitem[Hoekstra(2001)]{2001A&A...370..743H} Hoekstra, H.\ 2001, \aap, 
370, 743 
\bibitem[Holden et al.(2005)]{2005ApJ...620L..83H} Holden, B.~P., et al.\ 
2005, \apjl, 620, L83 
\bibitem[Jeltema et al.(2001)]{2001ApJ...562..124J} Jeltema, T.~E., 
Canizares, C.~R., Bautz, M.~W., Malm, M.~R., Donahue, M., \& Garmire, 
G.~P.\ 2001, \apj, 562, 124 
\bibitem[Jee et al.(2005)]{2005ApJ...618...46J} Jee, M.~J., White, R.~L., 
Ben{\'{\i}}tez, N., Ford, H.~C., Blakeslee, J.~P., Rosati, P., Demarco, R., 
\& Illingworth, G.~D.\ 2005, \apj, 618, 46 
\bibitem[Joy et al.(2001)]{joy01} Joy, M., et al.\ 2001, \apjl, 551, L1
\bibitem[Kaastra \& Mewe(1993)]{1993A&AS...97..443K} Kaastra, J.~S.~\& 
Mewe, R.\ 1993, \aaps, 97, 443  
\bibitem[Kaiser(1992)]{1992ApJ...388..272K} Kaiser, N.\ 1992, \apj, 388, 
272 
\bibitem[Kaiser(1998)]{1998ApJ...498...26K} Kaiser, N.\ 1998, \apj, 498, 26 
\bibitem[Kaiser(2000)]{kaiser00} Kaiser, N.\ 2000, \apj, 537, 555
\bibitem[Katgert, Biviano, \& Mazure(2004)]{kbm04} Katgert, P., Biviano, A., \& Mazure, A.\ 2004, \apj, 600, 657
\bibitem[Kinney et al.(1996)]{kinney96} Kinney, A.~L., Calzetti, D., Bohlin, R.~C., McQuade, K., Storchi-Bergmann, T., \& Schmitt, H.~R.\ 1996, \apj, 467, 38
\bibitem[Krist(2003)]{krist03} Krist, J.\ 2003, $Instrument$ $Science$ $Report$ $ACS$ 2003-06
\bibitem[Liedahl, Osterheld, \& Goldstein(1995)]{1995ApJ...438L.115L} 
Liedahl, D.~A., Osterheld, A.~L., \& Goldstein, W.~H.\ 1995, \apjl, 438, 
L115 
\bibitem[Lombardi et al.(2005)]{2005astro.ph..1150L} Lombardi, M., et al.\ 
2005, \apj, in press 
\bibitem[Luppino \& Kaiser(1997)]{1997ApJ...475...20L} Luppino, G.~A.~\& 
Kaiser, N.\ 1997, \apj, 475, 20 
\bibitem[Markevitch et al.(2002)]{2002ApJ...567L..27M} Markevitch, M., 
Gonzalez, A.~H., David, L., Vikhlinin, A., Murray, S., Forman, W., Jones, 
C., \& Tucker, W.\ 2002, \apjl, 567, L27 
\bibitem[Markevitch(2003)]{m03} Markevitch, M 2003, http://hea-www.harvard.edu/$\sim$maxim/axaf/acisbg/
\bibitem[Markevitch et al.(2004)]{2004ApJ...606..819M} Markevitch, M., 
Gonzalez, A.~H., Clowe, D., Vikhlinin, A., Forman, W., Jones, C., Murray, 
S., \& Tucker, W.\ 2004, \apj, 606, 819  
\bibitem[Marshall et al.(2002)]{2002MNRAS.335.1037M} Marshall, P.~J., 
Hobson, M.~P., Gull, S.~F., \& Bridle, S.~L.\ 2002, \mnras, 335, 1037 
\bibitem[Maughan et al.(2003)]{2003ApJ...587..589M} Maughan, B.~J., Jones, 
L.~R., Ebeling, H., Perlman, E., Rosati, P., Frye, C., \& Mullis, C.~R.\ 
2003, \apj, 587, 589 
\bibitem[McGaugh \& de Blok(1998)]{1998ApJ...499...66M} McGaugh, S.~S.~\& 
de Blok, W.~J.~G.\ 1998, \apj, 499, 66 
\bibitem[Meurer et al.(2003)]{meurer03} Meurer, G.~R., et al.\ 2003, \procspie, 4854, 507
\bibitem[Milgrom(1983)]{1983ApJ...270..365M} Milgrom, M.\ 1983, \apj, 270, 
365 
\bibitem[Milgrom \& Sanders(2003)]{2003ApJ...599L..25M} Milgrom, M.~\& 
Sanders, R.~H.\ 2003, \apjl, 599, L25 
\bibitem[Neumann \& Arnaud(2000)]{2000ApJ...542...35N} Neumann, D.~M.~\& 
Arnaud, M.\ 2000, \apj, 542, 35 
\bibitem[Pavlovsky et al(2004)]{pavlovsky04} Pavlovsky, C., et al. 2004, ACS Instrument Handbook, Version 5.0,
(Baltimore: STScI)
\bibitem[Peacock \& Dodds(1996)]{1996MNRAS.280L..19P} Peacock, J.~A., \& 
Dodds, S.~J.\ 1996, \mnras, 280, L19 
\bibitem[Postman et al.(2005)]{2005ApJ...623..721P} Postman, M., et al.\ 
2005, \apj, 623, 721 
\bibitem[Refregier(2003)]{refregier03} Refregier, A.\ 2003, \mnras, 338, 35 
\bibitem[Ricker(1998)]{1998ApJ...496..670R} Ricker, P.~M.\ 1998, \apj, 496, 
670 
\bibitem[Ricker \& Sarazin(2001)]{2001ApJ...561..621R} Ricker, P.~M., \& 
Sarazin, C.~L.\ 2001, \apj, 561, 621 
\bibitem[Rowley et al.(2004)]{2004MNRAS.352..508R} Rowley, D.~R., Thomas, 
P.~A., \& Kay, S.~T.\ 2004, \mnras, 352, 508 
\bibitem[Sanders(2003)]{2003MNRAS.342..901S} Sanders, R.~H.\ 2003, \mnras, 
342, 901 
\bibitem[Scharf et al.(2004)]{2004HEAD....8.0102S} Scharf, C.~A., Zurek, 
D., \& Bureau, M.\ 2004, AAS/High Energy Astrophysics Division, 8 
\bibitem[Schindler \& Mueller(1993)]{1993A&A...272..137S} Schindler, S., \& 
Mueller, E.\ 1993, \aap, 272, 137 
\bibitem[Schlegel, Finkbeiner, \& Davis(1998)]{sfd98} Schlegel, D.~J., Finkbeiner, D.~P., \& Davis, M.\ 1998, \apj, 500, 525
\bibitem[Schneider et al.(1998)]{1998MNRAS.296..873S} Schneider, P., van 
Waerbeke, L., Jain, B., \& Kruse, G.\ 1998, \mnras, 296, 873 
\bibitem[Stanford et al.(2001)]{2001ApJ...552..504S} Stanford, S.~A., 
Holden, B., Rosati, P., Tozzi, P., Borgani, S., Eisenhardt, P.~R., \& 
Spinrad, H.\ 2001, \apj, 552, 504 
\bibitem[Takizawa(1999)]{1999ApJ...520..514T} Takizawa, M.\ 1999, \apj, 
520, 514 
\bibitem[Takizawa(2000)]{2000ApJ...532..183T} Takizawa, M.\ 2000, \apj, 
532, 183 
\bibitem[Tozzi et al.(2003)]{2003ApJ...593..705T} Tozzi, P., Rosati, P., 
Ettori, S., Borgani, S., Mainieri, V., \& Norman, C.\ 2003, \apj, 593, 705 
\bibitem[Tran et al.(1999)]{1999ApJ...522...39T} Tran, K.~H., Kelson, 
D.~D., van Dokkum, P., Franx, M., Illingworth, G.~D., \& Magee, D.\ 1999, 
\apj, 522, 39 
\bibitem[van Dokkum et al.(2000)]{2000ApJ...541...95V} van Dokkum, P.~G., 
Franx, M., Fabricant, D., Illingworth, G.~D., \& Kelson, D.~D.\ 2000, \apj, 
541, 95 
\bibitem[van Dokkum \& Stanford (2003)]{vs03} van Dokkum, P.~G.~\& Stanford, S.~A.\ 2003, \apj, 585, 78

\bibitem[Vikhlinin et al.(2002)]{2002ApJ...578L.107V} Vikhlinin, A., 
VanSpeybroeck, L., Markevitch, M., Forman, W.~R., \& Grego, L.\ 2002, 
\apjl, 578, L107 
\bibitem[Willick(2000)]{2000ApJ...530...80W} Willick, J.~A.\ 2000, \apj, 
530, 80 
\bibitem[Wu et al.(1998)]{wu98} Wu, X., Chiueh, T., Fang, L., \& Xue, Y.\ 1998, \mnras, 301, 861


\end{thebibliography}
\end{document}